%% file: ms.tex
\pdfoutput=1
\documentclass[12pt]{article}

\usepackage[T1]{fontenc}
\usepackage[utf8]{inputenc}
\usepackage[english]{babel}

\usepackage{amsmath}
\usepackage{amssymb}
\usepackage{amsthm}
\usepackage{mathtools}
\mathtoolsset{showonlyrefs,showmanualtags}

\allowdisplaybreaks

\usepackage[natbibapa,nodoi]{apacite}
\usepackage{authblk}

\usepackage{bbm}
\newcommand{\bbone}{\mathbbm{1}}
\usepackage{booktabs}
\usepackage[format=plain,font=small,labelfont=bf,textfont=up]{caption}
\usepackage{enumitem}
\setlist[enumerate]{itemsep=0.1ex}
\setlist[itemize]{itemsep=0.1ex}
\newcommand\expcommand\newcommand
\usepackage[letterpaper,margin=1.05in,bottom=1.45in]{geometry}
\usepackage{graphicx}
\usepackage[hidelinks]{hyperref}
\usepackage[nodisplayskipstretch]{setspace}
\usepackage{subfigure}
\usepackage{makecell}

\usepackage{blkarray}
\usepackage[linesnumbered,ruled,vlined]{algorithm2e}
\usepackage{xcolor}

\usepackage[nohints,tight]{minitoc}

\newcommand\mainref\ref
\newcommand\suppref\ref

\newcommand{\simspath}{./sims}

\input{stddef.tex}

\input{\texpath/commands.tex}

\title{\textbf{A General Design-Based Framework and Estimator for Randomized Experiments}}

\author[1]{Christopher Harshaw}
\author[2]{Fredrik Sävje}
\author[3]{Yitan Wang}

\affil[1]{Columbia University}
\affil[2]{Uppsala University}
\affil[3]{Yale University}

\date{\today}

\begin{document}

\makeatletter%
\gdef\@thefnmark{}\@footnotetext{\hspace{-1em}%
	The authors thank
	P.M. Aronow,
	Haoge Chang,
	Xiaohong Chen,
	Jon Erickson,
	Avi Feller,
	Guido Imbens,
	Laurent Lessard,
	Joel Middleton,
	Benjamin Recht,
	Jamie Robins,
	Dominik Rothenhaeusler,
	Rahul Singh,
	Daniel A.\ Spielman,
	Vasilis Syrgkanis,
	Johan Ugander,
	Stefan Wager,
	and Angela Zhou
	for insightful discussions which helped to shape this work.
	Authors were partially supported by NSF award No.\ 2316335, ONR Award N00014-20-1-2335, Swedish Research Council grant agreement No.\ 2022-06725, and Tom Hedelius \& Tore Browaldh foundations under Grant No.\ P25-0067, and the Simons Institute for Theory of Computing.%
}%
\makeatother%

\begin{singlespace}
\maketitle

\begin{abstract}
\noindent\input{\texpath/abstract.tex}
\end{abstract}
\end{singlespace}

\doparttoc
\faketableofcontents

\clearpage

\input{\texpath/introduction.tex}

\input{\texpath/overview.tex}

\input{\texpath/previous-work.tex}

\input{\texpath/framework.tex}

\input{\texpath/estimator.tex}

\input{\texpath/consistency.tex}

\input{\texpath/estvar.tex}

\input{\texpath/illustration-total.tex}

\input{\texpath/conclusions.tex}

\begin{singlespace}
\bibliography{ms}
\end{singlespace}

\clearpage
\newcommand{\suppletter}{S}
\newcommand{\supptitle}{Supplement}
\newcommand{\supplabel}{supp:proofs}

\renewcommand{\numberline}[1]{#1\ \ \ }

\mtcaddpart[Supplement]

\newcommand\resetsuppcounters{%
\setcounter{section}{0}%
\setcounter{table}{0}%
\setcounter{figure}{0}%
\setcounter{equation}{0}%
\setcounter{theorem}{0}%
\setcounter{assumption}{0}%
\setcounter{condition}{0}%
\setcounter{definition}{0}%
\setcounter{remark}{0}%
\setcounter{example}{0}%
\setcounter{algocf}{0}%
}

\newcommand\updatesuppcounters[1]{%
\renewcommand\thesection{#1\arabic{section}}%
\renewcommand\thetable{#1\arabic{table}}%
\renewcommand\thefigure{#1\arabic{figure}}%
\renewcommand\theequation{#1\arabic{equation}}%
\renewcommand\theconjecture{#1\arabic{theorem}}%
\renewcommand\thecorollary{#1\arabic{theorem}}%
\renewcommand\thelemma{#1\arabic{theorem}}
\renewcommand\theproposition{#1\arabic{theorem}}
\renewcommand\thetheorem{#1\arabic{theorem}}
\renewcommand\theassumption{#1\arabic{assumption}}
\renewcommand\thecondition{#1\arabic{condition}}
\renewcommand\thedefinition{#1\arabic{definition}}
\renewcommand\theremark{#1\arabic{remark}}
\renewcommand\theexample{#1\arabic{example}}
\renewcommand\thealgocf{#1\arabic{algocf}}
}

\newcommand\updatesuppanchors[1]{
\renewcommand\theHsection{#1.\arabic{section}}
\renewcommand\theHtable{#1.\arabic{table}}
\renewcommand\theHfigure{#1.\arabic{figure}}
\renewcommand\theHequation{#1.\arabic{equation}}
\renewcommand\theHtheorem{#1.\arabic{theorem}}
\renewcommand\theHassumption{#1.\arabic{assumption}}
\renewcommand\theHcondition{#1.\arabic{condition}}
\renewcommand\theHdefinition{#1.\arabic{definition}}
\renewcommand\theHremark{#1.\arabic{remark}}
\renewcommand\theHexample{#1.\arabic{example}}
\renewcommand\theHalgocf{#1.\arabic{algocf}}
}

\resetsuppcounters
\updatesuppanchors{\suppletter}
\updatesuppcounters{\suppletter}

\makeatletter
\immediate\write\@auxout{\string\newlabel{\supplabel}{{\suppletter}{\thepage}{\supptitle}{supplement.\suppletter}{}}}
\makeatother
\hypertarget{supplement.\suppletter}{}

\begin{center}
	\LARGE \textbf{Supplement}
\end{center}
\bigskip

\parttoc

\input{\texpath/supp-illustrative-worked-examples}

\input{\texpath/supp-general-constructions}

\input{\texpath/supp-general-theoretical-results}

\input{\texpath/supp-simulations}

\end{document}

%% file: stddef.tex
\DeclarePairedDelimiter\paren\lparen\rparen
\DeclarePairedDelimiter\bracket\lbrack\rbrack
\DeclarePairedDelimiter\braces\lbrace\rbrace

\DeclarePairedDelimiter\abs\lvert\rvert

\providecommand{\bbone}{\mathbf{1}}
\DeclarePairedDelimiterXPP\indicator[1]{\bbone}{\lbrack}{\rbrack}{}{#1}

\DeclarePairedDelimiterXPP\expf[1]{\exp}{\lparen}{\rparen}{}{#1}
\DeclarePairedDelimiterXPP\logf[1]{\log}{\lparen}{\rparen}{}{#1}
\DeclarePairedDelimiterXPP\maxf[1]{\max}{\lparen}{\rparen}{}{#1}
\DeclarePairedDelimiterXPP\minf[1]{\min}{\lparen}{\rparen}{}{#1}

\DeclarePairedDelimiterXPP\func[2]{#1}{\lparen}{\rparen}{}{#2}

\DeclarePairedDelimiter\setb\lbrace\rbrace

\newcommand{\Reals}{\mathbb{R}}

\newcommand{\Naturals}{\mathbb{N}}

\DeclareMathOperator*{\argmin}{arg\,min}
\DeclarePairedDelimiter\card\lvert\rvert

\newcommand{\mat}[1]{\boldsymbol{#1}}
\renewcommand{\vec}[1]{\boldsymbol{#1}}

\makeatletter
\newcommand*{\tran}{{\mathpalette\@tran{}}}
\newcommand*{\@tran}[2]{\raisebox{\depth}{$\m@th#1\intercal$}}
\makeatother

\DeclarePairedDelimiter\iprod\langle\rangle

\DeclarePairedDelimiter\norm\lVert\rVert
\DeclarePairedDelimiterXPP\enorm[1]{}{\lVert}{\rVert_{2}}{}{#1}
\DeclarePairedDelimiterXPP\inorm[1]{}{\lVert}{\rVert_{\infty}}{}{#1}
\DeclarePairedDelimiterXPP\pnorm[2]{}{\lVert}{\rVert_{#1}}{}{#2}
\DeclareMathOperator{\opsym}{op}
\DeclarePairedDelimiterXPP\opnorm[1]{}{\lVert}{\rVert_{\opsym}}{}{#1}

\DeclarePairedDelimiterXPP\detf[1]{\det}{\lparen}{\rparen}{}{#1}

\DeclarePairedDelimiterXPP\kerf[1]{\ker}{\lparen}{\rparen}{}{#1}

\DeclareMathOperator{\trsym}{tr}
\DeclarePairedDelimiterXPP\tr[1]{\trsym}{\lparen}{\rparen}{}{#1}

\DeclareMathOperator{\vspansym}{span}
\DeclarePairedDelimiterXPP\vspan[1]{\vspansym}{\lparen}{\rparen}{}{#1}

\DeclareMathOperator{\diagsym}{diag}
\DeclarePairedDelimiterXPP\diag[1]{\diagsym}{\lparen}{\rparen}{}{#1}

\DeclareMathOperator{\ranksym}{rank}
\DeclarePairedDelimiterXPP\rank[1]{\ranksym}{\lparen}{\rparen}{}{#1}

\DeclareMathOperator{\vectorizesym}{vec}
\DeclarePairedDelimiterXPP\vectorize[1]{\vectorizesym}{\lparen}{\rparen}{}{#1}

\DeclareMathOperator*{\esssup}{ess\,sup}
\DeclarePairedDelimiterXPP\esssupf[1]{\esssup}{\lparen}{\rparen}{}{#1}

\let\Prsym\Pr
\let\Pr\relax
\DeclarePairedDelimiterXPP\Pr[1]{\Prsym}{\lparen}{\rparen}{}{%
	#1}

\DeclarePairedDelimiterXPP\Prsub[2]{\Prsym_{#1}}{\lparen}{\rparen}{}{%
	#2}

\DeclareMathOperator{\Esym}{E}
\DeclarePairedDelimiterXPP\E[1]{\Esym}{\lbrack}{\rbrack}{}{%
	#1}

\DeclarePairedDelimiterXPP\Esub[2]{\Esym_{#1}}{\lbrack}{\rbrack}{}{%
	#2}

\DeclareMathOperator{\Varsym}{Var}
\DeclarePairedDelimiterXPP\Var[1]{\Varsym}{\lparen}{\rparen}{}{%
	#1}

\DeclarePairedDelimiterXPP\Varsub[2]{\Varsym_{#1}}{\lparen}{\rparen}{}{%
	#2}

\DeclarePairedDelimiterXPP\EstVar[1]{\widehat{\Varsym}}{\lparen}{\rparen}{}{%
	#1}

\DeclareMathOperator{\Covsym}{Cov}
\DeclarePairedDelimiterXPP\Cov[1]{\Covsym}{\lparen}{\rparen}{}{%
	#1}

\DeclarePairedDelimiterXPP\Covsub[2]{\Covsym_{#1}}{\lparen}{\rparen}{}{%
	#2}

\DeclareMathOperator{\Corrsym}{Corr}
\DeclarePairedDelimiterXPP\Corr[1]{\Corrsym}{\lparen}{\rparen}{}{%
	#1}

\makeatletter
\newcommand{\indep}{\protect\mathpalette{\protect\@indep}{\perp}}
\newcommand*{\@indep}[2]{\mathrel{\rlap{$#1#2$}\mkern3mu{#1#2}}}
\makeatother

\newcommand{\bigOsym}{\mathcal{O}}
\DeclarePairedDelimiterXPP\bigO[1]{\bigOsym}{\lparen}{\rparen}{}{#1}

\newcommand{\littleOsym}{o}
\DeclarePairedDelimiterXPP\littleO[1]{\littleOsym}{\lparen}{\rparen}{}{#1}

\newcommand{\bigOpsym}{\bigOsym_p}
\DeclarePairedDelimiterXPP\bigOp[1]{\bigOpsym}{\lparen}{\rparen}{}{#1}

\newcommand{\littleOpsym}{\littleOsym_p}
\DeclarePairedDelimiterXPP\littleOp[1]{\littleOpsym}{\lparen}{\rparen}{}{#1}

\newcommand{\bigOmegasym}{\Omega}
\DeclarePairedDelimiterXPP\bigOmega[1]{\bigOmegasym}{\lparen}{\rparen}{}{#1}

\newcommand{\littleOmegasym}{\omega}
\DeclarePairedDelimiterXPP\littleOmega[1]{\littleOmegasym}{\lparen}{\rparen}{}{#1}

\newcommand{\bigThetasym}{\Theta}
\DeclarePairedDelimiterXPP\bigTheta[1]{\bigThetasym}{\lparen}{\rparen}{}{#1}

\newcommand{\sumin}{\sum_{i=1}^n}
\newcommand{\sumjn}{\sum_{j=1}^n}

\newcommand{\quadtext}[1]{\quad\text{#1}\quad}
\newcommand{\qquadtext}[1]{\qquad\text{#1}\qquad}

\newcommand{\quadand}{\quadtext{and}}
\newcommand{\qquadand}{\qquadtext{and}}

\theoremstyle{plain}
\newtheorem{theorem}{Theorem}[section]

\newtheorem{corollary}[theorem]{Corollary}
\newtheorem{lemma}[theorem]{Lemma}
\newtheorem{proposition}[theorem]{Proposition}

\newtheorem{innerrefcorollary}{Corollary}
\newenvironment{refcorollary}[1]
{\renewcommand\theinnerrefcorollary{#1}\innerrefcorollary}
{\endinnerrefcorollary}

\newtheorem{innerreflemma}{Lemma}
\newenvironment{reflemma}[1]
{\renewcommand\theinnerreflemma{#1}\innerreflemma}
{\endinnerreflemma}

\newtheorem{innerreftheorem}{Theorem}
\newenvironment{reftheorem}[1]
{\renewcommand\theinnerreftheorem{#1}\innerreftheorem}
{\endinnerreftheorem}

\theoremstyle{definition}
\newtheorem{assumption}{Assumption}

\newtheorem{definition}{Definition}

%% file: tex/commands.tex
\newcommand{\est}{\widehat{\tau}}
\DeclarePairedDelimiterXPP\estv[1]{\est}{\lparen}{\rparen}{}{#1}
\newcommand{\ate}{\tau}
\newcommand{\ite}[1]{\tau_{#1}}
\newcommand{\itei}{\ite{i}}

\newcommand{\Zspace}{\mathcal{Z}}
\newcommand{\Zalgebra}{\Sigma}
\newcommand{\Zprob}{P}

\newcommand{\offpol}{\mathcal{Q}}

\newcommand{\Lts}{\mathcal{L}^2}
\newcommand{\Lt}{L^2}
\DeclarePairedDelimiter\ec\lbrack\rbrack

\newcommand{\po}[1]{y_{#1}}
\newcommand{\poi}{\po{i}}
\newcommand{\poj}{\po{j}}
\DeclarePairedDelimiterXPP\pov[2]{\po{#1}}{\lparen}{\rparen}{}{#2}
\DeclarePairedDelimiterXPP\povi[1]{\poi}{\lparen}{\rparen}{}{#1}

\newcommand{\oo}[1]{Y_{#1}}
\newcommand{\ooi}{\oo{i}}
\newcommand{\ooj}{\oo{j}}

\newcommand{\Mspace}[1]{\mathcal{M}_{#1}}
\newcommand{\Mspacei}{\Mspace{i}}
\newcommand{\Mspacej}{\Mspace{j}}

\newcommand{\Ospace}[1]{M_{#1}}
\newcommand{\Ospacei}{\Ospace{i}}
\newcommand{\Ospacej}{\Ospace{j}}

\newcommand{\ef}[1]{\theta_{#1}}
\newcommand{\efi}{\ef{i}}
\DeclarePairedDelimiterXPP\efv[2]{\ef{#1}}{\lparen}{\rparen}{}{#2}
\DeclarePairedDelimiterXPP\efvi[1]{\efi}{\lparen}{\rparen}{}{#1}

\newcommand{\aef}{\tau}
\DeclarePairedDelimiterXPP\aefv[1]{\aef}{\lparen}{\rparen}{}{#1}

\newcommand{\rr}[1]{R_{#1}}
\newcommand{\rri}{\rr{i}}
\newcommand{\rrj}{\rr{j}}

\newcommand{\crr}[2]{ \Psi_{#1,#2} }
\newcommand{\crrij}{\crr{i}{j}}

\newcommand{\oef}[1]{\Theta_{#1}}
\newcommand{\oefi}{\oef{i}}
\DeclarePairedDelimiterXPP\oefv[2]{\oef{#1}}{\lparen}{\rparen}{}{#2}
\DeclarePairedDelimiterXPP\oefvi[1]{\oefi}{\lparen}{\rparen}{}{#1}

\newcommand{\obf}[2]{B_{#1,#2}}
\newcommand{\obfi}[1]{\obf{i}{#1}}
\newcommand{\obfik}{\obf{i}{k}}

\newcommand{\Mprodspace}{\Mspace{(n)}}
\newcommand{\Oprodspace}{\Ospace{(n)}}
\newcommand{\vpo}{\vec{y}}
\DeclarePairedDelimiterXPP\vpov[1]{\vpo}{\lparen}{\rparen}{}{#1}
\newcommand{\voo}{\vec{Y}}

\newcommand{\vu}{\vec{u}}
\newcommand{\vv}{\vec{v}}
\newcommand{\vU}{\vec{U}}
\newcommand{\vV}{\vec{V}}

\newcommand{\varmap}{\mathcal{V}_n}
\DeclarePairedDelimiterXPP\varmapv[1]{\varmap}{\lparen}{\rparen}{}{#1}

\newcommand{\covf}[2]{C_{#1,#2}}
\newcommand{\covfij}{\covf{i}{j}}
\DeclarePairedDelimiterXPP\covfv[3]{\covf{#1}{#2}}{\lparen}{\rparen}{}{#3}
\DeclarePairedDelimiterXPP\covfijv[1]{\covfij}{\lparen}{\rparen}{}{#1}

\newcommand{\bouf}[2]{B_{#1,#2}}
\newcommand{\boufij}{\bouf{i}{j}}
\DeclarePairedDelimiterXPP\boufv[3]{\bouf{#1}{#2}}{\lparen}{\rparen}{}{#3}
\DeclarePairedDelimiterXPP\boufijv[1]{\boufij}{\lparen}{\rparen}{}{#1}

\newcommand{\pairedOspace}[2]{\Ospace{#1,#2}}
\newcommand{\pairedOspaceij}{\pairedOspace{i}{j}}

\newcommand{\vb}{\mathrm{VB}}
\DeclarePairedDelimiterXPP\vbv[1]{\vb}{\lparen}{\rparen}{}{#1}
\newcommand{\vbn}[1]{\mathrm{VB}_{#1}}
\DeclarePairedDelimiterXPP\vbnv[2]{\vbn{#1}}{\lparen}{\rparen}{}{#2}

\newcommand{\vbpartf}[2]{B_{#1,#2}}
\newcommand{\vbpartij}{\vbpartf{i}{j}}
\DeclarePairedDelimiterXPP\vbpartv[3]{\vbpartf{#1}{#2}}{\lparen}{\rparen}{}{#3}
\DeclarePairedDelimiterXPP\vbpartijv[1]{\vbpartij}{\lparen}{\rparen}{}{#1}

\newcommand{\evb}{\widehat{\vb}}

\newcommand{\tensor}[1]{\mathfrak{#1}}

\newcommand{\tU}{\tensor{u}}
\newcommand{\tV}{\tensor{v}}

\newcommand{\uerr}{\mathcal{R}_n}
\newcommand{\uerrv}[1]{\uerr\paren{#1}}
\newcommand{\uerrC}{\uerrv{C}}

\newcommand{\neigh}[1]{N_{#1}}
\newcommand{\extneigh}[1]{\widetilde{N}_{#1}}
\newcommand{\neighi}{\neigh{i}}
\newcommand{\extneighi}{\extneigh{i}}
\newcommand{\dmax}{d_{\max}}

\newcommand{\depneigh}[1]{\mathcal{N}_{#1}}
\newcommand{\depneighi}{\depneigh{i}}
\newcommand{\depmax}{D_{\max}}

\newcommand{\expmap}[1]{d_{#1}}
\newcommand{\expmapi}{\expmap{i}}
\newcommand{\expset}[1]{\Delta_{#1}}
\newcommand{\expseti}{\expset{i}}

%% file: tex/abstract.tex
We describe a widely applicable design-based framework for drawing causal inference in randomized experiments.
Causal effects are defined as linear functionals evaluated at unit-level potential outcome functions.
Assumptions about the potential outcome functions are encoded as function spaces.
This makes the framework expressive, allowing experimenters to formulate and investigate a wide range of causal questions that previously could not be investigated with design-based methods.
The framework is particularly suited for complex, non-discrete interventions and causal interference.
We describe a class of estimators for estimands defined using the framework and investigate their properties.
We provide necessary and sufficient conditions for unbiasedness and consistency.
We also describe a class of conservative variance estimators, which facilitate the construction of confidence intervals.
In order to demonstrate the value of our approach in practice, we provide several illustrative examples of causal investigations which can be handled within our framework, but that could not be addressed using conventional design-based methods.

%% file: tex/introduction.tex
\section{Introduction}\label{sec:intro}

In the conventional design-based framework for formulating causal effects and analyzing randomized experiments, the intervention is individual assignment of categorical treatments to the experimental units.
The treatments can take small number of possible values, typically binary.
The units are presumed to be causally isolated in the sense that a unit's outcome is affected only by its own treatment.
Causal effects are formulated as aggregated contrasts of the two individual potential outcomes.
The structure of this framework limits the scope of causal questions that researchers can investigate using experimental methods.

This paper describes a new design-based experimental framework for causal estimation under arbitrary treatments and interference structures.
The purpose of the framework is to be expressive, allowing experimenters to define and investigate a wide range of causal questions involving essentially any type of treatments under rich and complex interference.
At the same time, the framework is sufficiently tractable to admit precise estimation and inference of the estimands defined with it.
The proposed framework unifies and generalizes most previously developed design-based frameworks.
However, we see its primary value as offering formulations of and statistical methods for causal questions that can not be expressed within existing design-based frameworks.

Recent work in design-based causal inference has also aimed to broaden the applicability of the conventional framework, but this has been achieved by showing that some slightly more complex type of experiment can be translated into a version of the conventional structure.
Nearly all of this recent work can be understood as special cases or variations of the method of \emph{exposure mappings} introduced by \citet{Aronow2017Estimating}.
The purpose of exposure mappings is to relax the assumption of causally isolated units, thus allowing for interference.
The underlying idea is that some types of experiments with interference can be reinterpreted as experiments without interference but with a more complex experimental design at the level of discrete exposures.
The method thereby translates the empirical problem into a familiar form, making it possible to solve it using conventional techniques.
While this translation is insightful and has been exceptionally impactful both for theory and practice, its scope is still limited, and there are many causal questions and experiments for which this translation is not possible.
The current state of affairs therefore forces experimenters to abandon important causal questions, or to artificially discretize them for the sole purpose of having them fit the exposure mapping framework.
The framework we describe in this paper is created to fill this gap.

There are three main contributions of the paper.
\begin{enumerate}
\item
	In Section~\ref{sec:framework}, we describe our design-based framework for randomized experiments, allowing experimenters to formalize and pose a wide range of empirically relevant causal questions.
	We provide two illustrative examples of the framework in Section~\ref{sec:overview} as well as numerical illustrations in Section~\ref{sec:num-illustrations}.

\item
	In Section~\ref{sec:estimator}, we describe a new class of treatment effect estimators for estimands defined in the framework, which we call the \emph{Riesz estimator}.
	Because causal questions posed in the framework generally cannot be translated into a problem of the familiar discrete structure, conventional estimation techniques cannot be used, and this necessitates us to develop this new estimation approach.
	In Section~\ref{sec:consistency}, we develop both finite- and large-sample theory for the estimator, including necessary and sufficient conditions for unbiasedness and consistency.

\item
	In Section~\ref{sec:variance-estimation}, we describe a new conservative estimator for the variance of the Riesz estimator.
	The variance estimator is constructed by applying the same techniques used to construct the point estimator, after a tensorization of the quadratic variance functional.
	The tensorization technique we develop for the variance estimator is useful in other contexts and might be of independent interest.
\end{enumerate}

%% file: tex/overview.tex

\section{Illustrative Examples}\label{sec:overview}

\subsection{Framework Overview}

This section presents two illustrative examples of causal questions that cannot be expressed and investigated in existing design-based frameworks, but can be accommodated in our framework.
We start with a brief overview of the framework that brushes over most of the technical aspects, describing the full framework in the next section.

We use $\Zspace$ to denote the set of interventions available to the experimenter.
A key feature of our framework is that we impose essentially no restrictions on this intervention set.
The experimental design $\Zprob$ is a probability measure over $\Zspace$ which is selected by and thus known to the experimenter.
There are $n$ experimental units, each of which has an associated potential outcome function $\poi : \Zspace \to \Reals$ describing what its outcome would have been had a particular, potentially counterfactual, intervention been performed.
The experimenter selects a random intervention $Z \sim \Zprob$ to perform and records the observed outcomes $\ooi = \povi{Z}$.\footnote{%
The formal description of the framework in Section~\ref{sec:framework} makes distinctions between measurable functions and equivalence classes. We eschew this distinction here in favor of understanding and familiarity.
}

Our framework consists of two main components an experimenter must specify:

\begin{enumerate}
	\item \textbf{Effect functional.} The experimenter must specify what causal aspect is the target of the study.
	This is specified through a linear functional $\efi$ of the potential outcome function for each unit $i \in [n]$, which we refer to as \emph{effect functionals}.
	The evaluation of the functional $\itei = \efvi{\poi}$ defines the causal aspect of interest for an individual unit, providing a scalar summary of its response to the interventions.
	The overall estimand of interest is the average of the individual effects: $\ate = n^{-1} \sumin \efvi{\poi}$.

	\item \textbf{Model spaces.}
	Meaningful causal inference is impossible without leveraging existing knowledge into well-justified assumptions on the causal structure in the experiment.
	In our framework, these assumptions are encoded in \emph{model spaces}.
	For each unit $i \in [n]$, the experimenter posits a linear space of functions $\Mspacei$ such that they are comfortable asserting that it contains the true potential outcome function: $\poi \in \Mspacei$.
\end{enumerate}

In Section~\ref{sec:estimator}, we describe the Riesz estimation approach, named after the Riesz representation theorem that acts as its basis.
The primary purpose of the Riesz estimator is to provide a unified approach to unbiased estimation within our framework, facilitating causal inference in settings where conventional estimators cannot be used.
The construction of the estimator is conceptually simple.
We seek a random variable $R_i$ that can act as a weighting of the observed outcome with the property $\E{R_iY_i} = \itei$.
An estimator $\est = n^{-1} \sumin R_iY_i$ based on such random variables is unbiased for the average causal effect $\ate$.
While many random variables have this unbiasedness property, the Riesz estimator is based on the random variables with minimal norm $\E{R_i^2}$.
The Riesz estimator is in this sense the simplest unbiased weighting estimator.

\paragraph*{Exposure Mapping Framework}
Our framework contains the the conventional no-interference (SUTVA) setting with binary treatments as a special instance.
More generally, it encapsulates the exposure mapping framework of \citet{Aronow2017Estimating} mentioned above.
In this framework, the intervention set consists of vectors of binary treatments assigned to all units: $\Zspace = \setb{0,1}^n$.
For each unit $i \in [n]$, the model $\Mspacei$ contains those functions that are constant on level sets of an \emph{exposure mapping function} $d_i : \Zspace \to \Delta$, where $\Delta$ is referred to as the \emph{exposure set}.
The effect functionals are point-wise contrasts of the potential outcome function at two unit-specific interventions, or expectations thereof.
For example, the Global Average Treatment Effect, which captures the effect of giving all subjects treatment versus control, is encoded by the functional $\efvi{y} = y\paren{\mathbf{1}} - y\paren{\mathbf{0}}$.
In this framework, the Riesz estimator is the exposure-level Horvtiz--Thompson estimator.

The purpose of our framework is to go beyond this discrete setting to facilitate the investigation of a wider range of causal questions.
Detailed derivations of the Riesz estimator for these illustrative examples appear in supplement Section~\suppref{sec:supp-worked-examples}.

\subsection{Continuous Treatments and Marginal Spillover Effects}\label{sec:continuous-treatments}

The first example considers estimation of marginal spill-over effects in a network experiment with continuous treatments.
An empirical study with a similar setup is the unconditional cash transfer experiment by \citet{Egger2022General}, where cash payments in varying amounts are given to $n$ households.
The households are presumed to interact through an observed network.
The intervention set is here the collection of individual cash transfers given to each unit: $z = \paren{z_1, \dotsc, z_n} \in \Zspace$ where $\Zspace = [0,1]^n$ is normalized to the unit interval for simplicity.
We consider independent assignment of the cash payment assignments here to ease the exposition, meaning that $Z_1, Z_2, \dotsc, Z_n$ are independent but not necessarily identically distributed.
Our framework and estimation approach can accommodate designs with severe dependencies, but it complicates the exposition.

The first step of the recipe for using the framework is to specify the effect functional.
Consider the aggregate marginal spill-over effect at fixed transfer level $s \in [0,1]^n$, defined as
\[
\ate
= \frac{1}{n} \sum_{i=1}^n \braces[\Big]{ \frac{\partial}{\partial t} \povi{s + t \cdot e_{-i}} \mid_{t=0}
}
\enspace,
\]
where $e_{-i}$ is the vector with $0$ in coordinate $i$ and $1$ in all other coordinates.
The effect functional takes a differential form: $\efvi{y} = \frac{\partial}{\partial t} y\paren{s + t \cdot e_{-i}} \mid_{t=0}$.
In the context of the cash transfer experiment, the effect describes how units are affected by marginal increases in the cash transfer given to all other units, holding the unit's own transfer fixed.
Note that this causal estimand cannot be expressed within the conventional exposure mapping framework, as the intervention is not discrete and cannot be discretized in a way that doesn't fundamentally change what the estimand is capturing.

The second step of the framework recipe is to specify the model spaces, capturing what we know about the causal setting before running the experiment.
A common assumption in the interference literature with binary treatments is that a unit is affected by other units' treatment only through the share of its neighbors that are treated.
We could use a similar linear-in-means model containing all functions of the form
\[
\povi{z}
= \alpha_{i,0}
+ \alpha_{i,1} \cdot z_i
+ \alpha_{i,2} \cdot \frac{1}{\abs{N_i}} \sum_{j \in N_i} z_j,
\]
where $N_i \subset [n]$ are the neighbors of unit $i$ in the observed interference network and $(\alpha_{i,0}, \alpha_{i,1}, \alpha_{i,2}) \in \Reals^3$ is a parametrization of the model space.

The concern with this model is that it requires the spillover effect to be the same no matter the treatment $z_i$ the unit itself received.
This will be unrealistic in many settings, including the current cash transfer example.
For example, a participant might be envious of other participants when they receive large transfer, but only when the participant receives a small (or no) transfer.
Similarly, in a vaccine study, spillover effects and herd immunity are relevant primarily for participants who receive low (or no) vaccine doses; adequately vaccinated participant will not get sick regardless of the doses given to their neighbors.
We do not want to rule out these aspects a priori and must therefore use a larger model space.

In Section~\mainref{sec:num-illustrations}, we consider a large model space accommodating complex interactions and spillover patterns.
For simplicity, we here consider only a slightly larger model than the one above, namely, all functions of the form
\[
\povi{z}
= \alpha_{i,0}
+ \alpha_{i,1} \cdot z_i
+ \alpha_{i,2} \cdot \frac{1}{\abs{\neighi}} \sum_{j \in \neighi} z_j
+ \alpha_{i,3} \cdot z_i \cdot \paren[\Big]{\frac{1}{\abs{\neighi}} \sum_{j \in \neighi} z_j}.
\]
Note that this model space permits interactions between a unit's own treatment and that of its neighbors through the final interaction term.

Once we have specified the causal effects and the model spaces, we can then move to deriving the Riesz estimator.
As discussed above, this requires deriving the (unique) random variable $R_i$ satisfying $\E{R_iY_i} = \itei$ with minimal norm.
In this case, that random variable is
\[
R_i = \frac{ \abs{N_i} \cdot \sum_{j \in N_i} (Z_j - \E{Z_j})}{ \sum_{j \in N_i} \Var{Z_j}}
+ \frac{\abs{N_i} \cdot (s_i - \E{Z_i}) (Z_i - \E{Z_i}) \cdot \sum_{j \in N_i} (Z_j - \E{Z_j}) }{\Var{Z_i} \sum_{j \in N_i} \Var{Z_j}},
\]
where $s = (s_1, s_2, \dotsc, s_n)$ is the benchmark transfer level used in the specification of the effect functionals.
The Riesz estimator in this setting is thus $\est = \frac{1}{n} \sum_{i=1}^n \rri \ooi$.
While the idea behind the estimator is conceptually straightforward, the derivation of $R_i$ is somewhat intricate, and we defer that technical discussion to later in the paper.

If either the effect functionals or model spaces are changed, the estimator will typically change as well.
In Section~\mainref{sec:num-illustrations}, we consider the same effect functionals as here but considerably more complex model spaces, and the estimator is different and also more complex.
The fact that the estimation approach adapts to and can accommodate such changes to the experimental setting, including the experimental design, is one of its strengths.

\subsection{Spatial Interference and Point Process Experiments}\label{sec:spatial-example}

The second example considers a spatial experiment in which $n$ units are located within a spatial region and the intervention is the deployment of some treatment to $m$ intervention sites within the region.
The treatment sites are not a finite set of fixed points in the spatial region, but can be placed continuously in the region.
The example therefore departs from the familiar structure both because the treatments are not discrete and because there is no direct one-to-one association between the treatments and the units.
An empirical study with a similar setup is the study by \citet{Manacorda2020Liberation}, who investigate the effect of access to mobile phone service in Africa on political mobilization, using the fact that phone coverage requires a close-by cell phone tower.
The placement of the phone towers are here the treatments.

The experimental units are located at points $u_1, u_2, \dotsc, u_n$ in a spatial region $\mathcal{X} \subseteq \Reals^2$.
The experimental design is given by a \emph{point process}, which is a random selection of intervention sites $z_1, z_2, \dotsc, z_m \in \mathcal{X}$.
For simplicity in this illustration, we presume that the number of intervention sites $m$ is fixed and collect random assignments in a vector $Z = (Z_1, \dotsc, Z_m)$.

As before, the first step for an experimenter using the framework is to specify the effect functional.
We here consider when the experimenter is interested in the outcomes under a counterfactual policy.
We use $\offpol$ to denote the distribution of $Z$ under the counterfactual policy, and as above, use $\Zprob$ to denote actual, status quo distribution in the experiment.
The \emph{counterfactual policy effect} is defined as
\[
\ate
= \frac{1}{n} \sum_{i=1}^n \braces[\big]{\Esub{Z \sim \offpol}{\povi{Z}} - \Esub{Z \sim \Zprob}{\povi{Z}}},
\]
where we use subscripts to denote the probability measure under consideration.
The effect functional is therefore the difference in the expected outcome between the counterfactual and status quo distributions: $\efvi{y} = \Esub{Z \sim \offpol}{y\paren{Z}} - \Esub{Z \sim \Zprob}{y\paren{Z}}$.

In the context of the study of the effect of mobile phone service on political mobilization, the counterfactual policy effect captures the expected change in political mobilization under a counterfactual policy regulating the placement of cell phone towers compared to status quo.
For example, if the actual distribution $\Zprob$ tends to place towers close to urban centers and the counterfactual distribution $\offpol$ places them in rural areas to a larger degree, the estimand captures the effect on political mobilization of a greater focus on rural cell phone coverage.
As before, this causal question cannot be formulated in the conventional, discrete framework.

The next step is to specify the model spaces.
There are two features of the causal setting we want to capture here.
First, because of the inverse-square law, the intensity of radio waves from a cell tower decrease with the distance from the tower.
We capture this with the unit-specific function
\[
h_{i,1}(z) = \sum_{t=1}^m \expf[\Big]{ - \frac{\norm{u_i - z_t}^2}{\sigma^2} }.
\]
Note that $h_{i,1}(z)$ is large for a unit $i$ when there is one or more cell towers close to unit $i$'s position $u_i$.
Cell towers far away, meaning $\norm{u_i - z_t} \gg \sigma$, have a negligible effect on $h_{i,1}(z)$.

Second, two cell towers close to each other are likely to lead to just a small or no improvement in coverage compared to a single tower at the same position.
In other words, if there are more than one tower close to a unit, we expect well-dispersed towers to be more efficacious.
We capture this with another unit-specific function:
\[
h_{i,2}(z) = \sum_{j=1}^n \sum_{k=1}^n
\expf[\Big]{ - \frac{\norm{u_i - z_j}^2}{\sigma^2} }
\expf[\Big]{ - \frac{\norm{u_i - z_k}^2}{\sigma^2} }
\bracket[\Big]{
    1
    -
    \expf[\Big]{ - \frac{\norm{z_j - z_k}^2}{\sigma^2} }
}.
\]
Note that $h_{i,2}(z)$ is large when there are two (or more) cell towers close to unit $i$'s position $u_i$ that are also far from each other.
When $z_j \approx z_k$, the third factor is close to zero, so the corresponding term in the double sum is negligible even when $z_j$ and $z_k$ are close to $u_i$.

We here consider the model spaces that simply combine these two aspects.
That is, the model spaces consist of all functions of the form
\[
\povi{z} =
\alpha_{i,0}
+ \alpha_{i,1} \cdot h_{i,1}(z)
+ \alpha_{i,2} \cdot h_{i,2}(z),
\]
where, as above, the vector $(\alpha_{i,0}, \alpha_{i,1}, \alpha_{i,2}) \in \Reals^3$ is a parametrization of the model spaces.
The first term captures a pure closeness effect of the towers and the second term captures potential interactive effects between towers.

As before, the estimation approach first constructs the random variable $\rri$ with minimal norm that is unbiased.
Let $H_{i,1} = h_{i,1}(Z)$ and $H_{i,2} = h_{i,1}(Z)$ be the random evaluations of the two functions defined above.
The random variable used in the estimator is then
\[
\rri
=
\sum_{k\in \setb{1,2}} \frac{ (\Esub{\offpol}{H_{i,k}} - \Esub{\Zprob}{H_{i,k}})( H_{i,k} - \Esub{\Zprob}{H_{i,k}} ) }{\Varsub{P}{H_{i,k}}(1-\rho^2)}
- \frac{\rho}{(1 - \rho^2)}
\frac{M}{\sqrt{ \Varsub{\Zprob}{H_{i,1}} \Varsub{\Zprob}{H_{i,2}} }},
\]
where
$
M = \braces[\big]{ H_{i,1} - \Esub{\Zprob}{ H_{i,1} } }
    \braces[\big]{ \Esub{\offpol}{H_{i,2}} - \Esub{\Zprob}{ H_{i,2} } }
    +
    \braces[\big]{ \Esub{\offpol}{H_{i,1}} - \Esub{\Zprob}{ H_{i,1} } }
    \braces[\big]{ H_{i,2} - \Esub{\Zprob}{ H_{i,2} } },
$
and $\rho = \Covsub{\Zprob}{H_{i,1}, H_{i,2}} / \sqrt{\Varsub{\Zprob}{H_{i,1}} \Varsub{\Zprob}{H_{i,2}}}$ is the correlation coefficient of $H_{i,1}$ and $H_{i,2}$ under the experimental design $\Zprob$.

In Section~\suppref{sec:worked-spatial-point-process} of the supplement, we derive a richer class of \emph{determinantal models} that can account for higher-order interactions.
As the model increases in complexity, so too does the estimator.

%% file: tex/previous-work.tex
\section{Related Work}

Our paper contributes to the literature on design-based causal inference using potential outcomes first formulated by \citet{Neyman1923Application}.
The subsequent literature is large and wide-reaching.
The strand of the literature closest to the current paper concerns causal inference under interference.
Early papers establishing key ideas in this literature were \citet{Sobel2006What} and \citet{Hudgens2008Toward}.
Much of the current literature can be understood as applications of the idea of exposure mappings or effective treatments introduced by \citet{Aronow2017Estimating} and \citet{Manski2013Identification}.
A large literature extending and building on this idea has followed \citep[see, e.g.,][]{Athey2018Exact,Basse2019Randomization,Forastiere2021Identification,Leung2022Causal,Li2022Random}.

The extension of causal inference beyond a discrete structure has to our knowledge only been considered in a super-population framework, including the work by \citet{Hirano2004Propensity}, \citet{Kennedy2017Nonparametric}, and \citet{Rothenhausler2019Incremental}.
We are the first to consider arbitrary, including continuous and uncountable, intervention spaces without discretization in a design-based setting .
On a conceptual level, the closest previous paper is \citet{Borusyak2023Nonrandom}, who consider settings where the treatment of interest (or an instrumental variable) contains multiple sources of variation that can be leveraged for inference.
However, the papers differ both in their frameworks and estimation approaches, and the methods fill complementary roles and are not easily compared.
Our paper addresses the setting of prospective experimental studies, and does not consider observational studies.

The Riesz estimator can be seen as a generalization of the Horvitz--Thompson estimator, which has been used extensively in the causal inference literature \citep{Aronow2013Class}.
The estimators described by \citet{Harshaw2023Design} and \citet{CortezRodriguez2023Mayleen} are extensions of the Horvitz--Thompson estimator that are special cases of the estimator we describe here.

Many existing frameworks and settings used in causal inference are special instances of our framework.
For example,  \citet{Zigler2021Bipartite} consider bipartite experiments, in which the units receiving treatment are different from the units for which we measure outcomes, and there is no one-to-one mapping between the two sets of units.
A second example is so-called group formation experiments, where the treatment is the assignment of units into groups \citet{Basse2024Randomization}.
A third example is spatial experiments, where the possible interventions are geographical locations or interference is spatially mediated \citep{Leung2022Rate,Papadogeorgou2022Causal,Pollmann2023Causal,Wang2025Design}.
Current design-based analyses of spatial experiments have used conventional discretizations.

Our framework bears some resemblance to the semiparametric causal inference framework in a super-population setting.
For example, similar to the semiparametric framework but unlike most the design-based literature, we define our causal effects using arbitrary linear functionals.
Furthermore, representation theorems from functional analysis \citep{Frechet1907Ensembles,Riesz1907Espece} play an important role in our work, and such theorems have taken an increasingly prominent position in the recent semiparametric literature.
\citet{Newey1994Asymptotic} and \citet{Robins1994Estimation} are early examples of the use of representation theorems in this literature, and \citet{Chernozhukov2022Locally,Chernozhukov2022Automatic} and \citet{Hirshberg2021Augmented} are more recent examples.

While there are similarities between the current paper and this strand of the semiparametric literature, stemming from the fact that both use insights from functional analysis, the current paper should not been seen as contributing to or building on the semiparametric literature.
A key difference, among others, is that functionals in the semiparametric literature operate on a single conditional expectation function in a super-population whereas the functionals in our framework operate directly on $n$ individual potential outcome functions.
As a consequence the super-population framework implicitly imposes between-unit homogeneity, while our framework does not impose any type of homogeneity in line with the design-based tradition.
This means that the corresponding representors and resulting estimators differ both in interpretation and construction.

%% file: tex/framework.tex
\section{The Framework}\label{sec:framework}

\subsection{Experimental Designs}\label{sec:experimental-design}

We use $\Zspace$ to denote the set of all interventions accessible to the experimenter.
The experimenter will select one intervention $z \in \Zspace$ to perform, which is the realized intervention.
We require only that the $\Zspace$ has an associated topology under which it is a completely separable space.
Complete separability is the minimal structure we require for our analysis, and it is typically an innocuous condition.
For example, any finite set is completely separable with respect to the discrete topology, and $\Reals^k$ is completely separable with respect to the usual Euclidean topology.
Let $\Zalgebra$ be the Borel sigma-algebra of $\Zspace$ generated by its topology.

A probability measure $\Zprob : \Zalgebra \to [0, 1]$ describes the mechanism by which the experimenter randomly selects an intervention from $\Zspace$ to implement.
The \emph{experimental design} is the probability space $(\Zspace, \Zalgebra, \Zprob)$.
The only randomness under consideration in this paper is that which is induced by the design.
All other aspects of the experiment are considered fixed and non-random.

\subsection{Lebesgue Spaces}

Lebesgue spaces with respect the experimental design are central to our framework.
We briefly introduce our notation and review standard constructs.
Let $\Lts$ be the set of all real-valued, square-integrable functions on $\Zspace$, meaning that $\E{u^2} < \infty$ for all $u \in \Lts$, where the expectation is taken with respect to the design.
Define a bilinear form $\iprod{\cdot, \cdot}$ on $\Lts$ as $\iprod{u, v} = \E{uv}$.
Let $\norm{\cdot}$ be the seminorm on $\Lts$ induced by the bilinear form: $\norm{u} = \sqrt{\iprod{u, u}}$.
Let $\mathcal{N} = \setb{ u \in \Lts : \norm{u} = 0 }$ be the null space of $\Lts$.

Let $\Lt = \Lts / \mathcal{N}$ be the quotient space of $\Lts$ by $\mathcal{N}$.
An equivalence class in $\Lt$ contains all functions that are observationally indistinguishable from one another, in the sense of being equal almost surely under the design.
The equivalence class to which a function $u \in \Lts$ belongs to is denoted $\ec{u} = \setb{v \in \Lts : \norm{u - v} = 0}$.
We denote elements of $\Lts$ and $\Lt$ with lowercase and uppercase letters, respectively: $u \in \Lts$ and $U = \ec{u} \in \Lt$.

We will often use the shorthand $\E{U} = \E{u}$ to denote the expectation of the functions in $U \in \Lt$.
This leads to no ambiguity because $\E{u}$ is constant over $u \in U$.
Define an inner product $\iprod{\cdot, \cdot}$ on $\Lt$ as $\iprod{U, V} = \E{UV}$, and let $\norm{\cdot}$ be the norm on $\Lt$ induced by the inner product.
Note that $\Lt$ is the standard construction of the Lebesgue space of square-integrable functions, and that $\Lt$ together with $\iprod{\cdot, \cdot}$ is a Hilbert space.

\subsection{Potential Outcome Functions}\label{sec:potential-outcomes}

There are $n$ experimental units for which outcomes are measured, indexed by integers $i \in [n]$.
Each unit has an associated potential outcome function $\poi : \Zspace \to \Reals$ that describes what the measured outcome would have been under a particular, potentially counterfactual, intervention.
It is assumed that all units have well-defined potential outcome functions, in the sense that an unambiguous, measurable outcome is produced by each intervention.
We require that the potential outcome functions are square-integrable with respect to the experimental design, meaning that they are elements of $\Lts$.

Let $\ooi = \ec{\poi}$ be the equivalence class in $\Lt$ of unit $i$'s potential outcome function.
This equivalence class captures the information that the experimental design provides about the potential outcome function, in the sense that all functions in $\ooi$ are indistinguishable under the design.
We can therefore interpret $\ooi$ as capturing the observed outcome.

It will prove useful to describe all $n$ potential outcomes together using a vector function.
Define the \emph{combined potential outcomes function} $\vpo : \Zspace \to \Reals^n$ as $\vpov{z} = \paren{\pov{1}{z}, \dotsc, \pov{n}{z}} $ and the \emph{combined observed outcomes} as $\voo = \ec{\vpo} = \paren{\oo{1}, \dotsc, \oo{n}}$.
Moments are most naturally defined at the level of the combined potential outcomes.
Let $\norm{\vpo}^2 = n^{-1} \sumin \norm{\poi}^2$ and $\norm{\voo}^2 = n^{-1} \sumin \norm{\ooi}^2$ denote extensions of the seminorm on $\Lts$ and the norm on $\Lt$ to their combined versions.
Note that both norms are the finite population second moment of the observed outcomes in the experiment: $\norm{\vpo}^2 = \norm{\voo}^2 = n^{-1} \sumin \E{ \ooi^2 }$.

\subsection{Model Spaces}\label{sec:model-spaces}

Knowledge the experimenter has about the potential outcome functions prior to running the experiment is encoded as a function space $\Mspacei$ for each unit $i$, which we call \emph{model space}.
The framework does not require the model spaces to take any particular form other than being subspaces of square-integrable functions.
The structure imposed on the potential outcome functions by their model spaces is presumed to be correctly specified.

\begin{assumption}[Correctly specified model spaces]\label{ass:correctly-specified}
	Each unit's potential outcome function is in the model space specified for that unit: $\poi \in \Mspacei$ for all $i \in [n]$.
\end{assumption}

Similar assumptions are widely used in the causal inference literature, but the conventional structure imposed on the potential outcome functions tend to be quite restrictive.
Our framework can accommodate arbitrary and large model spaces, making the assumption more tenable.
However, working with large model spaces can be challenging, requiring experimenters to select these model with care.

Following the construction of the observed outcomes above, we construct a space describing the observable information for each model space: $\Ospacei = \textrm{cl}\paren[\big]{ \setb{ \ec{u} \in \Lt : u \in \Mspacei } }$.
We refer to these spaces as the \emph{outcome spaces}.
These spaces are defined using a closure to ensure that they are Hilbert spaces with respect to the norm on $\Lt$.
When a model space has finite dimensions, the set $\setb{ \ec{u} \in \Lt : u \in \Mspacei }$ already contains all its limit points (and is therefore closed), but this might not be the case with infinite dimensions.

Mirroring the combined potential outcomes, we define the \emph{combined model space} as $\Mprodspace = \Mspace{1} \times \dotsb \times \Mspace{n}$ and the \emph{combined outcome space} as $\Oprodspace = \Ospace{1} \times \dotsb \times \Ospace{n}$.

The model spaces differ from conventional statistical models in a super-population framework both when it comes to complexity and unit heterogeniety.
A $d$-dimensional conventional statistical model can be parametrized with $d$ parameters irrespectively of the sample size, severely restricting both complexity and heterogeniety.
In contrast, if the model spaces in our framework have dimension $d$, then the combined model space is parameterized with $nd$ parameters.
Thus, the framework is inherently non-parametric, even when each individual model space has finite dimension.
For example, the conventional experimental setting with binary treatments and no interference corresponds to model spaces with two dimensions.

\subsection{Effect Functionals}\label{sec:effects}

The experimenter specifies a linear functional $\efi : \Mspacei \to \Reals$ for each unit to capture some aspect of interest of its potential outcome function.
We call them \emph{effect functionals}.
This is more expressive than conventional approaches to defining causal effects in the design-based literature and allows us to be agnostic about the structure of the intervention set $\Zspace$.

The causal effect for a unit is given by the unit's effect functional evaluated at the true potential outcome function: $\itei = \efvi{\poi}$.
The effect functionals $\efi$ do not need to be the same for all units; indeed, they will typically be different.
However, the units' effect functionals will share a similar interpretation in typical applications.
The aggregated effect functional $\aef : \Mprodspace \to \Reals$ is the average of the unit-level functionals: $\aefv{\vu} = n^{-1} \sumin \efvi{u_i}$.
This can be extended to any linear combination of the unit-level functionals, as the coefficients of the combination can be absorbed into the unit-level functionals.

The target estimand is the aggregated effect functional evaluated at the true potential outcome functions: $\aefv{\vpo} = n^{-1} \sumin \itei$.
When clear from context, we use $\ate$ as a shorthand for $\aefv{\vpo}$.
Because this class of estimands is large and includes essentially all causal effects previously considered in the design-based causal inference literature as special cases, we do not find it useful to give $\ate$ a particular name.
For convenience, we will refer to it as the aggregate causal effect.

%% file: tex/estimator.tex
\section{The Riesz Estimator}\label{sec:estimator}

The task ahead is to construct an estimator of $\ate$ using information about the observed outcomes $\ooi$.
This is challenging because $\itei = \efvi{\poi}$ depends on the whole potential outcome function $\poi$, but we only observe this function evaluated at a single point: $\povi{z}$.
When $\poi$ takes a small number of discrete values, we can use the Horvitz--Thompson estimator or one of its variants to confront this challenge.
This estimator combines direct imputation of the potential outcomes (when observed) with inverse probability weighting to adjust for the fact that the outcomes are not always observed.
This is not generally possible in our framework, because $\poi$ may be non-discrete, i.e. taking infinitely many values.

\subsection{Positivity}\label{sec:positivity}

We can only learn aspects of a potential outcome function that the experiment provides information about.
The following condition ensures that the experimental design is informative about all aspects of the potential outcomes that are relevant for the effect functionals.

\begin{assumption}[Positivity]\label{ass:positivity}
	For each $i \in [n]$, there exists $C < \infty$ such that $\abs{ \efvi{u} - \efvi{v} } \leq C \cdot \norm{u - v}$ for all $u, v \in \Mspacei$.
\end{assumption}

Positivity is a continuity condition on the effect functional with respect to the model space and experimental design, which can be understood as the analogue to the super-population concept of identifiability.
To see this, note that if two functions are indistinguishable, $\norm{u - v} = 0$, then there is no information in the experiment to discern whether the true effect is $\efvi{u}$ or $\efvi{v}$.
Therefore, the effect is not identified unless $\efvi{u} = \efvi{v}$.
Positivity implies that this indeed is the case: any two potential outcome functions that are indistinguishable under the design must yield the same effect.

Assumption~\ref{ass:positivity} coincides with the standard positivity condition in conventional experimental settings with discrete treatments.
When targeting a contrastive effect using exposure mappings, it is identical to assuming that each unit is assigned to both exposures in the effect contrast with some positive probability.
For example, with binary treatments and no interference, Assumption~\ref{ass:positivity} reduces to the usual positivity condition: $0 < \Pr{Z_i = 1} < 1$.
We elaborate on this connection and consider positivity in each of the illustrative examples in Section~\suppref{sec:supp-worked-exposure-mapping} of the supplement.

The technical function of the positivity condition is to facilitate an extension of the effect functional to the outcome spaces, as captured by the following lemma.
The underlying insight is that positivity implies that the equivalence classes in the outcome space $\Ospacei$ contain the same information about the effect functional $\efi$ as the functions in the model space $\Mspacei$.

\expcommand{\lemextendedfunctionals}{%
Given positivity, there exist continuous linear functionals $\oefi : \Ospacei \to \Reals$ for all $i \in [n]$ such that $\efvi{u} = \oefvi{\ec{u}}$ for all $u \in \Mspacei$.
}

\begin{lemma}\label{lem:extended-functionals}
\lemextendedfunctionals
\end{lemma}

As we shall see in the next subsection, this extension alone facilitates the construction of the Riesz estimator, making positivity a sufficient condition for the existence of an unbiased estimator of any causal effect defined in the framework.
Positivity is in a sense also a necessary condition for the existence of an unbiased estimator.
More precisely, positivity is required for the existence of an unbiased estimator in the class of estimators satisfying Lipschitz continuity.
An estimator $\est$ is \emph{Lipschitz continuous} with respect to $\Mprodspace$ if there exists $K < \infty$ such that $\sqrt{\E{\paren{\estv{\vu} - \estv{\vv}}^2}} \leq K \cdot\norm{\vu - \vv}$ for all $\vu, \vv \in \Mprodspace$, where we write $\estv{\vu}$ to denote the estimator under potential outcomes $\vu \in \Mprodspace$.

\expcommand{\thmpositivityisnecessary}{%
	Suppose that positivity does not hold for an effect $\ate$.
	Then, among potential outcome functions with bounded second moments, any Lipschitz continuous estimator $\est$ of $\ate$ has unbounded absolute bias.
	That is, for all $C > 0$,
	\[
	\sup \setb[\bigg]{
		\textstyle
		\abs[\big]{ \E{ \estv{\vu} } - \aefv{\vu} }
		\;:\;
		\vu \in \Mprodspace
		\;\text{ with }\;
		\sqrt{ \frac{1}{n} \sumin \E{U_i^2}} \leq C
	}
	=
	\displaystyle
	\infty.
	\]
}

\begin{theorem}\label{thm:positivity-is-necessary}
\thmpositivityisnecessary
\end{theorem}

A corollary of Theorem~\ref{thm:positivity-is-necessary} is that Lipschitz continuous estimators cannot be mean square consistent unless positivity holds.
We see the restriction to continuous estimators in the theorem as innocuous.
An estimator is Lipschitz continuous if small changes in the potential outcome functions result in only small changes in the behavior of the estimator.
An estimator that is not continuous would therefore be sensitive to small perturbations to the input data, typically making it imprecise also in large samples.
We conjecture that the use of a non-continuous estimator would not be a way to address violations to the positivity condition, but this conjecture escapes straightforward analysis.
Regardless, all linear estimators and, to our knowledge, all estimators considered in the design-based literature are Lipschitz continuous.

It is possible to investigate whether positivity holds before running the experiment, because the assumption depends only on the experimental design, the model spaces and the effect functionals, which are all known to the experimenter.
Importantly, the condition does not depend on the true potential outcome functions.
In Section~\suppref{supp:positivity-holds} of the supplement, we describe a procedure for determining whether positivity holds in a particular experiment.

\subsection{Definition and Unbiasedness}\label{sec:def-estimator}

The Riesz representation theorem states that any continuous linear functional on a Hilbert space can be represented with an element of the space with respect to its inner product.
Each outcome space $\Ospacei$ is a Hilbert space, and the extended functional $\oefi$ is continuous given positivity.
Thus, in our context, the representation theorem tells us that there exists a unique element $\rri \in \Ospacei$, which is called the \emph{Riesz representor}, such that $\oefvi{U} = \iprod{\rri, U} = \E{\rri U}$ for all $U \in \Ospacei$.
Note that positivity is required for the existence of the extended functional, and thus also required for the existence of the representors.

We use the representors to construct an estimator of the aggregated causal effect $\ate$.
When the model spaces are correctly specified, we have $\ooi \in \Ospacei$, so the Riesz representor has the property $\efvi{\poi} = \oefvi{\ooi} = \E{\rri \ooi}$ for the true (unobserved) potential outcome function.
The weighted outcome $\rri \ooi$ thereby acts as a direct (but typically very noisy) observation of the unit-level causal effect $\efvi{\poi}$, in the sense that it is unbiased.
An unbiased estimator of the aggregated effect is therefore formed by the average of the weighted outcomes.

\begin{definition}\label{def:riesz-estimator}
Given positivity, let $\rri \in \Ospacei$ be the Riesz representor of $\oefi$ in $\Ospacei$ for each $i \in [n]$.
The \emph{Riesz estimator} of the aggregated causal effect $\ate = n^{-1} \sumin \efvi{\poi}$ is
\begin{equation}
	\est = \frac{1}{n} \sumin \rri \ooi.
\end{equation}
\end{definition}

\expcommand{\thmRieszunbiased}{%
Given correctly specified model spaces and positivity (Assumptions~\mainref{ass:correctly-specified} and~\mainref{ass:positivity}), the Riesz estimator is unbiased: $\E{\est} = \ate$.
}

\begin{theorem}\label{thm:Riesz-unbiased}
\thmRieszunbiased
\end{theorem}

While the Riesz representor is the only element in $\Ospacei$ guaranteed to satisfy $\efvi{\poi} = \E{\rri \ooi}$ under correctly specified model spaces and positivity, there will typically be infinitely many random variables $W_i$ outside $\Ospacei$ with the property $\efvi{\poi} = \E{W_i \ooi}$.
In principle, any of these could also be used to produce unbiased weighting estimators.
However, the Riesz representor $\rri$ is special among all such unbiased weightings because it is the (unique) unbiased weighting with minimal norm.
That is, $\E{\rri^2} < \E{W_i^2}$ for all such $W_i$ outside $\Ospacei$.
We show this formally in Section~\suppref{sec:supp-riesz-minimal-norm} of the supplement.
Informally, this means that the Riesz estimator is the simplest unbiased weighting estimator.

The primary purpose of the Riesz estimator presented in this paper is to provide a general approach to unbiased estimation within our framework.
We have not shown that the Riesz estimator has any statistical optimality guarantees.
Indeed, in the conventional setting with discrete exposure mappings, the Riesz estimator is exactly the conventional Horvitz--Thompson estimator, whose variance can be further improved via Hájek-style normalizations.
A hurdle when studying optimality in this setting is that an optimality concept has not yet been rigorously defined in the design-based regime.
This is in contrast to the super-population framework, which has a well-developed efficiency theory.
While more efficient estimators are likely to exist, a rigorous investigation is beyond the scope of the current paper, as it would require the development of a design-based efficiency theory.

\subsection{Construction and Computation}\label{sec:construction}

The Riesz representors depend solely on the model spaces, the effect functionals, and the experimental design, all of which are known to the experimenter.
This means that the Riesz estimator can be constructed.
However, constructing and evaluating the Riesz representors can be non-trivial computational tasks in practice.
We here describe an approach for constructing the estimator based on basis representations of the outcome spaces.

By the fact that the intervention set is completely separable, each outcome space $\Ospacei$ has a countable orthonormal basis $\obfi{1}, \obfi{2}, \dotsc$ with respect to the $\Lt$ norm.
Given this basis, a unit's Riesz representor can be written in closed form as
\begin{equation}\label{eq:riesz-ortho-basis}
	\rri = \sum_{k=1}^\infty \oefvi{\obfik} \obfik.
\end{equation}
The sum in the expression for the representors will have finite terms when the outcome spaces have finite dimensions.
When the outcome spaces have infinite dimensions, the sum may have infinitely many non-zero terms, making it infeasible to exactly compute the representor.
In this case, experimenters should truncate the sum at some large but finite number of basis functions.
In Section~\suppref{supp:infinite-Riesz} of the supplement, we show that this can be done so that the Riesz representor is captured to arbitrary precision, thereby capturing all distributional properties of the estimator that are relevant for practical purposes.
It can sometimes be advantageous to truncate the sum earlier, possibly reducing variance at the cost of bias in a manner reminiscent of a sieve estimator.
It is beyond the scope of the current paper to investigate such bias--variance trade-offs.

Experimenters will often specify their model spaces with non-orthogonal basis functions, as we did in the illustrative examples of Section~\ref{sec:overview}.
In Section~\suppref{supp:orthogonalization-procedure} of the supplement, we describe an orthogonalization procedure that produces an orthonormal basis of $\Ospacei$ from an arbitrary basis of $\Mspacei$.
The procedure requires that the experimenter can calculate the expectation of products of the specified basis functions.
If expectations cannot be derived analytically, they could be computed to arbitrary precision using Monte Carlo estimation.

%% file: tex/consistency.tex
\section{Precision and Consistency}\label{sec:consistency}

The key benefit of the Riesz estimation approach is that it is applicable whenever an empirical problem can be cast in our framework and positivity holds: unbiasedness is ensured by construction, irrespectively of the circumstances of the particular application.
However, other aspects of the estimator, including its variance and limiting distribution, will depend on those circumstances.
Thus, a universal characterization of all properties of the estimator is not possible.
Our goal in this section is to provide general theory that can be used as a starting point for further analysis in a given setting.
In Section~\suppref{sec:supp-worked-examples} of the supplement, we discuss how the theory is applied to the illustrative examples in Section~\mainref{sec:overview}.

Following the convention in the design-based literature, we use triangular array asymptotics, where essentially all aspects of the experiment depend on the index $n \in \Naturals$ in the asymptotic sequence.
Statements regarding limiting behavior are with respect to these sequences.

\subsection{Uniform Consistency in Mean Square}\label{sec:uniform-consistency}

The conventional notion of consistency in the design-based causal inference literature is implicitly a uniform notion.
This is in contrast to a pointwise consistency, which sometimes is considered in super-population frameworks.
To our knowledge, the design-based, uniform notion of consistency has not previously been formally defined, and we provide such a definition here for clarity and completeness.

For an experiment indexed by $n$ in the asymptotic sequence, the \emph{uniform root mean square error} of an arbitrary estimator $\est$ of effect $\ate$ is defined as
\[
\uerrC = \sup \setb[\bigg]{
	\textstyle
	\sqrt{ \E[\big]{ \paren{ \estv{\vu} - \aefv{\vu} }^2 }}
	\;:\;
	\vu \in \Mprodspace
	\;\text{ with }\;
	\sqrt{ \frac{1}{n} \sumin \E{U_i^2}} \leq C
	}.
\]
The quantity $\uerrC$ is uniform in the sense that it bounds the error of the estimator uniformly over all potential outcome functions in the model space $\Mprodspace$ whose second moment is bounded by $C$.
For linear estimators, including the Riesz estimator, $\uerrC$ is proportional to $C$, and it is then sufficient to consider only the normalized error: $\uerr \triangleq \uerrv{1}$.
A linear estimator is said to be \emph{uniformly consistent in mean square} if $\uerr = \bigO{r_n}$ for some sequence $r_n \to 0$, capturing the rate of convergence.
For general (non-linear) estimators, $\uerrC$ can be a complex function of $C$.
Such estimators are said to be consistent if $\uerrC = \bigO{r_n}$ for every fixed $C > 0$.

An estimator is said to have \emph{finite variance} for some $n \in \Naturals$ if $\uerrC < \infty$ for every $C > 0$.
The Riesz estimator will have finite variance in typical experimental settings, such as when all functions in the model spaces have finite fourth moments or when each individual Riesz representor has finite essential supremum.
However, while uncommon in practice, it is possible to construct settings where the Riesz estimator has infinite variance.

\subsection{Variance Characterizing Operator}

The variance of the Riesz estimator is captured by a linear operator on the combined model space $\Mprodspace$, as detailed in the following theorem.
An analysis of this operator thus provides a way to investigate the precision of the estimator.

\expcommand{\thmvclostatement}{%
If the Riesz estimator has finite variance, there exists a bounded linear operator $\varmap : \Oprodspace \to \Oprodspace$ such that, for all $\vU \in \Oprodspace$,
\[
	n \Var[\big]{ \estv{\vU} } = \norm{ \varmapv{ \vU } }^2.
\]
}

\begin{theorem}\label{thm:vclo-statement}
\thmvclostatement
\end{theorem}

The theorem states that the norm of the evaluation of $\varmap$ on the combined potential outcomes $\voo$ gives the exact, finite-sample variance.
The norm is the extension to the combined space defined above: $\norm{\vU}^2 = \frac{1}{n} \sumin \E{ U_i^2 }$.
Because of this property, we refer to $\varmap$ as the \emph{variance characterizing operator}.

While the evaluation $\varmapv{ \voo }$ is inaccessible because the potential outcome functions are unknown, the definition of the operator $\varmap$ itself only involves aspects of the experimental design, model spaces and effect functionals.
The operator is therefore known at the design stage, and can be studied.
Because the estimator is unbiased, the scaled operator norm of $\varmap$ coincides exactly with the uniform mean square error of the estimator.

\expcommand{\cororieszestconsistent}{%
When the Riesz estimator has finite variance, $\uerr = n^{-1/2} \opnorm{\varmap}$.
}

\begin{corollary}\label{coro:riesz-est-consistent}
\cororieszestconsistent
\end{corollary}

The corollary provides a recipe for studying consistency of the Riesz estimator in any setting: asymptotic rates of the operator norm $\opnorm{\varmap}$ directly translate to rates of uniform consistency for the Riesz estimator.
For example, the Riesz estimator is uniformly consistent if and only if $\opnorm{\varmap} = \littleO{n^{1/2}}$, and it is root-$n$ consistent if and only if $\opnorm{\varmap} = \bigO{1}$.
While analyzing the operator norm can be a challenging exercise in practice, it is conceptually straightforward.
In Section~\suppref{supp:uniform-consistency} of the supplement, we provide an explicit definition of the operator and provide examples of its construction.
When the model spaces have finite dimension, the operator can be represented by a positive semi-definite matrix, and the operator norm is its largest eigenvalue.

The idea of studying the operator norm of the variance characterizing operator is inspired by \citet{Efron1971Forcing} and \citet{Harshaw2024Balancing}, who use operator norms to characterize the variance and construct experimental designs in settings with binary treatments.
In Section~\suppref{sec:supp-dep-graph-var}, we compare the operator norm and the dependency graph analysis \citep{Chen2004Normal,Ross2011Fundamentals}, which is the dominant approach to prove consistency in the recent network causal inference literature \citep[see e.g.][]{Aronow2017Estimating, Li2022Random, Ogburn2024Causal}.
We show how a dependency graph analysis is implicitly bounding the operator norm, although the resulting bound is typically very conservative.
For this reason, an analysis which targets the operator norm directly will typically be tighter and more informative.
An example of this is the analysis in \citet{Kandiros2025Conflict}, who target the operator norm directly to obtain improved rates of convergence in network experiments.

%% file: tex/estvar.tex
\section{Variance Estimation and Inference}\label{sec:variance-estimation}

\subsection{Variance Estimation Through Tensorization}

We describe a general approach for estimating the variance of the Riesz estimator using the same estimation principle that we developed for point estimator itself.
But the variance estimation problem is not of the same structure as above, so a translation is needed.
In particular, in place of the outcome spaces, we must construct Hilbert spaces that are conducive to the construction of variance estimators.

The central insight that facilitates this construction is that the variance can be understood as a sum of bilinear forms evaluated at pairs of potential outcome functions:
\[
\Var[\big]{ \estv{\voo} }
= \Var[\Big]{ \frac{1}{n} \sum_{i=1}^n \rri \ooi }
= \frac{1}{n^2} \sum_{i=1}^n \sum_{j=1}^n \Cov{ \rri \ooi, \rrj \ooj }.
\]
The bilinear forms capturing the covariances are not themselves linear functionals, as they map from the Cartesian products $\Mspacei \times \Mspacej$, prohibiting us from directly applying the Riesz estimation principle.
We can, however, reinterpret the bilinear forms as linear functionals on the tensor product spaces $\Mspacei \otimes \Mspacej$ of pairs of units.
In particular, for every pair of units $(i,j) \in [n]^2$, there exists a unique linear functional $\covfij : \Mspacei \otimes \Mspacej \to \Reals$ such that
\[
\covfijv{ u_i \otimes u_j } = \Cov[\big]{ \rri u_i , \rrj u_j }
\quad
\text{for all}
\quad
u_i \in \Mspacei,\; u_j \in \Mspacej.
\]
The variance is thus the average of the covariance functionals $\covfijv{ \poi \otimes \poj }$ over pairs $(i,j) \in [n]^2$, where $\poi \otimes \poj$ is the tensor product of the true potential outcome functions.

To construct Riesz representors for these functionals, we must construct an appropriate Hilbert space associated with each tensor product $\Mspacei \otimes \Mspacej$.
The canonical construction of an inner product on a tensor product of Hilbert spaces is unsuitable for the current purpose because it does not correspond to an expectation of observable quantities.

The first step in constructing an appropriate Hilbert space is to define a bilinear form $\iprod{\cdot, \cdot}$ that is an expectation of observable quantities.
Every tensor $\tU \in \Mspacei \otimes \Mspacej$ can be written as a sum of simple tensors, $\tU = \sum_{k} u_{i,k} \otimes u_{j,k}$, where $u_{i,k} \in \Mspacei$ and $u_{j,k} \in \Mspacej$.
For any two tensors $\tU = \sum_{k} u_{i,k} \otimes u_{j,k}$ and $\tV = \sum_{\ell} v_{i,\ell} \otimes v_{j,\ell}$, we define $\iprod{\tU, \tV} = \sum_{k} \sum_{\ell} \E[\big]{ u_{i,k} u_{j,k} v_{i,\ell} v_{j,\ell}}$.
We can interpret this as if we are associating each tensor $\tU$ with a function $\sum_{k} u_{i,k} u_{j,k}$, and defining the bilinear form as the expectation of their products.
The bilinear form induces a seminorm: $\norm{\tU} = \sqrt{\iprod{\tU, \tU}}$.
We must ensure that $\norm{\tU} < \infty$ for all tensors.
This is not guaranteed by construction, but is ensured by assuming bounded fourth moments.

\begin{assumption}\label{ass:fourth-moments}
	For all $i \in [n]$ and $u \in \Mspacei$, the fourth moment exists: $\E{ u^4 } < \infty$.
\end{assumption}

The \emph{paired outcome space} for units $(i,j) \in [n]^2$ is $\pairedOspaceij = \textrm{cl}\paren[\big]{ \Mspacei \otimes \Mspacej / \mathcal{N}_{i,j}}$, where $\mathcal{N}_{i,j} = \setb{\tU \in \Mspacei \otimes \Mspacej: \norm{\tU} =  0}$.
This is the closure of the quotient space induced by the seminorm, similar to the outcome space in Section~\ref{sec:model-spaces}.
Equipped with the induced inner product, the paired outcome space $\pairedOspaceij$ is a Hilbert space of the requisite form.

\subsection{Second Order Positivity and Variance Bounds}\label{sec:second-order-positivity}

As before, we can only estimate aspects of paired potential outcome functions that the experimental design provides information about, as formalized in the following definition.

\begin{definition}\label{def:second-order-positivity}
	A linear functional $C_{i,j} : \Mspacei \otimes \Mspacej \to \Reals$ satisfies \emph{second order positivity} if there exists $K > 0$ such that
	$
	\abs{ \covfijv{ \tU } - \covfijv{ \tV } }
	\leq
	K \cdot \norm{\tU - \tV}
	$
	for all tensors $\tU, \tV \in \Mspacei \otimes \Mspacej$.
\end{definition}

\expcommand{\thmsecondorderrieszrep}{%
If linear functional $\covfij : \Mspacei \otimes \Mspacej \to \Reals$ satisfies second order positivity, there exists a unique $\crrij \in \pairedOspaceij$ such that $\covfijv{ u_i \otimes u_j } = \E{ \crrij u_i u_j }$ $\forall \ u_i \in \Mspacei$, $u_j \in \Mspacej$.
}

\begin{theorem}\label{thm:second-order-riesz-rep}
\thmsecondorderrieszrep
\end{theorem}

Theorem~\ref{thm:second-order-riesz-rep} shows that second order positivity is sufficient to ensure unbiased estimation of a linear functional $\covfij$ on the tensor products $\Mspacei \otimes \Mspacej$.
We here use $\crrij$ to denote both the tensor and its associated random variable.

One of the central challenges for variance estimation in a design-based setting is that there are some covariance functionals that do not satisfy second order positivity.
It is widely recognized that the variance is generally not point identified in a design-based setting \citep{Imbens2015Causal}.
The conventional solution to this problem is to construct an estimator of an upper bound on variance, acting as a conservative variance estimator, and we follow this approach here.
However, there are rare situations in which second order positivity holds for all covariance functionals so that the variance can be estimated without bias (and consistently).
For example,
\citet{Harshaw2023Design} describes one such setting under a linear response model.
In addition, the spatial point process illustration in Section~\ref{sec:overview} is another example, as show in Section~\suppref{sec:supp-spatial-experiments} of the supplement.

\begin{definition}\label{def:variance-bound}
	A \emph{variance bound} is a functional $\vb : \Mprodspace \to \Reals$ such that $\vbv{\vu} \geq \Var{ \estv{\vu} }$ for all $\vu \in \Mprodspace$.
	A variance bound is \emph{estimable} if it admits a decomposition
	$\vbv{\vu} = n^{-2}\sum_{i=1}^n \sum_{j=1}^n \boufijv{ u_i \otimes u_j }$,
	where $\boufij : \Mspacei \otimes \Mspacej \to \Reals$ are linear functionals satisfying second order positivity.
\end{definition}

We describe two approaches for constructing estimable variance bounds in Section~\suppref{supp:bound-functionals} of the supplement.
The first is a generalization of the variance bound described by \citet{Aronow2013Conservative} and the second is based on the operator norm of the variance characterizing linear operator.
The two variance bounds are incomparable, in the sense that there are situations where one will be more conservative than the other, but the first bound will be less conservative in typical settings.
The benefit of the bound based on the operator norm is that the expected width of the resulting confidence interval always shrinks at the same rate as the variance of the point estimator, which is by not guaranteed by the Aronow-Samii-type bound.
It is possible to improve both of these bounds using the techniques described by \citet{Harshaw2026Optimized}, at the cost of additional complexity and computation.

\subsection{Riesz Variance Estimator} \label{sec:riesz-var-est}

With a variance bound in hand, Theorem~\ref{thm:second-order-riesz-rep} can be applied to construct a Riesz representor $\crrij$ for each bound functional $\boufij$.
The average of the product of these representors with the observed outcomes is our estimator of the variance of the Riesz estimator:
\[
\evb = \frac{1}{n^2} \sum_{i=1}^n \sum_{j=1}^n \crrij \ooi \ooj.
\]
Because the bound functionals are constructed to satisfy second-order positivity, each term is unbiased for the evaluation of the bound functional on the tensor corresponding to the true potential outcomes $\boufijv{ \poi \otimes \poj }$, resulting in the following theorem.

\expcommand{\thmconservativevarest}{%
Given correctly specified model spaces, first order positivity and existence of fourth moments (Assumptions~\mainref{ass:correctly-specified},~\mainref{ass:positivity}~and~\mainref{ass:fourth-moments}), the variance bound estimator is conservative in expectation for the variance: $\E{ \evb } \geq \Var{ \est }$.
}

\begin{theorem}\label{thm:conservative-var-est}
\thmconservativevarest
\end{theorem}

The magnitude of the bias is not easily characterized.
When there are severe positivity violations among the covariance functionals, the bias can be sizable.
In Section~\suppref{supp:var-est-consistency} of the supplement, we define uniform consistency of variance (bound) estimators and discuss techniques to prove consistency of the variance estimator.

\subsection{Confidence Intervals}

As noted in the previous section, our framework does not on its own facilitate a succinct and general characterization of the full sampling distribution of the Riesz estimator.
As an illustration of how the limiting distribution can be characterization when more structure is imposed, we follow the recent literature and use Stein's method with dependency graphs to provide sufficient conditions for asymptotic normality of the Riesz estimator in Section~\suppref{supp:dependency-graph-clt} of the supplement.
Other standard techniques for proving asymptotic normality under triangular array asymptotics, such Lindeberg and martingale central limit theorems, can also be applied when appropriate.
In situations where the sampling distribution cannot be well-approximated, experimenters can construct confidence intervals based on Chebyshev's inequality.
Chebyshev-type intervals will generally be quite conservative, but their width shrinks towards zero at the same rate as Wald-type intervals based on normal approximations, meaning that they are similarly informative in sufficiently large samples.

%% file: tex/illustration-total.tex
\section{Numerical Simulations}\label{sec:num-illustrations}

To demonstrate the practicality of the framework, we provide two numerical simulations based on the two illustrative examples in Section~\ref{sec:overview}: network experiments with continuous treatments and spatial point process experiments.
We present the first simulation in full here, and present the second in Section~\suppref{sec:supp-spatial-experiments} of the supplement due to space considerations.

\input{\texpath/illustration-continuous.tex}

%% file: tex/illustration-continuous.tex

Following Section~\ref{sec:continuous-treatments}, we focus on the estimation of marginal spillover effects of real-valued treatments in a network experiment.
The intervention is a vector $z = \paren{z_1, \dotsc, z_n}$ of continuous treatments in $[-1,1]$ so that the intervention set is $\Zspace = [-1,1]^n$.
We fix the experimental design to be the uniform measure over $\Zspace$, meaning that the individual treatments are independent and uniformly distributed on $[-1,1]$.
The causal estimand is given by the effect functional $\efvi{\poi} = \frac{d}{dt} \povi{ t \cdot e_{-i}} \mid_{t=0}$, which captures the spillover effect of a marginal increase in the treatment assigned to all other units starting at zero.

In this numerical simulation, we focus on a considerably more flexible model than that presented in Section~\ref{sec:continuous-treatments}.
Given a network $G$, let $\neighi \subseteq [n]$ be the neighbors of unit $i$ in the network and define the extended neighborhood as $\extneighi = \neighi \cup \setb{i}$.
Let $\Mspacei$ contain all multivariate polynomials in the neighborhood treatments of degree at most $r$, meaning that the potential outcomes take the form
\[
\povi{z}
= \sum_{ \substack{
	k : \extneighi \to \setb{0,\dots r}
	\\
	\sum_{j \in \extneighi} k(j) \leq r
	}
	}
	\alpha_{i,k}
	\prod_{j \in \extneighi} z_j^{k(j)}
	\enspace,
\]
where each function $k : \extneighi \to \setb{0 \dots r}$ represents the monomial $\prod_{j \in \extneighi} z_j^{k(j)}$.
The restriction $\sum_{j \in \extneighi} k(j) \leq r$ ensures that each monomial has degree at most $r$.

By allowing for higher-order interactions between each unit and its neighbors, the model accommodates more complex spill-overs than those described in Section~\ref{sec:continuous-treatments}.
For example, the model space contains all possible $r$-wise interactions between individual treatments.
In addition, this model allows for non-anonymous interference in the sense that each treatment $z_j$ can affect the outcome differently.
Despite this added complexity, the Riesz estimator can be shown to still be root-$n$ consistent in this setting where $r$ and $d$ are constant.

{
\setstretch{1}
\begin{table}[t]
\centering
\caption{Simulation Results: Spillover Effects of Continuous Treatments}\label{tab:continuous-sim-res}
\begin{tabular}{rrrrrrrrrrr}
	\toprule
	$d$ & $r$ & $n$ & dim & MSE & Bias & $\Varsym$ & $\evb$ & CI $\Varsym$ & CI $\evb$ & Width \\ \midrule
	\input{\simspath/continuous/table-3-3.tex}\\[0.6em]
	\input{\simspath/continuous/table-4-3.tex}\\[0.6em]
	\input{\simspath/continuous/table-4-4.tex}\\ \bottomrule
\end{tabular}
\end{table}
}

We run the simulation for different values of $(d, r)$, being set to $(3, 3)$, $(4, 3)$ and $(4, 4)$.
For each value of $(d, r)$, we run three sample sizes $n$, being set to $10^2$, $10^3$ and $10^4$.
In each setting, we construct a single network exhibiting homophily as well as a set of potential outcome functions which are heterogeneous across units.
Details of this procedure are given in Section~\suppref{sec:supp-network-experiment} of the supplement.
The number of Monte Carlo rounds for each setting is $300,000$.
The results from the simulation study are presented in Table~\ref{tab:continuous-sim-res}.
The first three columns describe the studied setting and ``dim'' gives the corresponding dimension of the model spaces.

The column ``MSE'' gives the mean square error of the Riesz estimator relative to the average second moment of the outcome: $\E{\paren{\est - \ate}^2} / n^{-1} \sumin \E{\ooi^2}$.
We see that the mean square error is large for small sample sizes, but decreases linearly in $n$, as expected given root-$n$ consistency.
The column ``Bias'' gives the squared bias relative to the mean square error: $\paren{\E{\est} - \ate}^2 / \E{\paren{\est - \ate}^2}$.
This is zero within three digits of precision, confirming that the estimator indeed is unbiased.
The column ``$\Varsym$'' gives the variance relative to the mean square error: $\Var{\est} / \E{\paren{\est - \ate}^2}$, which, as expected, is one.

The column ``$\evb$'' gives the expectation of the variance estimator relative to the true variance: $\E[\big]{\evb} / \Var{\est}$.
For the smaller model spaces in the first six rows, the bias of the variance estimator is moderate, being between $40\%$ to $45\%$ greater than the true variance irrespectively of the sample size.
For the larger model spaces in the last three rows, the bias is larger at approximately $90\%$.
The large bias will hurt power and is therefore problematic, but the magnitude of the bias is not greater than similar design-based variance estimators for binary treatment under interference \citep{Harshaw2026Optimized}.

The next two columns give coverage rates for $95\%$ Wald-type confidence intervals.
The column ``CI $\Varsym$'' gives coverage rates of intervals based on the true variance.
This interval is infeasible, and its purpose is to give an indication of the appropriateness of the normal approximation underlying the confidence intervals.
We see that the interval covers at or very close to the nominal rate, indicating that the normal approximation is appropriate.
In Section~\suppref{sec:supp-network-experiment} of the supplement, we provide QQ plots of the sampling distributions that corroborate this finding.
The column ``CI $\evb$'' gives coverage rates of intervals based on the estimated variance.
Due to the bias of the variance estimator, these intervals are wider with higher coverage rates.
The coverage rates are between $98\%$ and $99\%$ for the smaller model spaces, and above $99\%$ for the larger model spaces.
However, the relative conservativeness does not increase with $n$, and the width of the confidence intervals shrinks at a root-$n$ rate, as shown in the last column, titled ``Width'', which gives the relative width of the intervals.

%% file: sims/continuous/table-3-3.tex
    3  &      3  &    100  &     20  &  1.799  &  0.000  &  1.000  &  1.428  &  0.950  &  0.990  &  1.000 \\ 
    3  &      3  &   1000  &     20  &  0.175  &  0.000  &  1.000  &  1.448  &  0.950  &  0.983  &  0.330 \\ 
    3  &      3  &  10000  &     20  &  0.018  &  0.000  &  1.000  &  1.443  &  0.950  &  0.981  &  0.106

%% file: sims/continuous/table-4-3.tex
    4  &      3  &    100  &     35  &  4.607  &  0.000  &  1.000  &  1.396  &  0.951  &  0.991  &  1.000 \\ 
    4  &      3  &   1000  &     35  &  0.458  &  0.000  &  1.000  &  1.401  &  0.951  &  0.981  &  0.338 \\ 
    4  &      3  &  10000  &     35  &  0.046  &  0.000  &  1.000  &  1.399  &  0.950  &  0.980  &  0.109

%% file: sims/continuous/table-4-4.tex
    4  &      4  &    100  &     70  &  5.448  &  0.000  &  1.000  &  1.900  &  0.953  &  1.000  &  1.000 \\ 
    4  &      4  &   1000  &     70  &  0.545  &  0.000  &  1.000  &  1.888  &  0.951  &  0.995  &  0.340 \\ 
    4  &      4  &  10000  &     70  &  0.055  &  0.000  &  1.000  &  1.887  &  0.950  &  0.993  &  0.111

%% file: tex/conclusions.tex
\section{Concluding Remarks}

The framework we have described in this paper and the associated Riesz estimator allow empirical researchers to investigate a wide range of causal question using design-based, experimental methods.

Several open questions remain.
The Riesz estimator can be seen as a generalization of the Horvitz--Thompson estimator, and the Riesz estimator inherits many of its drawbacks.
In particular, both estimators achieve unbiasedness at all costs, and they can therefore have large variance.
In the conventional setting with discrete treatments, experimenters often use the Hájek estimator or a covariate-adjusted estimator in place of the Horvitz--Thompson estimator.
This typically leads to a noticeable reduction in variance at the cost of introducing small and vanishing bias.
Developing such versions of the Riesz estimator is important future work.

Concerns about precision tend to be particularly pressing when the model spaces are large or have infinite dimensions.
Large model spaces by themselves do not imply that the Riesz estimator performs poorly, and the estimator can perform well also under infinite-dimensional model spaces.
However, in many settings with large model spaces, unbiasedness can be achieved only by accepting very large variance.
Future work should better delineate these situations, and describe alternatives in settings where the Riesz estimator is not useful due to being overly imprecise.
We believe the best candidate for such an alternative estimation is a sieve version of the Riesz estimator that represents the effect functional on growing subspaces of the model spaces.

Another important open question is how the Riesz estimator behaves when the assumption of correctly specified model spaces does not hold.
The fact that the model spaces can be large, possibly infinite-dimensional, means that the assumption of correct specification might be less problematic here than in the conventional setting.
But it is nevertheless a strong assumption.
An investigation of the Riesz estimator under misspecification is important future work.
We conjecture that this investigation will reveal connections to the sieve version of the estimator.

%% file: tex/supp-illustrative-worked-examples.tex
\section{Worked Illustrative Examples of the Framework} \label{sec:supp-worked-examples}

\subsection{Exposure Mapping Models} \label{sec:supp-worked-exposure-mapping}

In this subsection, we show how the exposure mapping framework of \citet{Aronow2017Estimating} fits within the more general design-based framework proposed here.
In particular, we show that a correctly specified exposure mapping assumption can be expressed in terms of a particular model space.
We show that in this setting, our general positivity condition reduces to the usual positivity condition and that the Riesz estimator is precisely the Horvitz--Thompson estimator.

\paragraph{Model Space}
In an exposure mapping model, each unit $i \in [n]$ is associated with an \emph{exposure mapping} $\expmapi : \Zspace \to \expseti$ that maps interventions $z \in \Zspace$ to an exposure $e$ in the \emph{exposure set} $\expseti$.
In typical scenarios, the exposure set $\expseti$ contains a small finite number of exposures, e.g. 4 or 5.
The corresponding \emph{exposure mapping model} $\Mspacei$ contains all functions $\poi : \Zspace \to \Reals$ satisfying the following condition: for any two interventions $z, z' \in \Zspace$ with $\expmapi(z) = \expmapi(z')$ then $\povi{z} = \povi{z'}$.

The exposure mapping model $\Mspacei$ can be equivalently formulated as containing all functions of the form
\[
\povi{z} = \sum_{e \in \expseti} \alpha_{i,e} \indicator{\expmapi(z) = e} \enspace.
\]
This is a parameterization of the model space using exposure indicators as basis functions and $\alpha_{i,e}$ as coefficients.

\paragraph{Effect Functionals}
Using this parameterization of an exposure mapping model, we can see that any linear functional $\efi : \Mspacei \to \Reals$ takes the form
\begin{equation} \label{eq:generic-functional-em}
\efi(\poi) = \sum_{e \in \expseti} \beta_{i,e} \cdot \alpha_{i,e}
\end{equation}
for some set of coefficients $\beta_{i,e}$ for $e \in \expseti$.
To the best of our knowledge, all causal effect which have been studied in the exposure mapping literature can be expressed in this way.
We list two types of effects below:

\begin{itemize}
	\item \textbf{Contrastive Effects}:
	Contrastive effects are expressed as the difference of the potential outcome function evaluated at two different interventions:
	\[
	\efvi{\poi} = \povi{z_{i,1}} - \povi{z_{i,0}}
	\]
	for interventions $z_{i,1}, z_{i,0} \in \Zspace.$
	The subscript denotes the possibility that the two different interventions under consideration may depend on the particular unit.
	For example, the direct effect of treatment may be expressed by comparing the interventions $z_{i,1} = \vec{e}_i$ and $z_{i,0} = \vec{0}$, where $\vec{e}_i$ is the vector with 1 in entry $i$ and 0 in all other entries, and $\vec{0}$ is the zero vector.
	Similarly, the global average treatment effect may be expressed by comparing the interventions $z_{i,1} = \vec{1}$ and $z_{i,0} = \vec{0}$ which correspond to treating everyone and treating no one, respectively.

	To express contrastive effects in the generic form \eqref{eq:generic-functional-em}, we set
	\[
	\beta_{i,e} =
		\left\{
			\begin{array}{lr}
			1 &\text{ if } \expmapi(z_{i,1}) = e\\
			-1 &\text{ if } \expmapi(z_{i,0}) = e \\
			0 &\text{ otherwise}
			\end{array}
		\right.
	\enspace.
	\]

	\item \textbf{Expected Policy Effects}:
	Policy effects are expressed as the expectation of the observed outcome under some distribution over interventions.
	Let $\mathcal{Q}$ be a probability measure over interventions $\Zspace$ and define the marginal effect functional as
	\[
	\efvi{\poi} = \Esub{z \sim \mathcal{Q}}{\povi{z}} \enspace.
	\]
	This type of marginal effect functional is used in off-policy evaluation, where $\mathcal{Q}$ represents a policy which is different from the experimental design.

	To express marginal effects in the generic form \eqref{eq:generic-functional-em}, we may set the coefficients $\beta_{i,e}$ for each exposure based on the probability of observing that exposure under $\mathcal{Q}$:
	\[
	\beta_{i,e} = \mathcal{Q}( \setb{ z : \expmapi(z) = e } )
	\enspace.
	\]
\end{itemize}

\paragraph{Positivity}
Within an exposure mapping model, the general positivity condition stated in Assumption~\mainref{ass:positivity} reduces to the usual positivity condition: that the probability of observing each estimand-relevant exposure is non-zero.
This is formalized in the following proposition:

\begin{proposition} \label{prop:positivity-in-em}
	Consider an exposure mapping model $\Mspacei$ and a linear functional $\efi$ parameterized with coefficients $\beta_{i,e}$ as above.
	Positivity holds if and only if for every experimental unit $i \in [n]$, $\Pr{\expmapi(z) = e} > 0$ for all exposures $e \in \expseti$ with $\beta_{i,e} \neq 0$.
\end{proposition}

\paragraph{Riesz Estimator}
Within an exposure mapping model, the Riesz estimator is precisely a Horvitz--Thompson estimator.
As discussed in Section~\suppref{sec:supp-riesz-construction}, there are many equivalent ways to derive the Riesz estimator.
For illustration purposes, we will choose the method of constructing an orthonormal basis described in Section~\suppref{supp:orthogonalization-procedure}.

Recall that the basis for the exposure mapping model $\Mspacei$ is given by
\[
b_{i,e} = \indicator{\expmapi(z) = e}
\text{ for all exposures } e \in \expseti
\enspace.
\]
Running the modified Gram-Schmidt procedure (Algorithm~\ref{alg:mod-gso} in Section~\suppref{supp:orthogonalization-procedure}), we obtain the two sets of functions:
\begin{align*}
O_i &= \setb[\Big]{ \frac{\indicator{\expmapi(z) = e}}{\sqrt{\Pr{\expmapi(z) = e}}} \text{ for } e \in \expseti \text{ such that }\Pr{\expmapi(z) = e} > 0 } \\
\text{ and }
N_i &= \setb[\Big]{ \indicator{\expmapi(z) = e} \text{ for } e \in \expseti \text{ such that }\Pr{\expmapi(z) = e} = 0  }
\enspace.
\end{align*}
Observe that the functions in $N_i$ are zero almost surely under the experimental design.
On the other hand, the functions in $O_i$ are orthonormal in the sense that for any pair $h_{i,e}, h_{i,e'} \in O_i$, we have that
\[
\E{ h_{i,e}(z) \cdot h_{i,e'}(z) }
= \E[\Bigg]{ \frac{\indicator{\expmapi(z) = e}}{\Pr{\expmapi(z) = e}} \cdot \frac{\indicator{\expmapi(z) = e'}}{\Pr{\expmapi(z) = e'}} }
= \indicator{e = e'}
  \enspace.
\]
Using the procedure for checking positivity in Section~\suppref{supp:positivity-holds}, we can see that it holds exactly when the effect evaluates to zero on functions in $N_i$, i.e. that $\beta_{i,e} = 0$ for those exposures $e$ with $\Pr{\expmapi(z) = e} = 0$.
In fact, this observation provides one proof of Proposition~\suppref{prop:positivity-in-em}.
For the rest of this section, let us now presume that positivity holds.

Following the recipe provided in Section~\suppref{supp:orthogonalization-procedure}, the Riesz representor is given by the explicit function
\[
r_i(z) = \sum_{h_{i,e} \in O_i} \efvi{h_{i,e}} \cdot h_{i,e}
= \sum_{e \in \expseti} \frac{\beta_{i,e}}{\sqrt{\Pr{\expmapi(z) = e}}} \cdot \frac{\indicator{\expmapi(z) = e}}{\sqrt{\Pr{\expmapi(z) = e}}}
= \sum_{e \in \expseti} \beta_{i,e} \cdot \frac{\indicator{\expmapi(z) = e}}{\Pr{\expmapi(z) = e}},
\]
which is precisely a Horvitz--Thompson estimator.

\paragraph{Asymptotic Properties}
Even within an exposure mapping model, the model and estimand alone are not enough information to characterize the asymptotic properties of the Horvitz--Thompson estimator.
In Section~6 of their paper, \citet{Aronow2017Estimating} provide generic dependency graph conditions which are sufficient to establish consistency and asymptotic normality of the Horvtiz--Thompson estimator.
In this paper, we have taken a similar approach by providing sufficient conditions based on dependency graphs.

\subsection{Continuous Treatment Network Experiment}

The primary goal in this section is to derive the Riesz estimator for the illustration of network marginal spill-over effects in continuous treatments presented in Section~\mainref{sec:continuous-treatments}.
Along the way, we will also determine the relevant positivity assumption.
Finally, we discuss applications of the dependency graph method to establishing consistency and a CLT.

Let us begin by recalling the setting of the illustration.
The set of interventions is $\Zspace = [0,1]^n$, so that each unit receives a continuous treatment $z_i \in [0,1]$.
We consider a generic experimental design $\Zprob$ under which the individual treatments $z = (z_1, \dots z_n)$ are mutually independent.
The units are vertices in a (possibly directed) network $G$ where their neighborhoods are denoted as $\neighi \subseteq [n]$.
We define the extended neighborhood as $\extneighi = \neighi \cup \setb{i}$.

The marginal spill-over effect of interest is given by the individual effect functional
\[
\efvi{y} = \frac{\partial}{\partial t} y\paren{s + t \cdot e_{-i}} \mid_{t=0}
\enspace,
\]
where $s = (s_1 \dots s_n) \in [0,1]^n$ is a fixed subsidy and $e_{-i} \in \Reals^n$ is a vector whose $i$th entry is 0 and all other entries are 1.

Recall that the model $\Mspacei$ considered in Section~\mainref{sec:continuous-treatments} consists of all functions $\poi$ of the form
\[
\povi{z} =
\alpha_{i,0}
+ \alpha_{i,1} \cdot z_i
+ \alpha_{i,2} \cdot \frac{1}{\abs{\neighi}} \sum_{j \in \neighi} z_j
+ \alpha_{i,3} \cdot z_i \cdot \paren[\Big]{\frac{1}{\abs{\neighi}} \sum_{j \in \neighi} z_j}
\enspace.
\]
This specification constitutes a basis for the model space $\Mspacei$ where the basis functions are given by
\[
b_{i,0}(z) = 1,
\quad
b_{i,1}(z) = z_i,
\quad
b_{i,2}(z) = \frac{1}{\abs{\neighi}} \sum_{j \in \neighi} z_j,
\quadand
b_{i,3}(z) = z_i \cdot \paren[\Big]{\frac{1}{\abs{\neighi}} \sum_{j \in \neighi} z_j}
\enspace.
\]
For notational simplicity, let us define the function $u_i(z) = \abs{\neighi}^{-1} \sum_{j \in \neighi} z_j$.
In a slight abuse of notation, we will write $u_i$ and remember that this is actually a function of $z$.

\paragraph{Derivation of Riesz Estimator}
In order to proceed with the derivation of the Riesz estimator and evaluation of the positivity condition, we use the procedures described in Section~\suppref{sec:supp-riesz-construction}.
The first step is to carry out the modified Gram-Schmidt orthogonalization procedure described in Section~\suppref{supp:orthogonalization-procedure}.
For simplicity, we suppose that the individual variances $\Var{z_i}$ and $\Var{u_i}$ are positive and return to this presumption and its connection to positivity later in the section.
In this case, the modified Gram--Schmidt procedure returns the set of functions
\begin{multline}
h_{i,0}(z) = 1,
\quad
h_{i,1}(z) = \frac{z_i - \E{z_i}}{\sqrt{\Var{z_i}}},
\quad
h_{i,2}(z) = \frac{u_i - \E{u_i}}{\sqrt{\Var{u_i}}},
\\
\quadand
h_{i,3}(z) = \frac{\paren{z_i - \E{z_i}} \cdot \paren{u_i - \E{u_i}}}{ \sqrt{ \Var{z_i} \cdot \Var{u_i} } }
\enspace.
\end{multline}
Using the fact that the individual treatments $z_1 \dots z_n$ are independent and
the variances $\Var{z_i}$ and $\Var{u_i}$ are positive, these functions are orthonormal in the sense that
$\E{ h_{i,k}(z) \cdot h_{i,\ell}(z) } = \indicator{k = \ell}$.

The next step is to calculate the functional on these basis functions.
Simple calculations show that
\[
\efvi{h_{i,0}} =
\efvi{h_{i,1}} = 0,
\quad
\efvi{h_{i,2}} = \frac{1}{\sqrt{\Var{u_i}}},
\quadand
\efvi{h_{i,3}} = \frac{s_i - \E{z_i}}{\sqrt{\Var{z_i} \Var{u_i}}}
\enspace.
\]

Following the recipe provided in Section~\suppref{supp:orthogonalization-procedure}, the Riesz representor will be given by the explicit function
\begin{align*}
	r_i(z)
	&= \sum_{k=0}^3 \efvi{h_{i,k}} \cdot h_{i,k}(z)
	\\
	&= \frac{1}{\sqrt{\Var{u_i}}} \cdot \frac{(u_i - \E{u_i})}{\sqrt{\Var{u_i}}}
	+ \frac{(s_i - \E{z_i})}{\sqrt{ \Var{z_i} \Var{u_i} }}
	\cdot \frac{(z_i - \E{z_i})(u_i - \E{u_i})}{\sqrt{ \Var{z_i} \Var{u_i} }}
	\\
	&= \frac{(u_i - \E{u_i})}{\Var{u_i}} + \frac{(s_i - \E{z_i})(z_i - \E{z_i})(u_i - \E{u_i})}{\Var{z_i} \Var{u_i}}
	\enspace.
\end{align*}
Substituting $u_i = \abs{\neighi}^{-1} \sum_{j \in \neighi} z_j$ into the above expression yields the Riesz estimator presented in Section~\mainref{sec:continuous-treatments}.

\paragraph{Positivity}
In the above derivation, we presumed that $\Var{z_i}$ and $\Var{u_i}$ were positive.
Not only did this simplify the modified Gram--Schmidt orthogonalization, but it actually implies that positivity held.
Now, we turn our attention to examining the positivity condition more closely.
The following proposition derives the exact positivity conditions required under this model.

\begin{proposition} \label{prop:positivity-continuous-network}
	In the example above, positivity holds if and only if for each unit $i \in [n]$, the following condition holds:
	\[
	\Var{u_i} > 0 \quadand \textrm{ if } s_i \neq \E{z_i} \textrm{ then } \Var{z_i} > 0
	\enspace.
	\]
\end{proposition}

The proof may be demonstrated by the arguments in Section~\suppref{supp:positivity-holds}.
In particular, positivity holds if and only if effect functional $\efi$ evaluates to zero on all functions which are $0$ almost surely under the design.
By going through the Gram--Schmidt procedure, we see that the effect functional is non-zero on $u_i - \E{u_i}$, which means that $\Var{u_i} > 0$ is required.
Similarly, we see that if $s_i \neq \E{z_i}$, then the effect functional is non-zero on $(u_i = \E{u_i})(z_i - \E{z_i})$, which means that $\Var{z_i} > 0$ is required in this case.

\paragraph{Asymptotic Properties}
The asymptotic behavior of the Riesz estimator in this context will depend on the underlying network $G$ and the experimental design $\Zprob$.
In this final part of the subsection, we provide a sketch of how the dependency graph analysis derived in Sections~\ref{sec:supp-dep-graph-var} and \ref{supp:dependency-graph-clt} may be used to analyze the consistency and asymptotic normality of the estimator in this setting.
Throughout this section, let $\dmax = \max_{i \in [n]} \abs{\neighi}$ be the largest degree in the network $G$.
For simplicity, we suppose that the network $G$ is undirected.

To analyze consistency, we will use Proposition~\suppref{prop:dep-graph-op-norm} which gives general consistency analysis via dependency graph techniques.
The first step is to determine the largest size of the dependency neighborhoods.
Consider a function $u_i \in \Mspacei$ and a function $u_j \in \Mspacej$.
Because
\begin{enumerate}
 \item $u_i$ is a function of the individual treatments $\setb{ z_k : k \in \extneighi }$ and $u_j$ is a function of the individual treatments $\setb{ z_k : k \in \extneigh{j} }$ and
 \item we are only considering experimental designs under which the individual treatments $z_1 \dots z_n$ are mutually independent,
\end{enumerate}
we arrive at the conclusion that the dependency neighborhoods are at most $\depneighi \subseteq \setb{ j \in [n] : \extneighi \cap \extneigh{j} \neq \emptyset }$.
In this way, we have the maximum size of a dependency neighborhood is $\depmax \leq \dmax^2$.

Next, we need to establish the essential supremum of the Riesz Representors.
To this end, suppose that we have an experimental design under which each individual treatment satisfies $\Var{z_i} \geq c$ for some constant $c > 0$.
Thus, the variance of $\Var{u_i}$ is lower bounded as
\[
\Var{u_i}
= \Var[\Big]{ \frac{1}{\abs{\neighi}} \sum_{j \in \neighi} z_j }
= \frac{1}{\abs{\neighi}} \cdot \paren[\Big]{ \frac{1}{\abs{\neighi}} \sum_{j \in \neighi} \Var{z_j} }
\geq \frac{c}{\abs{\neighi}}
\enspace.
\]
Thus, the essential supremum of the Riesz representor can be upper bounded as
\begin{align*}
	\norm{\rri}_{\infty}
	&= \norm[\Big]{ \frac{(u_i - \E{u_i})}{\Var{u_i}} + \frac{(s_i - \E{z_i})(z_i - \E{z_i})(u_i - \E{u_i})}{\Var{z_i} \Var{u_i}} }_{\infty}
	\\
	&\leq
	\norm[\Big]{ \frac{(u_i - \E{u_i})}{\Var{u_i}} }_{\infty}
	+ \norm[\Big]{ \frac{(s_i - \E{z_i})(z_i - \E{z_i})(u_i - \E{u_i})}{\Var{z_i} \Var{u_i}} }_{\infty}
	\\
	&\leq \Var{u_i}^{-1} + \Var{u_i}^{-1} \Var{z_i}^{-1}
	\\
	&\leq \frac{\abs{\neighi}}{c(1+c)}
	\\
	&\leq \frac{\dmax}{c(1+c)}
	\enspace,
\end{align*}
where we used that $\abs{u_i - \E{u_i}}, \abs{z_i - \E{z_i}}, \abs{s_i - \E{z_i}} \leq 1$ by virtue of the fact that they are in the interval $[0,1]$.

Using these two facts together, we can use Proposition~\suppref{prop:dep-graph-op-norm} to obtain that the uniform root mean square error $\uerr$ of the Riesz estimator goes to zero at the rate $\bigO{ \dmax^2 / \sqrt{n} }$.
This means that $\dmax = \littleO{n^{1/2}}$ is a sufficient condition for consistency.
Using the same techniques, one can use the Theorem~\ref{thm:clt-dep-graph} to obtain a Central Limit Theorem (under standard bounded 4th moment and non-superefficiency assumptions) for graphs which satisfy $\dmax = \littleO{n^{1/14}}$.
These calculations highlight the practicality of the generic dependency graph analysis developed in Sections~\ref{sec:supp-dep-graph-var} and \ref{supp:dependency-graph-clt}.

\subsection{Spatial Point Processes} \label{sec:worked-spatial-point-process}

Our third illustrative example is a spatial point process experiment, as described in Section~\mainref{sec:spatial-example}.
The intervention is here a set of sites placed in some geographical space, and a unit's response to such interventions may exhibit complex interactions depending on the spatial configuration of the intervention sites.

A point process experiment may be formalized using a compact subset $\mathcal{X} \subset \Reals^d$, where $d = 2$ would be used to describe planar geographies.
Each unit $i \in [n]$ is associated with a location $u_i \in \mathcal{X}$ related to the corresponding outcome $\ooi$.
The intervention is $m$ randomly chosen locations $z_1, \dotsc, z_m \in \mathcal{X}$, which we refer to as the \emph{treatment points}.
In the example in the main paper, phone towers would be built at the treatment points.
Each treatment point $z_j$ can take any of the uncountably many values in $\mathcal{X}$, and the number of points $m$ may itself be random.
This makes the experimental design a \emph{point process} on the underlying space $\mathcal{X}$, giving the experimental design its name.
Point processes can be understood as random discrete measures on a Polish space, as discussed in more detail by \citet{Hough2006Determinantal}.
The intervention space $\Zspace$ is therefore the set of all such discrete measures.

In Section~\mainref{sec:spatial-example}, we focused on the counter-factual policy effect defined via the individual effect functionals
\[
\efvi{y} = \Esub{Z \sim \offpol}{y\paren{Z}} - \Esub{Z \sim \Zprob}{y\paren{Z}} \enspace,
\]
where $\offpol$ is the counterfactual policy of interest and $\Zprob$ is the experimental design under consideration.
We presented a model space which accommodated direct effects and pair-wise interactions of the intervention sites.
In particular, each model space $\Mspacei$ contained all functions of the form
\[
\povi{z}
= \alpha_{i,0} b_{i,0}(z) + \alpha_{i,1} b_{i,1}(z) + \alpha_{i,2} b_{i,2}(z) \enspace,
\]
where the basis functions are given by
\begin{align*}
b_{i,0}(z) &= 1 \enspace,
\quad
b_{i,1}(z) = \sum_{t=1}^m \expf[\Big]{ - \frac{\norm{u_i - z_t}^2}{\sigma^2} },
\quad \text{ and }
\\
b_{i,2}(z) &= \sum_{j=1}^n \sum_{k=1}^n
\expf[\Big]{ - \frac{\norm{u_i - z_j}^2}{\sigma^2} }
\expf[\Big]{ - \frac{\norm{u_i - z_k}^2}{\sigma^2} }
\bracket[\Big]{
    1
    -
    \expf[\Big]{ - \frac{\norm{z_j - z_k}^2}{\sigma^2} }
}.
\end{align*}
For notational simplicity, we will sometimes write $b_{i,1}$ or $b_{i,2}$ without explicit reference to $z$, but keeping in mind that they are indeed functions.

We will keep this three-dimensional model space fixed throughout our worked example and in the simulation section.
However, we describe a class of determinantal models which express higher order interactions in Section~\ref{sec:determinantal-generalizations}.

\paragraph{Derivation of Riesz Representor}
In order to proceed with the derivation of the Riesz estimator and evaluation of the positivity condition, we use the procedures described in Section~\suppref{sec:supp-riesz-construction}.
The first step is to carry out the modified Gram-Schmidt orthogonalization procedure described in Section~\suppref{supp:orthogonalization-procedure}.
For simplicity, we will suppose that all denominator terms are positive and return to this presumption and its connection to positivity later in the section.
In this case, the modified Gram--Schmidt procedure returns the set of functions:
\[
h_{i,0}(z) = 1,
\quad
h_{i,1}(z) = \frac{b_{i,1} - \E{b_{i,1}}}{\sqrt{\Var{b_{i,1}}}},
\quad
h_{i,2}(z) = \frac{ (b_{i,2} - \E{ b_{i,2} }) - \frac{\Cov{b_{i,1}, b_{i,2}}}{\Var{b_{i,1}}} (b_{i,1} - \E{ b_{i,1} }) }{\sqrt{ \Var{b_{i,2}} - \frac{\Cov{b_{i,1}, b_{i,2}}^2}{\Var{b_{i,1}}} }}
\enspace.
\]
One can verify that these are orthogonal in the sense that $\E{ h_{i,k}(z) h_{i,\ell}(z) } = \indicator{k = \ell}$.
Using the definition of the correlation coefficient, i.e. $\Cov{X,Y} = \rho \sqrt{\Var{X} \Var{Y}}$, we may also express this last function more compactly as
\[
h_{i,2}(z) = \frac{ (b_{i,2} - \E{ b_{i,2} }) - \rho \sqrt{\frac{\Var{b_{i,2}}}{\Var{b_{i,1}}}} (b_{i,1} - \E{ b_{i,1} }) }{\sqrt{ (1 - \rho^2)\Var{b_{i,2}} }}
\enspace.
\]
The next step in constructing the Riesz representor is to evaluate these basis functions under the linear functional of interest.
To this end, we obtain that
\begin{align*}
\efvi{h_{i,0}} &= 0  \enspace,
\\
\efvi{h_{i,1}} &= \frac{ \Esub{\offpol}{b_{i,1}} - \Esub{\Zprob}{b_{i,1}} }{\sqrt{ \Var{b_{i,1}} }} \enspace,
\\
\efvi{h_{i,2}} &= \frac{ (\Esub{\offpol}{b_{i,2}} - \Esub{\Zprob}{ b_{i,2} }) - \rho \sqrt{\frac{\Var{b_{i,2}}}{\Var{b_{i,1}}}} ( \Esub{\offpol}{b_{i,1}} - \Esub{\Zprob}{ b_{i,1} }) }{\sqrt{ (1 - \rho^2)\Var{b_{i,2}} }}
\enspace,
\end{align*}
where we have used the subscript in the expectations to denote the off-policy measure $\offpol$ and the experimental design $\Zprob$.
Note that the variance and covariance terms are only with respect to $\Zprob$, so we do not add the subscript here for notational convenience.

Now, we may use the expression presented in Section~\suppref{supp:orthogonalization-procedure} to derive the Riesz representor as the explicit function
\[
r_i(z)
= \sum_{k=0}^2 \efvi{h_{i,k}} \cdot h_{i,k}(z)
= \efvi{h_{i,1}} \cdot h_{i,1}(z)
+ \efvi{h_{i,2}} \cdot h_{i,2}(z)
\enspace,
\]
where we have used that $\efvi{h_{i,0}} = 0$.
Now, the first term may be worked out as
\[
\efvi{h_{i,1}} \cdot h_{i,1}(z)
= \frac{(b_{i,1}(z) - \Esub{\Zprob}{b_{i,1}})(\Esub{\offpol}{b_{i,1}} - \Esub{\Zprob}{ b_{i,1} })}{\Var{b_{i,1}}}
\enspace.
\]
The second term can be derived as
\begin{align*}
	&\efvi{h_{i,2}} \cdot h_{i,2}(z)
	\\
	&=
	\frac{ (\Esub{\offpol}{b_{i,2}} - \Esub{\Zprob}{ b_{i,2} }) - \rho \sqrt{\frac{\Var{b_{i,2}}}{\Var{b_{i,1}}}} ( \Esub{\offpol}{b_{i,1}} - \Esub{\Zprob}{ b_{i,1} }) }{\sqrt{ (1 - \rho^2)\Var{b_{i,2}} }}
	\\
	&\quad \times
	\frac{ (b_{i,2} - \E{ b_{i,2} }) - \rho \sqrt{\frac{\Var{b_{i,2}}}{\Var{b_{i,1}}}} (b_{i,1} - \E{ b_{i,1} }) }{\sqrt{ (1 - \rho^2)\Var{b_{i,2}} }}
	\\
	&=
	\frac{
		(\Esub{\offpol}{b_{i,2}} - \Esub{\Zprob}{ b_{i,2} })
		(b_{i,2} - \Esub{\Zprob}{ b_{i,2} })
	}{(1-\rho^2)\Var{b_{i,2}}}
	+ \frac{
	\rho^2 \frac{\Var{b_{i,2}}}{\Var{b_{i,1}}}
	( \Esub{\offpol}{b_{i,1}} - \Esub{\Zprob}{ b_{i,1} })
	(b_{i,1} - \Esub{\Zprob}{ b_{i,1} })
	}{(1-\rho^2)\Var{b_{i,2}}}
	\\
	&\quad
	- \frac{
		\rho \sqrt{ \frac{\Var{b_{i,2}}}{\Var{b_{i,1}}} }
		\braces[\Big]{
			(\Esub{\offpol}{b_{i,2}} - \Esub{\Zprob}{ b_{i,2} })
			(b_{i,1} - \Esub{\Zprob}{ b_{i,1} })
			+
			( \Esub{\offpol}{b_{i,1}} - \Esub{\Zprob}{ b_{i,1} })
			(b_{i,2} - \Esub{\Zprob}{ b_{i,2} })
		}
	}{(1-\rho^2)\Var{b_{i,2}}}
	\\
	&=
	\frac{
		(\Esub{\offpol}{b_{i,2}} - \Esub{\Zprob}{ b_{i,2} })
		(b_{i,2} - \Esub{\Zprob}{ b_{i,2} })
	}{(1-\rho^2)\Var{b_{i,2}}}
	+ \frac{
	\rho^2
	( \Esub{\offpol}{b_{i,1}} - \Esub{\Zprob}{ b_{i,1} })
	(b_{i,1} - \Esub{\Zprob}{ b_{i,1} })
	}{(1-\rho^2)\Var{b_{i,1}}}
	\\
	&\quad
	- \frac{
		\rho
		\braces[\Big]{
			(\Esub{\offpol}{b_{i,2}} - \Esub{\Zprob}{ b_{i,2} })
			(b_{i,1} - \Esub{\Zprob}{ b_{i,1} })
			+
			( \Esub{\offpol}{b_{i,1}} - \Esub{\Zprob}{ b_{i,1} })
			(b_{i,2} - \Esub{\Zprob}{ b_{i,2} })
		}
	}{(1-\rho^2)\sqrt{ \Var{b_{i,1}} \Var{b_{i,2}} } }
\end{align*}
Combining this with the first term, we have that the Riesz representor is given by the explicit function
\begin{align*}
r_i(z)
&=
\frac{(\Esub{\offpol}{b_{i,1}} - \Esub{\Zprob}{ b_{i,1} })(b_{i,1}(z) - \Esub{\Zprob}{b_{i,1}})}{(1-\rho^2)\Var{b_{i,1}}}
+
\frac{
		(\Esub{\offpol}{b_{i,2}} - \Esub{\Zprob}{ b_{i,2} })
		(b_{i,2} - \Esub{\Zprob}{ b_{i,2} })
	}{(1-\rho^2)\Var{b_{i,2}}}
\\
&\quad + \frac{\rho}{(1-\rho^2)\sqrt{\Var{b_{i,1}} \Var{b_{i,2}}}}
M
\enspace,
\end{align*}
where
\[
M =
\braces[\Big]{
		(\Esub{\offpol}{b_{i,2}} - \Esub{\Zprob}{ b_{i,2} })
		(b_{i,1} - \Esub{\Zprob}{ b_{i,1} })
		+
		( \Esub{\offpol}{b_{i,1}} - \Esub{\Zprob}{ b_{i,1} })
		(b_{i,2} - \Esub{\Zprob}{ b_{i,2} })
	}
\enspace,
\]
which recovers the Riesz representor derived in the main body.

\paragraph{Positivity}
In the above derivation, we assumed that all relevant denominators were positive.
In fact, this ensured that the positivity condition held.
However, by carefully reviewing the output of the modified Gram--Schmidt procedure, we can work out the explicit condition required for positivity, given below as Table~\ref{tab:positivity-for-spatial-pp}:

\begin{table}[ht]
\centering
\begin{tabular}{|c|c|c|c|}
\hline
\textbf{$\Esub{\offpol}{b_{i,1}} = \Esub{\Zprob}{ b_{i,1} }$} & \textbf{$\Esub{\offpol}{b_{i,2}} = \Esub{\Zprob}{ b_{i,2} }$} & \textbf{Positivity} & \textbf{Riesz Representor $\rri$} \\ \hline
Yes     & Yes    & --     & $\rri = 0$     \\ \hline
No     & Yes     & $\Var{b_{i,1}} > 0$     & $\rri = \frac{(\Esub{\offpol}{b_{i,1}} - \Esub{\Zprob}{b_{i,1}})( b_{i,1} - \Esub{\Zprob}{b_{i,1}} )}{\Var{b_{i,1}}}$    \\ \hline
Yes     & No     & $\Var{b_{i,2}} > 0$     & $\rri = \frac{(\Esub{\offpol}{b_{i,2}} - \Esub{\Zprob}{b_{i,2}})( b_{i,2} - \Esub{\Zprob}{b_{i,2}} )}{\Var{b_{i,2}}}$    \\ \hline
No     & No     & \makecell{$\Var{b_{i,1}}> 0$ \\ $\Var{b_{i,2}}>0$ \\ $\rho^2 < 1$ } & as above \\ \hline
\end{tabular}
\caption{Positivity Conditions for Spatial Point Process Experiments}
\label{tab:positivity-for-spatial-pp}
\end{table}

Table~\ref{tab:positivity-for-spatial-pp} gives four different conditions based on whether or not the two non-constant basis functions $b_{i,1}$ and $b_{i,2}$ have the same expectation under the off-policy $\offpol$ and experimental design $\Zprob$.
If both expectations are matching, then the effect is zero.
If only one expectation is matching, then it suffices to have variation in the non-matching basis function under design $\Zprob$.
If both expectation are not matching, then the two basis functions must have positive variation and must also be not perfectly correlated.

\paragraph{Asymptotic Analysis}
The spatial point process experiment is one in which the standard dependency graph analysis presented in Sections~\ref{sec:supp-dep-graph-var} and \ref{supp:dependency-graph-clt} is inappropriate without modification.
More precisely, $(\ooi, \rri)$ and $(\ooj, \rrj)$ will be correlated for every pair of units $i, j \in [n]$ and thus each dependency neighborhood will contain all units.
This correlation happens because each of the $m$ randomly placed intervention sites will affect each individual unit.

Nevertheless, it should be possible to develop an appropriately modified dependency-graph type of analysis for this example. 
The key insight is that the influence of an intervention site on an experimental unit decays exponentially once the site is beyond distance $\sigma$ from the unit.
Thus, in order to provide meaningful asymptotic analysis, one must distinguish between pairs of units which are ``strongly correlated'' by virtue of being near each other from those pairs of units which are ``weakly correlated'' because they are far away.
Only once these different pairs of units are distinguished would it be appropriate to apply dependency graph analyses to the strongly correlated pairs of units.
We believe that similar ideas may be used to develop a CLT, though that is beyond the scope of this worked example.

\subsubsection{Determinantal Generalizations} \label{sec:determinantal-generalizations}

In this section, we show how the previous model which allows for pair-wise interactions may be extended to allow for higher order interactions.
We refer to these as \emph{determinantal interaction model}.
Determinantal functions have been widely used in machine learning to model different forms of dispersion and diversity.

Let $r$ be a fixed integer, which we refer to as the rank of the model.
In a determinantal interaction model of rank $r$, each potential outcome function can be written as
\[
y_i(\vec{z}) = \sum_{k=0}^r \alpha_i^{(k)} \sum_{ \substack{ S \subset [m] \\ |S| = k} } f_i^{(k)}(\vec{z}_S) \enspace,
\]
where $\vec{z}$ is a vector collecting all treatment points and $\vec{z}_S$ is the vector of treatment points with indices in the set $S$.
The functions $f_i^{(k)} : \mathcal{X}^k \to \Reals$ are called \emph{$k$th order interaction functions} and $\alpha_i^{(k)}$ are the associated coefficients indexing the model space.
This model space is thus $(r+1)$-dimensional and the choice of $r$ reflects the largest order of possible interactions.
The interaction functions are determinantal, meaning that they are of the form $f_i^{(k)}(z_S) = \det( K_i(z_S) )$, where $K_i(z_S)$ is an $|S|$-by-$|S|$ matrix whose $(s,t)$ entries are given by a kernel function: $\mathcal{K}(u_i - z_s, u_i - z_t)$.

Determinantal functions have been used in the machine learning literature to encode dispersion and diversity because a determinantal function becomes larger when the collection of points being evaluated are more dispersed or diverse.
This makes them a natural candidate for modelling potential outcomes in spatial experiment.
By including higher order interactions with $r > 2$, one models the possibility of having interactions among $r$-tuples of intervention sites affecting a unit's potential outcome.

To illustrate this perspective, it may be helpful to see that the pair-wise model considered in the previous section is in fact determinantal of order $r=2$ with the following kernel:
\[
\mathcal{K}(u_i - z_s, u_i - z_t) = \expf[\bigg]{ -\frac{1}{2\sigma^2} \braces[\big]{
	\norm{u_i - z_s}^2
	+ \norm{u_i - z_t}^2
	+ \norm{z_s - z_t}^2
 } }.
\]
To see this, observe that the first two interaction functions are
\begin{align*}
	f_i^{(1)}(z_t) &= \expf[\bigg]{ - \frac{\norm{u_i-z_t}^2}{\sigma^2} }, \\
	f_i^{(2)}(z_t, z_s) &= \expf[\bigg]{ - \frac{\norm{u_i-z_t}^2}{\sigma^2} } \cdot \expf[\bigg]{ - \frac{\norm{u_i-z_s}^2}{\sigma^2} } \cdot \braces[\bigg]{ 1 - \expf[\bigg]{ - \frac{\norm{z_s - z_t}^2 }{\sigma^2} } }.
\end{align*}
The first function, $f_i^{(1)}$, captures the effect of a single treatment point on the outcome of an experimental unit, in this case through an exponentially decaying function of its distance.
In contrast, the second function, $f_i^{(2)}$, captures the effect of pairs of treatment points on an experimental unit, and specifically the role of the dispersion of the points.
If two treatment points are close, $z_s \approx z_t$, then $f_i^{(2)}$ is close to zero, and the treatment points contribute little in addition to their effect through the first function, $f_i^{(1)}$.
On the other hand, if two treatment points are close to $u_i$ but relatively far away from each other, then they will affect the outcome of unit $i$ in addition to their effect through $f_i^{(1)}$.
The same type of behavior holds true for higher order interactions terms $k > 2$.

%% file: tex/supp-general-constructions.tex
\section{General Construction of Estimators}

\subsection{The Riesz Estimator} \label{sec:supp-riesz-construction}

In this section, we demonstrate how to construct the Riesz estimator and determine whether positivity holds. 
We highlight two different perspectives: the first perspective is based on the construction of an orthogonal basis and the second perspective is based on solution to a system of linear equations.

We presume that the experimenter has specified each model space $\Mspacei$ by a set of basis functions.
In particular, these basis functions $\setb{ b_{i,k} }_{k=1}^{d_i}$ have signature $b_{i,k} : \Zspace \to \Reals$ and constitute the model space in the sense that $\Mspacei = \vspan{ b_{i,1} \dots b_{i,d_i} }$.
We show in Section~\ref{supp:infinite-Riesz} that if the model space $\Mspacei$ is infinite dimensional (so that $d_i = \infty$), then the experimenter can apply a truncation of the model space so that the distributional properties of the Riesz estimator is conserved up to arbitrary tolerance. 
For this reason, it is assumed that if the model space has infinite dimensions, then the experimenter has already applied the truncation discussed in  by selecting $d$ basis functions to include in the truncated model space in which case $\Mspacei$ below refers to the truncated model space.
For notational simplicity, we suppose that all dimensions are equal across units, i.e. $d_1 = \dots = d_n = d$.

\subsubsection{Model Space Orthogonalization}\label{supp:orthogonalization-procedure}

In this section, we describe a modification of the Gram--Schmidt orthogonalization procedure that will produce an orthonormal basis for the outcome space $\Ospacei$ from a basis $b_{i,1} \dots b_{i,d}$ for the model space $\Mspacei$.
The output of this orthogonalization procedure will be used to construct the Riesz representor and determine whether positivity holds.

The input to the procedure are the basis function $b_{i,1} \dots b_{i,d}$ which constitute the model space $\Mspacei$.
The procedure operates in a recursive manner which is similar to Gram--Schmdit orthogonalization.
A key difference, however, is that the bilinear form $(u,v) \to \E{u, v}$ may not form a valid inner product on the model space $\Mspacei$, so we need to additionally keep track of which functions are zero almost surely under the design.
The procedure is given formally as Algorithm~\ref{alg:mod-gso}.

\begin{algorithm}[H]
\SetAlgoLined
\caption{Modified Gram--Schmidt Orthogonalization}\label{alg:mod-gso}
\SetKwInOut{Input}{Input}
\SetKwInOut{Output}{Output}
\Input{Functions $b_{i,1}, b_{i,2}, \dotsc, b_{i,d}$ from $\Lts$ that forms a basis for $\Mspacei$.}
\Output{Partition of basis for $\Mspacei$ into two sets.}

Initialize sets $O_i \gets \emptyset$ and $N_i \gets \emptyset$.\\
\For{$k \in [d]$}{
	$u_k \gets b_{i,k} - \sum_{h \in O_i} \iprod{b_{i,k}, h} h$.\\
	\uIf{$\norm{u_k} = 0$}{
		$N \gets N \cup \setb{u_k}$.\\
	}
	\uElse{
		$O \gets O \cup \setb{u_k / \norm{u_k}}$.\\
	}
}

\Return Sets $O_i$ and $N_i$.
\end{algorithm}

The set $O_i$ is a collection of functions which can be understood as forming an orthonormal basis for the outcome space $\Ospacei$; strictly speaking, it is the equivalence classes of these functions that form the basis, i.e. $\Ospacei = \vspan{ \ec{u} : h \in O_i }$.
Hence, the orthonormal basis $\obfi{1}, \obfi{2}, \dotsc$ used in the construction of the Riesz representor in Section~\mainref{sec:construction} in the main paper is exactly $\setb{ \ec{h} : h \in O_i }$.
On the other hand, the set $N_i$ is a basis for the null space $\setb{ u \in \Mspacei : \norm{u} = 0 } \subseteq \Mspacei$.
These are the functions which are identically $0$ under the experimental design.

We can use the output set of functions $O_i$ to construct the Riesz representor.
In particular, construct the function $r_i : \Zspace \to \Reals$ as the linear combination
\[
r_i = \sum_{h \in O_i} \efvi{h} \cdot h
\enspace,
\]
where the coefficients $\efvi{h}$ are the evaluation of the effect functional on each $h \in O_i$.
The function $r_i$ is one of the members of the $\Lts$ equivalence class corresponding to the Riesz representor, i.e. $\rri = \ec{r_i}$.

When computing the Riesz representors by hand, evaluation of $\efi$ on each $h \in O_i$ is usually straightforward.
It may also be helpful to note that each evaluation $\efvi{h}$ for $h \in O_i$ can be expressed as a linear combination of the evaluations on the original basis functions $\efvi{b_{i,1}}$, \dots $\efvi{b_{i,d}}$.
Thus, it is only necessary to evaluate the effect functional on the original set of basis functions for $\Mspacei$.
This is particularly helpful when the Riesz representors are computed in an automated fashion, as in our simulations.

The procedure listed in Algorithm~\ref{alg:mod-gso} presumes that all relevant expectations under the experimental design can be exactly calculated.
These expectations could also be calculated to arbitrary precision using Monte Carlo procedures.

\subsubsection{Determining Whether Positivity Holds}\label{supp:positivity-holds}

Given the effect functionals, model spaces, and experimental design, an experimenter can determine whether positivity holds.
In this section, we provide a computationally simple procedure for determining whether positivity holds.
The key insight is that positivity can be equivalently formulated as follows: $\efvi{u} = 0$ for all $u \in \Mspacei$ such that $\norm{u} = 0$.
For finite dimensional model spaces, Algorithm~\ref{alg:mod-gso} produces a basis $N_i$ for the subspace $\{ u \in \Mspacei : \norm{u} = 0 \}$.
To determine whether positivity holds, Algorithm~\ref{alg:check-positivity} proceeds by determining whether $\efvi{b} = 0$ for each basis function $b \in N_i$.

\begin{algorithm}[H]
\SetAlgoLined
\caption{Checking whether positivity holds for unit $i$}\label{alg:check-positivity}
\SetKwInOut{Input}{Input}
\SetKwInOut{Output}{Output}
\Input{Effect functional $\efi$ and set $N_i$ produced by Algorithm~\ref{alg:mod-gso}.}
\Output{Returns \textsf{true} if positivity holds, otherwise \textsf{false}.}

\For{$u \in N_i$}{
	\uIf{$\efvi{u} \neq 0$}{
		\Return \textsf{false}
	}
}
\Return \textsf{true}
\end{algorithm}

\subsubsection{Riesz Representors as Solutions to Matrix Equations}\label{supp:matrix-construction}

The Riesz representors can also be understood as a solution to a system of linear equations.
This perspective is insightful both from the linear algebraic and computational viewpoints.
In this section, we describe this alternative perspective and give an alternative construction of the Riesz representors.

Throughout the remainder of the section, we fix an individual unit $i \in [n]$.
Recall that the model space $\Mspacei$ is represented in terms of a basis $b_{i,1} \dots b_{i,d}$.
We have implicitly presumed that the model space $\Mspacei$ is finite dimensional, or that the experimenter has already appropriate truncated an infinite basis.
The Riesz representor $\rri$ can be expressed as a function $r_i \in \Mspacei$ which can in turn be written in terms of the basis:
\[
r_i(z) = \sum_{k=1}^d \gamma_{i,k} \cdot b_{i,k}(z)
\]
We collect the coefficients in the vector $\vec{\gamma}_i = (\gamma_{i,1} \dots \gamma_{i,d})$.
Recall that there may be many different functions $r_i$ which correspond to the Riesz representor $\rri$, and thus many choices of coefficients $\gamma_{i,1} \dots \gamma_{i,d}$.
Regardless, our goal will be to show how to construct such coefficients.

By unpacking the definition of Riesz representor, we have that a collection of coefficients $\vec{\gamma}_i$ corresponds to the Riesz representor if and only if it is the solution to the system of linear equations:
\[
\mat{S}_i \vec{\gamma}_i = \vec{\ate}_i
\enspace,
\]
where $\mat{S}_i$ is a $d$-by-$d$ matrix with entries given by expected product of basis functions, i.e. $\mat{S}_i(k,\ell) = \E{ b_{i,k} b_{i,\ell} }$, and $\vec{\ate}_i$ is a $d$-length vector whose entries are the effect functional evaluated on the basis functions, i.e. $\vec{\ate}_i(k) = \efvi{b_{i,k}}$.
Thus, positivity holds if and only if a solution exists and the Riesz representor may be obtained directly by solving this linear system.

When a solution exists, it may be found using the psuedo-inverse matrix of $\mat{S}_i$, denoted $\mat{S}_i^+$.
In particular, the solution may be obtained as $\vec{\gamma}_i = \mat{S}_i^+ \vec{\ate}_i$.
This provides not only another method for computing the Riesz representor, but also another way to verify whether positivity holds.
More precisely, positivity holds if and only if $\mat{S}_i \mat{S}_i^+ \vec{\ate}_i = \vec{\ate}_i$.

\subsubsection{Truncation for Infinite-Dimensional Model Spaces}\label{supp:infinite-Riesz}

In the case the model spaces have infinite dimensions, truncated Riesz representors are used to construct the Riesz estimator.
We here describe this truncation, and show that it approximates the full infinite-dimensional estimator to arbitrary precision.

Fix a unit $i \in [n]$.
Let $\setb{\obfi{k}}_{k=1}^\infty$ be an orthonormal Schauder basis of $\Ospacei$, which we have ordered in a particular way.
Such a basis exists because the underlying topology of the intervention space $\Zspace$ is assumed to be separable.
Recall that the individual Riesz representor for unit $i$ is given as
\[
\rri = \sum_{k=1}^\infty \oefvi{\obfik} \obfik.
\]
Given a positive integer $d$, we define the \emph{truncated Riesz representor} for unit $i$ to be the truncation of this series according to the $d$ first basis elements:
\begin{equation}
	\rri(d) = \sum_{k=1}^d \oefvi{\obfik} \obfik.
\end{equation}
Note that $\rri(d)$ is the projection of the full infinite-dimensional Riesz representor onto the span of $\setb{\obfi{1}, \obfi{2}, \dotsc, \obfi{d}}$.
Let $\est(d)$ denote the truncated Riesz estimator, constructed using the truncated Riesz representors:
\begin{equation}
	\est(d) = \frac{1}{n} \sumin \rri(d) \ooi.
\end{equation}
It is possible to use different truncations for different units, but we keep it the same for all units here for notational simplicity.
The following result demonstrates that it is possible to choose a truncation point so the truncated Riesz estimator retains all relevant distributional properties of the full Riesz estimator, up to an arbitrarily small approximation.

\begin{proposition}
There exists a truncated Riesz estimator that approximates the full infinite-dimensional estimator to arbitrary precision.
That is, for any $\varepsilon > 0$, there exists a positive integer $d$ such that $\E{\abs{\est(d) - \est}} \leq \varepsilon$.
\end{proposition}

\begin{proof}
	Fix an experimental unit $i \in [n]$.
	By orthonormality, the norm of each individual Riesz representor can be expressed as
	$
	\norm{\rri}^2 = \sum_{k=1}^\infty \oefvi{\obfik}^2,
	$
	so that $\sum_{k=1}^\infty \oefvi{\obfik}^2$ is a convergent series.
	Likewise, the norm of the difference between a Riesz representor and its truncation at $d$ terms may be expressed as
	$
	\norm{\rri - \rri(d)}^2 = \sum_{k=d+1}^\infty \oefvi{\obfik}^2.
	$
	By the Cauchy criterion, this series converges to zero as $d$ grows.
	Thus, for each unit $i \in [n]$, there exists $d_i$ sufficiently large so that 
	\[
	\norm{\rri - \rri(d_i)} \leq \varepsilon / \norm{\ooj}
	\enspace.
	\]
	Set the truncation index to be $d = \max_{i \in [n]} d_i$.
	The expected absolute difference between the Riesz estimator and the truncated Riesz estimator using $d$ as given above may be bounded as
	\begin{align*}
		\E[\big]{\abs{\est(d) - \est}}
		&\leq
		\frac{1}{n} \sumin \abs[\Big]{\E{(\rri(d) - \rri) \ooi}} 
			&\text{(triangle inequality)}\\
		&\leq
		\frac{1}{n} \sumin \norm{\rri(d) - \rri} \norm{\ooi},
			&\text{(Cauchy-Schwarz)} \\
		&\leq \frac{1}{n} \sumin \frac{\varepsilon}{ \norm{\ooj} } \cdot \norm{\ooi}
			&\text{(choice of truncation)} \\
		&\leq \varepsilon
		\enspace.
		&\qedhere
	\end{align*}
\end{proof}

\subsubsection{Riesz Estimator as Minimal Norm} \label{sec:supp-riesz-minimal-norm}

In this section, we show that the Riesz estimator considered in the main paper is the simplest unbiased weighting estimator, in the sense that it has minimal norm.
For the remainder of this section, we will presume that positivity holds and we fix a particular unit $i \in [n]$.

Consider the extended linear functional $\oefi : \Ospacei \to \Reals$.
We say that a random variable $W_i \in \Lt$ is an \emph{unbiased weighting} if 
\[
\E{ \ooi \cdot W_i } = \oefvi{\ooi}
\text{ for all }
\ooi \in \Ospacei
\enspace.
\]
Note that this definition requires positivity in order for $\oefi$ to be well-defined.
Moreover, this definition is working entirely with random variables up to $\Lt$ equivalence.

\begin{proposition}
	Suppose positivity holds and let $\rri$ be the Riesz representor for $\oefi : \Ospacei \to \Reals$.
	If $W_i \in \Lt$ is an unbiased weighting estimator, then 
	\[
	\E{ W_i^2 } \geq \E{ \rri^2 }
	\enspace,
	\]
	where equality holds if and only if $W_i = \rri$.
\end{proposition}
\begin{proof}
	We begin by considering the unique orthogonal decomposition of $W_i$ into 
	\[
	W_i = U_i + V_i \enspace,
	\]
	where $U_i \in \Ospacei$ and $V_i$ is in the orthogonal complement $\Ospacei^{\perp}$.
	Note that for any $\ooi \in \Ospacei$, we have that
	\begin{align*}
		\oefvi{\ooi}
		&= \E{ W_i \ooi }
		&\text{(unbiased weighting)} 
		\\
		&= \E{ U_i \ooi }
		+ \E{ V_i \ooi }
		&\text{(orthogonal decomposition)} 
		\\
		&= \E{U_i \ooi}
		\enspace,
	\end{align*}
	where the last equality follows because $V_i$ is orthogonal to each function in $\Ospacei$, i.e. $\E{V_i \ooi} = 0$.
	Thus, we see that $U_i$ must be the Riesz representor: $U_i = \rri$.

	Next, let us calculate the norm of $W_i$:
	\begin{align*}
		\E{W_i^2}
		&= \E{ (\rri + V_i)^2 }
		&\text{(def of $W_i$)}
		\\
		&= \E{ \rri^2 } + \E{V_i^2}
		+ 2 \E{\rri V_i} 
		&\text{(expanding terms)}
		\\
		&= \E{\rri^2 } + \E{V_i^2}
		&\text{(orthogonality, $V_i \in \Ospacei^\perp$)}
		\\
		&\geq \E{\rri^2}
		\enspace.
	\end{align*}
	Thus, we have established that $\E{W_i^2} \geq \E{\rri^2}$.
	Moreover, the calculation above demonstrates that equality holds if and only if $\E{V_i^2} =0$, a condition which is equivalent to $W_i = \rri$.
\end{proof}

\subsection{Variance Estimators}

In this section, we present two approaches for constructing estimable variance bounds.
The first approach is based on the operator norm of the variance characterizing linear operator and the second approach may be seen as a generalization of the Aronow-Samii bound \cite{Aronow2013Conservative}.

\subsubsection{Approach 1: Operator Norm Bound}\label{supp:var-est-w-operator}

Our first approach to constructing estimable variance bounds goes through the variance characterizing linear operator.
We define this variance bound $\vbn{VCO}$ as
\[
\vbnv{VCO}{\vu} = \frac{\opnorm{ \varmap }^2}{n} \cdot \frac{1}{n} \sum_{i=1}^n \E{ U_i^2 }
\]

To verify that $\vbn{VCO}$ is a variance bound, we may use Theorem~\mainref{thm:vclo-statement} together with the definition of the operator norm to see that
\[
\Var{\estv{\vu}} 
= \norm{ \varmapv{ \vU } }^2
\leq \opnorm{ \varmap }^2 \cdot \norm{\vU}^2
= \frac{\opnorm{ \varmap }^2}{n} \cdot \frac{1}{n} \sum_{i=1}^n \E{ U_i^2 }
= \vbnv{VCO}{\vu} 
\enspace.
\]

Next, we verify that the bound is estimable.
To this end, observe that it may be written as the decomposition $\vbv{\vu} = n^{-2}\sum_{i=1}^n \boufv{i}{i}{ u_i \otimes u_i }$, where $\bouf{i}{i} : \Mspacei \otimes \Mspacej \to \Reals$ is defined for simple tensors as $\boufv{i}{j}{u \otimes v} = \opnorm{ \varmap }^2 \E{ U V }$, and extended to general tensors via linearity.
It is a straightforward exercise to verify that these functionals satisfy second order positivity.

A simple unbiased estimate of the variance bound is given by
$$
\evb_{\textsc{vco}} = \frac{\opnorm{\varmap}^2}{n^2} \sumin \ooi^2
\enspace.
$$
When each of the model spaces contains the constant function, this will correspond exactly to the Riesz variance estimator.
This variance estimator will be conservative in expectation, by construction, and will generally converge on a normalized scale under the same assumptions as the point estimator.

The advantage of this estimator is its simplicity.
The only significant challenge in its implementation is the computation of the operator norm of the variance characterizing operator.
The downside is that this variance bound may be overly conservative for some experimental designs. 
Indeed, the bound itself is based on taking a supremum over possible outcome functions in the model space.
In this way, the variance bound reflects the worst-case dependencies over $\vu \in \Mprodspace$, rather than the actual dependencies associated with the true potential outcome function $\vpo$.
For a typical experimental design, it is likely that the worst case variance over all $\vu \in \Mprodspace$ will be much bigger than the true variance for the actual potential outcome function $\vpo$.
The only exception is when the experimental design has been carefully constructed so as to minimize worst-case dependencies, in which case the worst case variance given by the variance bound will be closer to the true variance.

To overcome these types of concerns, the estimator we describe in the next section estimates part of this dependence for the actual potential outcome function $\vpo$.

\subsubsection{Approach 2: Generalized Aronow-Samii Bound}\label{supp:bound-functionals}

In this section, we present a second approach for constructing variance bounds within the general framework.
The variance bound may be understood as a generalization of the Aronow--Samii bound from the discrete exposure mapping framework \cite{Aronow2013Conservative,Aronow2017Estimating}.

To construct the variance bound, we proceed in three main steps.
The first step is to decompose the original covariance functionals into a part that is identified (i.e., that satisfy second-order positivity) and a part that is unidentified.
The second step is to construct an identified bound on the unidentified parts.
The third step is to combine these together to obtain the Aronow-Samii bound.
Each of the subsequent subsections focuses on one of these steps.

\paragraph{Step 1: Identified and Unidentified Parts} 
We start by defining a new inner product $\iprod{\cdot, \cdot}_{\otimes}$ on the tensor product $\Mspacei \otimes \Mspacej$.
For two simple tensors, the inner product is
\begin{equation}
\iprod{u_1 \otimes u_2, v_1 \otimes v_2}_{\otimes}
= \iprod{U_1, V_1} \iprod{U_2, V_2},
\end{equation}
where, on the right-hand side, $\iprod{U_1, V_1}$ and $\iprod{U_2, V_2}$ are the corresponding inner products on $\Ospacei$ and $\Ospacej$.
We extend this to the full tensor product $\Mspacei \otimes \Mspacej$ by bilinearity.
The corresponding norm is $\pnorm{\otimes}{\tU} = \sqrt{\iprod{\tU, \tU}_{\otimes}}$.
This is the canonical inner product and norm for tensor products of Hilbert spaces, and it is not the bilinear form and seminorm we defined in the main paper.

Let $\mathcal{N}_{ij} = \setb{ \tU \in \Mspacei \otimes \Mspacej : \norm{\tU} = 0}$ be the null space of $\Mspacei \otimes \Mspacej$ with respect to its seminorm.
Let $P_{ij} : \Mspacei \otimes \Mspacej \to \Mspacei \otimes \Mspacej$ be the orthogonal projection onto the null space $\mathcal{N}_{ij}$ with respect to the canonical inner product:
$$
P_{ij}(\tU) = \argmin_{\tV \in \mathcal{N}_{ij}} \pnorm{\otimes}{\tU - \tV}.
$$
As we discuss at the end of this section, it is possible to use oblique projections, but the orthogonal projection is optimal for the type of bound we consider.
Define $Q_{ij}(\tU) = \tU - P_{ij}(\tU)$ to be the projection onto the orthogonal complement of $\mathcal{N}_{ij}$.
For each covariance functional $\covfij$, define two new functionals as the composition of the covariance functional and the two projections: $C^U_{i,j} = \covfij \circ P_{ij}$ and $C^I_{i,j} = \covfij \circ Q_{ij}$.

\begin{lemma} \label{lemma:identified-sop}
The functional $C^I_{i,j}$ satisfies second-order positivity.
\end{lemma}

\begin{proof}
We need to show that $C^I_{i,j}(\tU) = 0$ holds for all $\tU \in \mathcal{N}_{ij}$.
By the properties of projections, we have $P_{ij}(\tU) = \tU$ for all $\tU \in \mathcal{N}_{ij}$.
This implies that, $Q_{ij}(\tU) = \tU - P_{ij}(\tU) = \tU - \tU = \tensor{0}$.
Therefore, for all $\tU \in \mathcal{N}_{ij}$,
$$
C^I_{i,j}(\tU)
= C_{i,j}(Q_{ij}(\tU))
= C_{i,j}(\tensor{0})
= 0.
\qedhere
$$
\end{proof}

\paragraph{Step 2: Bound on Unidentified Functionals}
Let $\tensor{R}_{i,j} = \rri \otimes \rrj$ be the tensor in $\Mspacei \otimes \Mspacej$ corresponding to the point estimator Riesz representors for pair $(i,j) \in [n]^2$.
Using the bilinear form defined on the tensor product and its canonical inner product, we can write the covariance functional as
$$
\covfijv{ \tU } = \iprod{\tU, \tensor{R}_{i,j}} - \iprod{\tU, \tensor{R}_{i,j}}_{\otimes}.
$$
Therefore,
$$
C^U_{i,j}(\tU) = \iprod{P_{ij}(\tU), \tensor{R}_{i,j}} - \iprod{P_{ij}(\tU), \tensor{R}_{i,j}}_{\otimes}.
$$
Note that $\iprod{P_{ij}(\tU), \tensor{R}_{i,j}} = 0$, because $P_{ij}(\tU)$ is in the null space $\mathcal{N}_{ij}$.
This means that
$$
C^U_{i,j}(\tU) = - \iprod{P_{ij}(\tU), \tensor{R}_{i,j}}_{\otimes}.
$$
Let $P^*_{ij}$ denote the adjoint of $P_{ij}$, meaning that we can write
$$
C^U_{i,j}(\tU) = - \iprod{\tU, P^*_{ij}(\tensor{R}_{i,j})}_{\otimes}.
$$
Note that $P_{ij}$ is the orthogonal projection, so we have $P^*_{ij} = P_{ij}$, but we still use the adjoint $P^*_{ij}$ here because we will consider other oblique projections in place of $P_{ij}$ in the next subsection.
By the Cauchy--Schwarz inequality,
$$
C^U_{i,j}(\tU)
\leq \abs{\iprod{\tU, P^*_{ij}(\tensor{R}_{i,j})}_{\otimes}}
\leq \pnorm{\otimes}{\tU} \cdot \pnorm{\otimes}{P^*_{ij}(\tensor{R}_{i,j})}.
$$
When $\tU = u_i \otimes u_j$ is a simple tensor, we have
$$
\pnorm{\otimes}{\tU}
= \pnorm{\otimes}{u_i \otimes u_j}
= \norm{U_i} \cdot \norm{U_j}
\leq \frac{\norm{U_i}^2 + \norm{U_j}^2}{2}.
$$
Therefore, for any set of $u_i \in \Mspacei$ for $i \in [n]$,
$$
\sumin \sumjn C^U_{i,j}(u_i \otimes u_j) \leq \sumin b_i \norm{U_i}^2
$$
where
$$
b_i = \sumjn \frac{\pnorm{\otimes}{P^*_{ij}(\tensor{R}_{i,j})} + \pnorm{\otimes}{P^*_{ij}(\tensor{R}_{j,i})}}{2}.
$$
Note that
$$
\norm{U_i}^2 = \E{U_i^2} = \iprod{u_i \otimes u_i, [f_1] \otimes [f_1]},
$$
where $f_1$ is the constant function $z \mapsto 1$, and $[f_1]$ is the corresponding equivalence class in $\Lt$.
If $f_1$ is in both $\Mspacei$ and $\Mspacej$, then $[f_1] \otimes [f_1]$ will be in $\Mspacei \otimes \Mspacei$, but the current argument applies even if that is not the case.
Define a linear functional $B^U_{i,j}$ on the tensor product $\Mspacei \otimes \Mspacei$ as
$$
B^U_{i}(\tU) = b_i \iprod{\tU, [f_1] \otimes [f_1]}.
$$
We then have that, for any set of $U_i \in \Ospacei$ for $i \in [n]$,
\begin{equation} \label{eq:main-vb-step}
\sumin \sumjn C^U_{i,j}(u_i \otimes u_j) \leq \sumin B^U_{i}(u_i \otimes u_i).
\end{equation}

\begin{lemma} \label{lemma:bound-is-sop}
The functional $B^U_{i}$ satisfies second-order positivity.
\end{lemma}

\begin{proof}
We have $B^U_{i}(\tU) < \infty$ from the fact that the tensor product $\Ospacei \otimes \Ospacei$ is built from a subspace of the $\Lt$ space.
We then need to show that $B^U_{i}(\tU) = 0$ holds for all $\tU \in \mathcal{N}_{ii}$.
When $\tU \in \mathcal{N}_{ii}$, we have $\norm{\tU} = 0$, and by the Cauchy--Schwarz inequality,
$$
\abs{\iprod{\tU, [f_1] \otimes [f_1]}} \leq \norm{\tU} \times \norm{[f_1] \otimes [f_1]} = 0 \times 1 = 0.
\qedhere
$$
\end{proof}

\paragraph{Step 3: Generalization of Aronow--Samii Bound}
We can now define the generalization of the Aronow--Samii bound, which we denote $\vbn{AS}$.
For all $i, j \in [n]$ such that $i \neq j$, define $\boufij = C^I_{i,j}$.
For all $i \in [n]$, define $\bouf{i}{i} = C^I_{i,i} + B^U_{i}$.
We formally define the Aronow--Samii generalization as follows:
\[
\vbnv{AS}{\vu} = \sumin \sumjn \boufijv{u_i \otimes u_j}.
\]

\begin{lemma}
	The generalized Aronow-Samii variance bound is an estimable variance bound.
\end{lemma}
\begin{proof}
The proof follows largely from lemmas in the previous section.
First, we show that $\vbn{AS}$ is indeed a variance bound.
We can write $u_i \otimes u_j = P_{ij}(u_i \otimes u_j) + Q_{ij}(u_i \otimes u_j)$.
By linearity of the covariance functional and the definitions of $C^U_{i,j}$ and $C^I_{i,j}$, we have
\begin{equation} \label{eq:cov-decomp}
\covfijv{u_i \otimes u_j}
= \covfijv{P_{ij}(u_i \otimes u_j) + Q_{ij}(u_i \otimes u_j)}
= C^U_{i,j}(u_i \otimes u_j) + C^I_{i,j}(u_i \otimes u_j).
\end{equation}
Hence, the variance can be bounded as
\begin{align*}
	\Var{\estv{\vu}}
	&= \sumin \sumjn \covfijv{u_i \otimes u_j} 
		&\text{(covariance functionals)}\\
	&= \sumin \sumjn C^U_{i,j}(u_i \otimes u_j)
	+ \sumin \sumjn C^I_{i,j}(u_i \otimes u_j) 
		&\text{(using \eqref{eq:cov-decomp})}\\
	&\leq \sumin B^U_{i}(u_i \otimes u_i)
	+ \sumin \sumjn C^I_{i,j}(u_i \otimes u_j) 
		&\text{(using \eqref{eq:main-vb-step})}\\
	&= \sumin \sumjn \boufijv{U_i \otimes U_j} 
		&\text{(definition of $\boufij$)}\\
	&= \vbnv{AS}{\vu}
	\enspace.
\end{align*}
The fact that $\vbn{AS}$ is estimable follows from the fact that the identified covariance functionals $C^I_{i,j}$ satisfy second order positivity (Lemma~\ref{lemma:identified-sop}) and that the constructed bound functionals $B^U_{i}$ satisfies second order positivity (Lemma~\ref{lemma:bound-is-sop}).
\end{proof}

\paragraph{Optimality of Orthogonal Projection}
In the decomposition into identified and unidentified parts in Step~1, we used the orthogonal projection onto the null space $\mathcal{N}_{ij}$.
It is possible to use other (oblique) projections in this step, in place of the orthogonal projection, to produce a valid bound.
However, the orthogonal projection is the optimal choice, in the sense that it produces the lowest bound among all projections when following the bound procedure described in the previous subsection.

Let $\tilde{P}_{ij} : \Ospacei \otimes \Ospacei \to \Ospacei \otimes \Ospacei$ be an alternative projection onto $\mathcal{N}_{ij}$, and let $\tilde{P}^*_{ij}$ be the adjoint operator.
We will now show that $\pnorm{\otimes}{\tilde{P}^*_{ij}(\tensor{R}_{i,j})} \geq \pnorm{\otimes}{P^*_{ij}(\tensor{R}_{i,j})}$.

Note that we have
$$
\pnorm{\otimes}{P^*_{ij}(\tensor{R}_{i,j})}
= \sup_{\substack{\tU \in \Ospacei \otimes \Ospacei \\ \pnorm{\otimes}{\tU} = 1}} \abs{\iprod{\tilde{P}^*_{ij}(\tensor{R}_{i,j}), \tU}_\otimes}
= \sup_{\substack{\tU \in \Ospacei \otimes \Ospacei \\ \pnorm{\otimes}{\tU} = 1}} \abs{\iprod{\tensor{R}_{i,j}, \tilde{P}_{ij}(\tU)}_\otimes}.
$$
Let $m_\tensor{R} = \pnorm{\otimes}{P_{ij}(\tensor{R}_{i,j})}$ be the norm of the orthogonal projection of $\tensor{R}_{i,j}$ onto $\mathcal{N}_{ij}$.
Note that $m_\tensor{R}^{-1} P_{ij}(\tensor{R}_{i,j}) \in \Ospacei \otimes \Ospacei$ and that it has norm one, so
$$
\sup_{\substack{\tU \in \Ospacei \otimes \Ospacei \\ \pnorm{\otimes}{\tU} = 1}} \abs{\iprod{\tensor{R}_{i,j}, \tilde{P}_{ij}(\tU)}_\otimes}
\geq \abs{\iprod{\tensor{R}_{i,j}, \tilde{P}_{ij}(m_\tensor{R}^{-1} P_{ij}(\tensor{R}_{i,j}))}_\otimes}
= \frac{1}{m_\tensor{R}} \abs{\iprod{\tensor{R}_{i,j}, P_{ij}(\tensor{R}_{i,j})}_\otimes},
$$
where the last equality follows from the fact that $m_\tensor{R}^{-1} P_{ij}(\tensor{R}_{i,j}) \in \mathcal{N}_{ij}$, so $\tilde{P}_{ij}(m_\tensor{R}^{-1} P_{ij}(\tensor{R}_{i,j})) = m_\tensor{R}^{-1} P_{ij}(\tensor{R}_{i,j})$.

Using orthogonal projections $P_{ij}$ and $Q_{ij}$, we can decompose any element of $\Ospacei \otimes \Ospacei$, including $\tensor{R}_{i,j}$, as a projection onto $\mathcal{N}_{ij}$ and its orthogonal complement:
$$
\tensor{R}_{i,j} = P_{ij}(\tensor{R}_{i,j}) + Q_{ij}(\tensor{R}_{i,j}),
$$
where $\iprod{P_{ij}(\tensor{R}_{i,j}), Q_{ij}(\tensor{R}_{i,j})} = 0$ due to orthogonality.
Therefore,
$$
\iprod{\tensor{R}_{i,j}, P_{ij}(\tensor{R}_{i,j})}_\otimes
= \iprod{P_{ij}(\tensor{R}_{i,j}), P_{ij}(\tensor{R}_{i,j})}_\otimes
+ \iprod{Q_{ij}(\tensor{R}_{i,j}), P_{ij}(\tensor{R}_{i,j})}_\otimes
= \pnorm{\otimes}{P_{ij}(\tensor{R}_{i,j})}^2.
$$
Putting this together, we have that for any projection $\tilde{P}_{ij}$ onto $\mathcal{N}_{ij}$,
$$
\pnorm{\otimes}{\tilde{P}^*_{ij}(\tensor{R}_{i,j})}
\geq \frac{1}{m_\tensor{R}} \pnorm{\otimes}{P_{ij}(\tensor{R}_{i,j})}^2
= \pnorm{\otimes}{P_{ij}(\tensor{R}_{i,j})},
$$
where the last equality follows from $m_\tensor{R} = \pnorm{\otimes}{P_{ij}(\tensor{R}_{i,j})}$.
Recall that the orthogonal projection is self-adjoint, $P^*_{ij} = P_{ij}$, so it attains the minimum: $\pnorm{\otimes}{P^*_{ij}(\tensor{R}_{i,j})} = \pnorm{\otimes}{P_{ij}(\tensor{R}_{i,j})}$.

%% file: tex/supp-general-theoretical-results.tex
\section{General Theoretical Results}

This section contains proof of the general theoretical results presented in the main body.
For the sake of completeness, we restate each result before giving its proof.

\subsection{Positivity and Unbiasedness}

\subsubsection{Positivity as a sufficient condition (Lemma~\mainref{lem:extended-functionals} and Theorem~\mainref{thm:Riesz-unbiased})}

In this section, we show that positivity is a sufficient condition for the existence of an unbiased estimator for the aggregate causal effect.
In particular, we show how positivity facilitates the construction of the Riesz estimator and verify that it is unbiased.

We begin by proving the following lemma, which relates the effect functional defined on the model space $\Mspacei$ to a related functional defined on the outcome space $\Ospacei$.

\begin{reflemma}{\mainref{lem:extended-functionals}}
\lemextendedfunctionals
\end{reflemma}

\newcommand{\roef}[1]{T_{#1}}
\newcommand{\roefi}{\roef{i}}
\DeclarePairedDelimiterXPP\roefv[2]{\roef{#1}}{\lparen}{\rparen}{}{#2}
\DeclarePairedDelimiterXPP\roefvi[1]{\roefi}{\lparen}{\rparen}{}{#1}

\begin{proof}
Let $E_i = \braces{ [u] : u \in \Mspacei}$ be the collection of all equivalence classes that can be built from functions in the model space.
Using the axiom of choice, let $p_i : E_i \to \Mspacei$ be a function that selects an element from each equivalence class in $E_i$, so that $p_i(U) \in U$ for each $U \in E_i$.
Define a functional $\roefi : E_i \to \Reals$ such that $\roefi = \efi \circ p_i$.

First, we will show $\efvi{u} = \roefvi{[u]}$ for all $u \in \Mspacei$.
Note that by linearity of $\efi$, we have
$$
\efvi{u}
= \efvi{p_i([u]) + u - p_i([u])}
= \efvi{p_i([u])} + \efvi{u - p_i([u])}
= \roefvi{[u]}
\enspace,
$$
where the last equality follows because the positivity condition.
in particular, positivity ensures that $\efi$ is zero for input with zero norm and $\norm{u - p_i([u])} = 0$ by construction.

Next, we will show that $\roefi$ is a continuous linear functional.
First, observe that the selection function $p_i: E_i \to \Mspacei$ is a linear map.
Because the composition of linear maps is linear, we have that $\roefi = \efi \circ p_i$ is linear.
Next, we establish continuity of $\roefi$.
For any $U \in E_i$ and $V \in E_i$,
\begin{equation}
\abs{\roefvi{U} - \roefvi{V}}
= \abs{\efvi{p_i(U)} - \efvi{p_i(V)}}
\leq C \norm{p_i(U) - p_i(V)}
= C \norm{U - V}.
\end{equation}
The inequality is positivity (Assumption~\mainref{ass:positivity}), in which $C$ is defined.
The final equality follows from $\norm{u} = \norm{[u]}$.

We have now shown that $\roefi$ is a bounded linear functional on $E_i$ that coincides with the effect functional on the model space.
Recall that the outcome space $\Ospacei$ is the closure of $E_i$, so $E_i$ is a subspace of $\Ospacei$.
The Hahn--Banach theorem states that there exists a bounded linear functional $\oefi : \Ospacei \to \Reals$ that coincides with $\roefi$ on $E_i$.
\end{proof}

\begin{reftheorem}{\mainref{thm:Riesz-unbiased}}
\thmRieszunbiased
\end{reftheorem}

\begin{proof}
We have
$$
\E{\est}
= \frac{1}{n} \sumin \E{\rri \ooi}
= \frac{1}{n} \sumin \iprod{\rri, \ooi}.
$$
Recall that, given positivity, $\oefvi{U} = \iprod{\rri, U}$ for all $U \in \Ospacei$.
Given correctly specified model spaces, $\ooi \in \Ospacei$, so $\iprod{\rri, \ooi} = \oefvi{\ooi}$.
By construction of the extended effect functional, $\oefvi{\ooi} = \efvi{\poi}$, so
$$
\E{\est}
= \frac{1}{n} \sumin \iprod{\rri, \ooi}
= \frac{1}{n} \sumin \oefvi{\ooi}
= \frac{1}{n} \sumin \efvi{\poi}
= \ate.
\qedhere
$$
\end{proof}

\subsubsection{Positivity as a necessary condition (Theorem~\mainref{thm:positivity-is-necessary})}

In this section, we show that positivity is a necessary condition for the existence of an unbiased Lipschitz continuous estimator.
Our argument relies on the following intuition: if positivity does not hold, then there are two possible potential outcome functions $\vu$ and $\vu'$ which look the same under the experimental design but for which the aggregate effects are very different.

\begin{reftheorem}{\mainref{thm:positivity-is-necessary}}
\thmpositivityisnecessary
\end{reftheorem}

\begin{proof}
	If positivity does not hold, then there exists a unit $j \in [n]$ such that for every $\beta \in \Reals$, there exists two functions $v, v' \in \Mspacej$ such that
	\[
	\abs{ \efv{j}{v} - \efv{j}{v'} } > \beta \cdot \norm{v - v'}
	\enspace.
	\]
	We will fix one $\beta \in \Reals$ and let it grow arbitrarily large later in the proof.
	
	Fix $C > 0$ as in the statement of the theorem.
	Define the two combined potential outcome functions $\vu, \vu' \in \Mprodspace$ as follows: $\vu$ is the zero function, i.e. $\vu(Z) = 0$ for all $Z \in \Zspace$, and $\vu'$ is given coordinate-wise $\vu' = (u'_1, \dots u'_n)$ where
	\[
	u_i' = \left\{
	\begin{array}{lr}
		0 & \text{if } i \neq j\\
		\alpha \cdot (v - v') & i = j
	\end{array}
	\right.
	\]
	where we will select $\alpha = \sqrt{n} \cdot C / \norm{v - v'}$.
	In other words, $\vu$ and $\vu'$ agree on all units except unit $j$, where $\vu'$ is the difference between $v$ and $v'$, scaled by $\alpha$.
	The choice of $\vu = 0$ is convenient, but not necessary for our proof; indeed, the initial choice of $\vu$ can be any function for which $\norm{\vu} < C$.
	
	Let us verify several properties of these two functions $\vu$ and $\vu'$.
	First, let us verify that $\norm{\vu} \leq C$ and $\norm{\vu'} \leq C$.
	That $\norm{\vu} \leq C$ follows because $\vu$ is the zero function.
	Next, observe that
	\begin{equation} \label{eq:up_bounded_2m}
	\norm{\vu'}^2 
	= \frac{1}{n} \sum_{i=1}^n \E{ u_i^2 }
	= \frac{1}{n} \alpha^2 \cdot \E{ (v - v')^2 }
	= \frac{1}{n} \alpha^2 \cdot \norm{v-v'}^2
	= C^2
	\enspace.
	\end{equation}
	
	Next, let us verify that any Lipschitz estimator will have similar expectations under $\vu$ and $\vu'$.
	\begin{align*}
		\abs[\Big]{ \E{ \estv{\vu} } - \E{ \estv{\vu'} } } 
		&= \abs[\Big]{ \E{ \estv{\vu} - \estv{\vu'} } } 
			&\text{(linearity)} \\
		&\leq \E[\big]{ \abs[\big]{ \estv{\vu} - \estv{\vu'} } }
			&\text{(Jensen's inequality)} \\
		&= \E[\big]{ \paren[\big]{ \estv{\vu} - \estv{\vu'} }^2 }^{1/2}
			&\text{(H\"older's inequality)} \\
		&\leq K \cdot \norm{\vu - \vu'} 
			&\text{(Lipschitz estimator)} \\
		&=K \cdot \paren[\Big]{ \frac{1}{n} \sum_{i=1}^n \norm{u_i - u_i'}^2  }^{1/2}
			&\text{(def of norm)} \\
		&= K \cdot \paren[\Big]{ \frac{\alpha^2}{n} \norm{v -v'}^2  }^{1/2}
		&\text{(by construction of $\vu, \vu'$)} \\
		&= K \frac{\alpha }{\sqrt{n}} \norm{v - v'} \\
		&= K C
	\end{align*}

	Finally, we need to establish that the causal estimand is different when evaluated at $\vu$ and $\vu'$.
	To this end, first observe that by linearity of the effect functionals and the construction of the two functions, the difference in the estimands can be expressed as
	\[
	\aefv{\vu} - \aefv{\vu'}
	= \aefv{\vu - \vu'}
	= \frac{1}{n} \sum_{i=1}^n \efvi{u_i - u_i'}
	= \frac{\alpha}{n} \efv{j}{ v - v' }
	= \frac{\alpha}{n} \cdot \paren[\big]{ \efv{j}{v} - \efv{j}{v'} }
	\enspace.
	\]
	Using the fact that positivity is violated, we can lower bound the absolute difference of the estimands:
	\[
	\abs[\big]{ \aefv{\vu} - \aefv{\vu'}  }
	= \frac{\alpha}{n} \cdot \abs[\big]{ \efv{j}{v} - \efv{j}{v'} }
	\geq \frac{\alpha}{n} \cdot \beta \cdot \norm{v - v'}
	= \beta \cdot \frac{C}{\sqrt{n}}
	\]
	The key insight is that by choosing the positivity violation $\beta$ to be sufficiently large, we can ensure that $\abs{\aefv{\vu} - \aefv{\vu'}} \gg \abs{\E{ \estv{\vu} } - \E{ \estv{\vu'} }}$.
	In this case, the estimator must be biased for at least one of the two functions.
	The following argument makes this intuition more precise.
	
	\begin{align*}
		\sup &\setb[\big]{ \abs[\big]{ \E{ \estv{\vu} - \aefv{\vu} } }  : \norm{\vu} \leq C } \\
		&\geq \max \setb[\big]{ 
				\abs[\big]{ \E{ \estv{\vu} - \aefv{\vu} } }, 
				\abs[\big]{ \E{ \estv{\vu'} - \aefv{\vu'} } } 
			} \\
		&\geq \frac{1}{2} \braces[\Big]{ 
				\abs[\big]{ \E{ \estv{\vu} } - \aefv{\vu} }
				+ \abs[\big]{ \E{ \estv{\vu'} } - \aefv{\vu'} }
	 		} 
 			&\text{(max $\geq$ average)}\\
 		&\geq \frac{1}{2} \abs[\Big]{ 
 				\braces[\big]{ \E{ \estv{\vu} } - \aefv{\vu} }
 				- \braces[\big]{ \E{ \estv{\vu'} } - \aefv{\vu'} } 
 	 		}
  			&\text{(triangle inequality)} \\
  		&=  \frac{1}{2} \abs[\Big]{ 
  			\braces[\big]{ \E{ \estv{\vu} } -  \E{ \estv{\vu'} } }
  			- \braces[\big]{ \aefv{\vu} - \aefv{\vu'} } 
  		}
  			&\text{(rearranging terms)} \\
  		&\geq \frac{1}{2} \abs[\Big]{ 
  			\abs[\big]{ \E{ \estv{\vu} } -  \E{ \estv{\vu'} } }
  			- \abs[\big]{ \aefv{\vu} - \aefv{\vu'} } 
  	 	} &\text{(reverse triangle inequality)} \\
   	&\geq \frac{1}{2} \paren[\Big]{ 
   		\beta \cdot \frac{C}{\sqrt{n}} - K C
   	 }
    \enspace,
	\end{align*}
	where the last line holds for sufficiently large $\beta$, i.e. $\beta \geq \sqrt{n} \cdot K$.
	Thus, the supremum is unbounded by letting $\beta$ grow arbitrarily large.
	
	A final remark is in order.
	Our proof implicitly assumed that $\norm{v - v'} > 0$ in order to define the scaling $\alpha$.
	If in fact it is the case that $\norm{v - v'} = 0$, then the proof is even simpler in the sense that (1) the second moment of $\vu'$ is unaffected by the choice of scaling $\alpha$, (2) the estimator will be equal almost surely under $\vu$ and $\vu'$ so that the means are equal reardless of the choice of $\alpha$, and (3) the estimands can be made arbitrarily far apart by choosing a sufficiently large $\alpha$.
\end{proof}

\subsection{Uniform Consistency}\label{supp:uniform-consistency}

In this section, we present results which relate to uniform consistency.
We begin by proving the existence of the variance characterizing linear operator and providing an explicit matrix representation for finite dimensional model spaces.
Then, we show that its operator norm exactly corresponds to uniform consistency and provide an explicit matrix representation.
Finally, we highlight how the dependency graph method can be understood in this context.

\subsubsection{Variance Characterizating Linear Operator (Theorem~\mainref{thm:vclo-statement})}

In this section, we prove the existence of the variance characterizing linear operator.
We begin by stating a standard lemma about the representation of bilinear forms in Hilbert spaces.
The proof may be found at the end of this section.

\newcommand{\hs}{\mathcal{H}}

\begin{lemma}\label{lem:bilinear-representation}
	Let $(\hs, \iprod{\cdot, \cdot})$ be a Hilbert space and let $V: \hs \times \hs \to \Reals$ be a bilinear form satisfying the following properties:
	\begin{itemize}
		\item Symmetric: $V(x,y) = V(y,x)$ for all $x, y \in \hs$,
		\item Positive semi-definite: $V(x,x) \geq 0$ for all $x \in \hs$,
		\item Bounded: $\abs{V(x,y)} \leq C \norm{x} \norm{y}$ for all $x,y \in \hs$ and some $C < \infty$.
	\end{itemize}
	Then, there exists a bounded linear operator $L : \hs \to \hs$ such that
	\begin{equation}
		V(x,y) = \iprod{ L x, L y }
		\qquadtext{for all}
		x, y \in \hs.
	\end{equation}
\end{lemma}

With this general lemma in hand, we are now ready to prove the existence of the variance characterizing linear operator.

\begin{reftheorem}{\mainref{thm:vclo-statement}}
  \thmvclostatement
\end{reftheorem}

\begin{proof}
Define the bilinear form $V : \Oprodspace \times \Oprodspace \to \Reals$ as
\begin{equation}
	V(\vU, \vV) = \frac{1}{n} \sumin \sumjn \Cov[\big]{\rri U_i , \rrj V_j},
\end{equation}
where $\vU = (U_1, \dotsc, U_n)$ and $\vV = (V_1, \dotsc, V_n)$.
Our proof will be to show that $V$ satisfies the conditions of Lemma~\ref{lem:bilinear-representation}, hence yielding the representing linear operator.
The fact that $V$ is symmetric and bilinear follows directly from symmetry and bilinearity of the covariance operator.
The fact that $V$ is positive semi-definite follows from non-negativity of the variance.

It remains to be shown that the bilinear form $V$ is also bounded.
To this end, recall that $\uerr$ is the uniform root mean square error of the Riesz estimator over the model space $\Mprodspace$.
Using this together with the unbiasedness of the Riesz estimator, we have that
$\Var{ \estv{\vu} } \leq \uerr^2 \cdot \norm{\vU}^2$.
Now, boundednes follows from the finite variance condition (i.e. $\uerr < \infty$) and
\begin{align*}
	V(\vU, \vV) 
	&= n \Cov[\Big]{ \frac{1}{n} \sumin \rri U_j , \frac{1}{n} \sumjn \rrj V_j } 
		&\text{(bilinearity)}\\
	&= n \Cov{ \estv{\vU} , \estv{\vV}  }
		&\text{(definition of estimator)} \\
	&\leq n \sqrt{ \Var{ \estv{\vU}}  \estv{\vV} }
		&\text{(Cauchy-Schwarz)} \\
	&\leq n \cdot \uerr^2 \cdot \norm{\vU} \norm{\vV}
		\enspace.
		&\text{(uniform MSE bound)}
\end{align*}

Applying Lemma~\ref{lem:bilinear-representation}, we have that there exists a bounded linear operator $\varmap : \Oprodspace \to \Oprodspace$ such that $V(\vU, \vV) = \iprod{ \varmapv{\vU}, \varmapv{\vV} }$.
Therefore,
\begin{equation}
	n \Var{\estv{\vU}} = V(\vU, \vU) = \iprod{ \varmapv{\vU}, \varmapv{\vU} } = \norm{\varmapv{\vU}}^2.
	\tag*{\qedhere}
\end{equation}
\end{proof}

We end this section by providing a proof of Lemma~\ref{lem:bilinear-representation}.

\begin{proof}[Proof of Lemma~\ref{lem:bilinear-representation}]
Observe that for a fixed $x \in \hs$, we have that $y \mapsto V(x,y)$ is a linear functional.
Moreover, this linear functional is bounded in the sense that
\begin{equation}
	\abs{V(x,y)} \leq C \norm{x} \norm{y} = C_x \norm{y},
\end{equation}
where $C_x = C \norm{x}$.
We have $C_x < \infty$ due to boundedness of the bilinear form.
Thus, by the Riesz representation theorem, there exists a unique vector $f_x \in \hs$ such that $V(x,y) = \iprod{f_x, y}$ for each fixed $x \in \hs$.
Define the operator $A: \hs \to \hs$ to be the mapping $x \mapsto f_x$, which is linear due to bilinearity of $V$.
This means that we can represent the bilinear form with respect to the inner product using the linear operator $A$:
\begin{equation}
	V(x, y) = \iprod{A x, y}.
\end{equation}

The next step is to show that $A$ is positive semi-definite, self-adjoint, and bounded.
The fact that $A$ is positive semidefinite follows directly from the fact that $V$ is positive semidefinite, as $\iprod{A x, x} = V(x, x) \geq 0$.
The fact that $A$ is self-adjoint follows from the symmetry of $V$, as
\begin{equation}
	\iprod{ A x, y }
	= V(x, y)
	= V(y, x)
	= \iprod{ A y, x }
	= \iprod{ x, A y }.
\end{equation}
Finally, boundedness of $A$ follows from boundedness of $V$, as
\begin{equation}
	\norm{A x}
	= \sup_{\norm{y} = 1} \iprod{A x, y}
	= \sup_{\norm{y} = 1} V(x, y)
	\leq C \norm{x}.
\end{equation}

Every bounded linear operator that is positive semi-definite and self-adjoint has a unique bounded square root, meaning that there exists a linear operator $L : \hs \to \hs$ such that $A = L^* L$.
Therefore,
\begin{equation}
	V(x, y)
	= \iprod{A x, y}
	= \iprod{L^* L x, y}
	= \iprod{L x, L^{**} y}
	= \iprod{L x, L y},
\end{equation}
where the last equality follows from involution of the adjoint: $L^{**} = L$.
\end{proof}

\subsubsection{Operator Norm and Uniform Consistency (Corollary~\mainref{coro:riesz-est-consistent})}

\begin{refcorollary}{\mainref{coro:riesz-est-consistent}}
\cororieszestconsistent
\end{refcorollary}

\begin{proof}
By Theorem~\mainref{thm:vclo-statement}, when the Riesz estimator has finite variance, the variance characterizing operator $\varmap$ exists.
Because the Riesz estimator is unbiased, we have
$$
n \uerr^2 = \sup_{\norm{\vU} = 1} n \Var{\estv{\vU}} = \sup_{\norm{\vU} = 1} \norm{\varmapv{\vU}}^2 = \opnorm{\varmap}^2,
$$
where $\opnorm{\varmap} = \sup_{\norm{\vU} = 1} \norm{\varmapv{\vU}}$ is by definition of the operator norm.
A direct consequence is $\uerr = n^{-1/2} \opnorm{\varmap}$, and that $\uerr \to 0$ if and only if $\opnorm{\varmap} = \littleO{n^{1/2}}$.
\end{proof}

\subsubsection{Matrix Representation of Variance Characterizing Linear Operator}

One can express the variance characterizing operator as a matrix using similar ideas to those found in Section~\ref{supp:matrix-construction}.
Recall from there that $b_{i,1} \dots b_{i,d}$ formed a basis for $\Mspacei$.

We begin by showing how to construct a matrix which represents the variance of the estimator.
Consider the combined outcome function $\vu \in \Mprodspace$, whose coordinate functions are represented in the given basis as
\[
u_i(z) = \sum_{k=1}^d \beta_{i,k} b_{i,k}(z) \enspace.
\]
The combined outcome function is $n \cdot d$ dimensional, where $d$ is the dimension of each of the model spaces.
We write the $n \cdot d$-dimensional parameter vector as $\vec{\beta} = (\vec{\beta}_1 \dots \vec{\beta}_n)$, where $\vec{\beta}_i = (\beta_{i,1} \dots \beta_{i,d})$.

The variance will be represented by an $nd \times nd$ matrix $\mat{C}$ whose entries are given by
\[
\mat{C}(i, k ; j, \ell)
= \Cov{ b_{i,k} R_i, b_{j,\ell} R_j }
\enspace.
\]
To see that this matrix represents the variance in the given basis, observe that
\begin{align*}
	n^2 \cdot \Var{\estv{\vu}}
	&=  \sum_{i=1}^n \sum_{j=1}^n \Cov{ U_i \rri , U_j \rrj } \\
	&= \sum_{i=1}^n \sum_{j=1}^n \Cov[\Big]{ \sum_{k=1}^d \beta_{i,k} b_{i,k} \rri , \sum_{j=1}^d \beta_{j,\ell} b_{j,\ell} \rri } \\
	&= \sum_{i=1}^n \sum_{j=1}^n \sum_{k=1}^d \sum_{\ell=1}^d \beta_{i,k} \beta_{j,\ell} \Cov{ b_{i,k} \rri , b_{j,\ell} \rrj } \\
	&= \sum_{i=1}^n \sum_{j=1}^n \sum_{k=1}^d \sum_{\ell=1}^d \beta_{i,k} \beta_{j,\ell} \Cov{ b_{i,k} R_i, b_{j,\ell} R_j } \\
	&= \vec{\beta}^\top \mat{C} \vec{\beta}
\end{align*}
While this matrix $\mat{C}$ represents the variance in the sense above, its largest eigenvalue does not correspond to the operator norm of the variance characterizing operator, $\opnorm{\varmap}^2$.
The reason is that the Euclidean norm of $\vec{\beta}$ does not correspond to the $L^2$ norm of the corresponding function $\vu$.
In order to compute $\opnorm{\varmap}^2$ as the eigenvalue of some matrix, we will have to consider a change of basis. 

To this end, observe that the $L^2$ norm of the combined function $\vu$ is given as
\[
\frac{1}{n} \sum_{i=1}^n \E{ U_i^2 }
= \frac{1}{n} \sum_{i=1}^n \E[\Big]{ \paren[\Big]{ \sum_{k=1}^d \beta_{i,k} b_{i,k} }^2 } 
= \frac{1}{n} \sum_{i=1}^n \sum_{k=1}^d \sum_{\ell=1}^d \beta_{i,k} \beta_{i,\ell} \E{ b_{i,k} b_{i,\ell} } 
= \frac{1}{n} \vec{\beta}^\top \mat{S} \vec{\beta}
\enspace,
\]
where $\mat{S}$ is the $nd \times nd$ diagonal block matrix whose $d \times d$ diagonal blocks are $\mat{S}_1 \dots \mat{S}_n$, where $\mat{S}_i$ is a $d$-by-$d$ matrix with entries given by expected product of basis functions, i.e. $\mat{S}_i(k,\ell) = \E{ b_{i,k} b_{i,\ell} }$.

Define the matrix $\mat{V} = \mat{S}^{+/2} \mat{C} \mat{S}^{+/2}$, where $\mat{S}^{+/2}$ is the square root of the pseudo-inverse of $\mat{S}$.
We claim that the largest eigenvalue of this matrix yields the square of the operator norm of the variance characterizing operator: $\lambda_{\max}(\mat{V}) = \opnorm{\varmap}^2$.
To see this, observe that
\begin{align*}
	\opnorm{\varmap}^2 
	&= \sup \setb[\Big]{ n \cdot \Var{ \estv{\vu} } : \frac{1}{n} \sum_{i=1}^n \E{ U^2 } \leq 1 } \\
	&= \sup \setb[\Big]{ \frac{1}{n} \vec{\beta}^\top \mat{C} \vec{\beta} : \frac{1}{n} \vec{\beta}^\top \mat{S} \vec{\beta} \leq 1 } \\
	&= \sup \setb[\Big]{ \widetilde{\vec{\beta}}^\top \mat{V} \widetilde{\vec{\beta}} :  \widetilde{\vec{\beta}}^\top \widetilde{\vec{\beta}} \leq 1 } \\
	&= \lambda_{\max}(\mat{V}) \enspace,
\end{align*}
where the third equality used the change of basis $\widetilde{\vec{\beta}} = \mat{S}^{1/2} \vec{\beta}$.
Thus, the operator norm of the variance characterizing operator may be explicitly computed by experimenters in this way.

\subsubsection{Comparison to Dependency Graph Method} \label{sec:supp-dep-graph-var}

The dominant approach to prove consistency in the recent design-based causal inference literature is the so-called dependency graph method \citep{Chen2004Normal,Ross2011Fundamentals}.
Examples include \citet{Aronow2017Estimating}, \citet{Li2022Random} and \citet{Ogburn2024Causal}.
In this section, we show that the dependency graph method can be viewed as bounding the operator norm of the variance characterizing operator.

At a high level, the dependency graph method associates a graph to the units based on the pattern of dependence dictated by the model spaces.
It then provides an upper bound on the variance of Horvitz--Thompson-type estimators depending, in part, on the degree distribution of the dependency graph.
The graph is constructed using a binary independence concept, where two units are dependent (and thus have an edge connecting them in the graph) if their model spaces are not strictly independent.

More formally, we define $\depneighi \subseteq [n]$ to be the dependency neighborhood of the outcome spaces $\Ospacei$ for all $i \in [n]$, according to the definition of \citet{Ross2011Fundamentals}.
That is, $\depneighi$ is the smallest set such that $\Ospacei$ is jointly independent of $\cup_{j \notin \depneighi} \Ospacej$.
This means that any collection of random variables in $\Ospacei$ will be jointly independent of any collection of random variables in $\cup_{j \in [n] \setminus \depneighi} \Ospacej$.
Let $\depmax = \max_{i \in [n]} \card{\depneighi}$ be the largest dependency neighborhood.

A typical bound produced by the dependency graph method consists of two parts.
The first part is the maximum degree of the dependency graph: $\depmax$.
The second part is a summary of the experimental design: $\gamma$.
In the conventional experimental setting with binary treatments and no interference, the design summary $\gamma$ is the inverse of the smallest treatment probability, see Section~\suppref{sec:supp-worked-exposure-mapping}.
In a general setting, $\gamma$ is the maximum essential supremum of the Riesz representors: $\gamma = \max_{i \in [n]} \pnorm{\infty}{\rri}$.

The following proposition presents the style of dependency graph analysis.
We defer its proof to the end of this section.

\begin{proposition} \label{prop:dep-graph-op-norm}
	Using the dependency graph method, the operator norm of the variance characterizing operator is bounded as $\opnorm{\varmap} \leq \gamma \depmax^{1/2}$.
	Thus, provided that $\gamma = \bigO{1}$, a sufficient (but not necessary) condition for consistency of the Riesz estimator is $\depmax = \littleO{n}$.
\end{proposition}

We emphasize that the use of the variance characterizing operator is not restricted to the dependency graph method; indeed, an important purpose of the operator is to provide alternative routes to study precision of design-based estimators.
The purpose of the current subsection was to show that existing approaches can be understood as implicitly bounding the operator norm.
The concern with the dependency graph method is that the upper bound it produces can be quite loose.
It is possible to sharpen the bound, but the method never provides necessary conditions for consistency unless one imposes strong auxiliary conditions.
In particular, the method does not differentiate between weak and strong dependencies between the units, making it overly conservative in settings with widespread but weak dependence.
Methods that more directly investigate the norm of the variance characterizing operator will typically be more informative.
For example, \citet{Kandiros2025Conflict} obtain improved rates of convergence in network experiments using techniques from spectral graph theory that account for both strong and weak dependencies.

Let us now provide a proof of Proposition~\ref{prop:dep-graph-op-norm}.

\begin{proof}[Proof of Proposition~\ref{prop:dep-graph-op-norm}]

Let $e_{ij} = \indicator{j \in \depneighi} \cdot \indicator{i \in N_j}$ be an indicator for whether $i$ is in $j$'s dependency neighborhood \emph{and} $j$ is in $i$'s dependency neighborhood.
Note that $e_{ij} = 0$ implies that $\Cov{\rri U_i, \rrj U_j} = 0$, because if the covariance is non-zero, $\rri U_i$ and $\rrj U_j$ are dependent, and $i$ and $j$ must be in each other's dependency neighborhoods.
Also note that the indicator is symmetric: $e_{ij} = e_{ji}$.

Using this indicator, we can write
$$
n \Var{\estv{\vU}}
= \frac{1}{n} \sumin \sumjn e_{ij} \Cov{\rri U_i, \rrj U_j}.
$$
By the Cauchy--Schwarz, Hölder's and the AM--GM inequalities,
\begin{align*}
	\Cov{\rri U_i, \rrj U_j}
	&\leq \sqrt{\Var{\rri U_i} \Var{\rrj U_j}} \\
	&\leq \sqrt{\E{\rri^2 U_i^2} \E{\rri^2 U_i^2}} \\
	&\leq \gamma^2 \sqrt{\E{U_i^2} \E{U_i^2}} \\
	&\leq \frac{\gamma^2}{2} \paren[\big]{ \E{U_i^2} + \E{U_j^2}}
	\enspace,
\end{align*}
where $\gamma = \max_{i \in [n]} \pnorm{\infty}{\rri}$ is the maximum essential supremum of the Riesz representors.
Using symmetry of the indicator, $e_{ij} = e_{ji}$, we have
$$
n \Var{\estv{\vU}}
\leq \frac{\gamma^2}{n} \sumin \sumjn \frac{e_{ij}}{2} \paren[\big]{ \E{U_i^2} + \E{U_j^2}}
= \frac{\gamma^2}{n} \sumin \sumjn e_{ij} \E{U_i^2}.
$$

Note that $e_{ij} \leq \indicator{j \in \depneighi}$, so $\sumjn e_{ij} \leq \card{\depneighi} \leq \depmax$.
We can therefore write
$$
n \Var{\estv{\vU}}
\leq \frac{\gamma^2}{n} \sumin \E{U_i^2} \sumjn e_{ij}
\leq \frac{\gamma^2 \depmax}{n} \sumin \E{U_i^2}
= \gamma^2 \depmax \norm{\vU}^2.
$$
Recall that $n \uerr^2 = \sup_{\norm{\vU} = 1} n \Var{\estv{\vU}}$, from which it follows that $n \uerr^2 \leq \gamma^2 \depmax$.
The proof is completed by Corollary~\mainref{coro:riesz-est-consistent}, showing that $n \uerr^2 = \opnorm{\varmap}^2$.
\end{proof}

\subsection{Asymptotically Valid Inference} 

In this section, we review theoretical results which facilitate the construction of asymptotically valid confidence intervals.

\subsubsection{Conservative Variance Estimation (Theorems~\mainref{thm:second-order-riesz-rep} and \mainref{thm:conservative-var-est})}

In this section, we show that the variance estimators described in the main paper are conservatively biased.
To this end, our first step will be to establish that second order positivity is a sufficient condition for the existence of an estimator for the covariance of two individual estimators.

\begin{reftheorem}{\mainref{thm:second-order-riesz-rep}}
\thmsecondorderrieszrep
\end{reftheorem}

\begin{proof}
A linear functional $\covfij : \Ospacei \otimes \Ospacej \to \Reals$ that satisfies second order positivity is continuous.
This means that we can use the same approach as in the proof of Lemma~\mainref{lem:extended-functionals} above to extend the functional $\covfij$ to the paired outcome space $\pairedOspaceij = \textrm{cl}\paren[\big]{ \Ospacei \otimes \Ospacej / N_{i,j}}$ discussed in the main paper.
Note that $\pairedOspaceij$ is a Hilbert space.
Therefore, by the Riesz representation theorem, there exists an element $\crrij \in \pairedOspaceij$ such that $\covfijv{\tU} = \iprod{\crrij, \tU}$ for all $\tU \in \pairedOspaceij$.
Overloading the notation as in the main paper, let $\crrij$ also denote the random variable associated with the tensor $\crrij$.
We then have $\iprod{\crrij, \tU} = \E{\crrij U}$, where $U$ is the random variable associated with the tensor $\tU$.
It follows that for all $U_i \in \Ospacei$ and $U_j \in \Ospacej$,
$$
\covfijv{ U_i \otimes U_j } = \iprod{\crrij, U_i \otimes U_j} = \E{ \crrij U_i U_j }.
\qedhere
$$
\end{proof}

The following theorem shows that the class of variance estimators considered in the main body are conservatively biased.

\begin{reftheorem}{\mainref{thm:conservative-var-est}}
\thmconservativevarest
\end{reftheorem}

\begin{proof}
First-order positivity and existence of fourth moments ensure that $\crrij$ and $\evb$ exist.
Correctly specified model spaces ensure that $\ooi \otimes \ooj \in \pairedOspaceij$, meaning that $\crrij$ represents the true potential outcome tensor:
$$
\E{ \crrij \ooi \ooj }
= \iprod{\crrij, \ooi \otimes \ooj}
= \boufijv{ \ooi \otimes \ooj }.
$$
It then follows that
$$
\E{\evb}
= \frac{1}{n^2} \sumin \sumjn \E{\crrij \ooi \ooj}
= \frac{1}{n^2} \sumin \sumjn \boufijv{ \ooi \otimes \ooj }
= \vbv{\voo}.
$$
We showed in Section~\ref{supp:bound-functionals} that $\vbv{\voo} \geq \Var{ \estv{\voo} }$ for all potential outcomes $\voo \in \Oprodspace$, which completes the proof.
\end{proof}

\subsubsection{Conservatively Consistent of Variance Estimation}
\label{supp:var-est-consistency}
 
In this section, we introduce the notion of uniformly consistent (conservative) variance estimation.
Given the connection to uniform mean square error of the point estimation (Section~\mainref{sec:uniform-consistency}), we keep this section brief.

Our goal is to understand when the variance estimator achieves high precision estimates of the variance bound.
Because the variance is decreasing at the rate $\opnorm{\varmap}^2 / n$, we normalize the variance by this quantity so that it stays as a constant rate, i.e. $\frac{n}{\opnorm{\varmap}^2} \cdot \Var{\est} = \bigTheta{1}$.
With this re-scaling in hand, we define the \emph{uniform mean square error of the variance estimator} as
\[
\mathcal{H}_n(C) =
\sup_{ \substack{ \vpo \in \Mprodspace \\ \frac{1}{n} \sum_{i=1}^n \E{\ooi^4} \leq C^4 } }
	\E[\Bigg]{ \paren[\Bigg]{ \frac{n}{\opnorm{\varmap}^2} \cdot \evb(\vpo) - \frac{n}{\opnorm{\varmap}^2} \cdot \vbv{\vpo}  }^2 }^{1/2}
\enspace,
\]
which is the largest mean square error attained by the (normalized) variance estimation to the (normalized) variance bound when the potential outcomes have fourth moment bounded by $C$.
There are two key differences that distinguish this notion of uniform MSE for variance estimators and the notion of uniform MSE for effect estimators, introduced in Section~\mainref{sec:uniform-consistency}.
First, the error in the variance estimator is normalized by $\opnorm{\varmap}^2 / n$ to account for the fact that the variance itself is decreasing with the sample size.
Second, the potential outcomes are now restricted by their fourth moment, rather than their second moment.
This difference reflects the fact that variance estimation{\textemdash}which depends on the square of the potential outcomes{\textemdash}will typically require higher order moment conditions than effect estimation{\textemdash}which depends only on the magnitude of the potential outcome functions.

We say that the variance estimator is \emph{uniformly consistent at rate $r_n$}, denoted $\mathcal{H}_n = \bigO{1 / r_n}$ if for all $C \geq 0$, $\limsup_{n \to \infty} r_n \mathcal{H}_n(C) < \infty$.
The variance estimator based on Riesz representors presented in Section~\mainref{sec:riesz-var-est} is a quadratic form in the observed outcomes.
For estimators of this form, the uniform mean square error $\mathcal{H}_n(C)$ scales proportionally with $C$, so that the choice of $C$ is irrelevant.
In usual experimental settings, the methods that may be used to establish rates of consistency for point estimation may also be used to establish rates of consistency for variance estimation.

Under fourth moment restrictions and non-superefficiency assumption, uniform consistency of the normalized variance estimator ensures stability of the variance estimator, i.e. $\vb / \evb \xrightarrow{p} 1$.
This stability is required for conventional confidence intervals to asymptotically cover at the nominal rates.
The details of this argument are well-known in the literature \citep[see e.g.,][]{Kandiros2025Conflict}, and we only briefly review them here.

The non-superefficiency assumption states that the asymptotic sequence of potential outcome functions is such that the variance cannot go to zero too quickly, e.g. $\liminf_{n \to \infty} \frac{n}{\opnorm{\varmap}^2} \cdot \Var{\estv{\vpo}} > 0$.
For example, this removes from consideration the case where all the potential outcome functions in the sequence are all identically zero in which case the variance would be equal to zero.
If the fourth moments of the potential outcome functions are asymptoticaly bounded (i.e. $\limsup_{n \to \infty} n^{-1} \sum_{i=1}^n \E{ \ooi^4 } < \infty$) and the variance estimator is uniformly consistent (i.e. $\mathcal{H}_n \to 0$), then the non-superefficiency assumption may be invoked together with the continuous mapping theorem to obtain that $\vb / \evb \xrightarrow{p} 1$.

\subsubsection{Central Limit Theorem using Dependency Graphs}\label{supp:dependency-graph-clt}

As discussed in the main body, the derivation of the limiting distribution of the Riesz estimator will depend on the situation at hand.
In this section, we demonstrate how the dependency graph method can be generically employed to obtain central limit theorems.

The following theorem provides a generic central limit theorem using the dependency graph method.
While the dependency graph method can be carried out in various ways, we adopt a general analysis that mirrors the conditions for consistency analyzed in Section~\ref{sec:supp-dep-graph-var}.
In particular, the condition for asymptotic normality depends on both the maximum dependency degree $\depmax$ as well as the essential supremum of the individual Reisz representors, $\gamma = \max_{i \in [n]} \norm{\rri}_{\infty}$.
We refer to Section~\ref{sec:supp-dep-graph-var} for a definition of the dependency graph neighborhoods.

\begin{theorem} \label{thm:clt-dep-graph}
	Let $\sigma^2 = \Var{\est}$ be the variance of the Riesz estimator.
	Suppose that positivity and correct specification are satisfied and further suppose that the asymptotic sequence of experiments that satisfy the following conditions:
	\begin{enumerate}
		\item \textbf{Bounded 4th Moment}: $ \frac{1}{n} \sum_{i=1}^n \E{ \ooi^4 } = \bigO{1} $
		\item \textbf{Bounded Dependence}: $\depmax^2 \gamma^3 = \littleO{n^{1/2}}$.
		\item \textbf{Non-Super-efficiency}: $\sigma^{-2} = \bigO{n}$ 
	\end{enumerate}
	Then, the Riesz estimator is asymptotically normal: $\frac{\est - \ate}{\Var{\est}} \xrightarrow{d} \mathcal{N}(0,1)$.
\end{theorem}

To prove the theorem, we will use the dependency graph analysis of Stein's method.
To this end, we use the following result which appears in \citet{Ross2011Fundamentals}.

\begin{proposition}[(Theorem 3.5 in \citet{Ross2011Fundamentals})] \label{prop:dep-graph-clt-general}
	Let $X_1 \dots X_n$ be random variables with $\E{ X_i^4 } < \infty$, $\E{X_i} = 0$.
	Let $\sigma^2 = \Var{\sum_{i=1}^n X_i}$ and define $W = \sigma^{-1} \sum_{i=1}^n X_i$.
	Suppose $(X_1, \dots X_n)$ have maximum dependency neighborhood $\depmax$.
	Then, the Wasserstein distance of $W$ to a standard normal random variable $Z$, is bounded as 
	\[
	d_{W}(W, Z) \leq 
	\frac{\depmax^2}{\sigma^3} \sum_{i=1}^n \E{ \abs{X_i}^3 }
	+
	\sqrt{ \frac{26}{\pi}} \frac{\depmax^{3/2}}{\sigma^2} \sqrt{ \sum_{i=1}^n \E{ X_i^4 } }
	\enspace.
	\]
\end{proposition}

At a high level, we will use Proposition~\ref{prop:dep-graph-clt-general} where we consider each $X_i$ to be the (scaled) error an individual treatment estimator.
The following lemma will use the essential supremum of the Riesz estimator to bound the moments of the individual estimator errors.

\begin{lemma} \label{lemma:individual-error-moments}
	Suppose that positivity and correct specification are satisfied.
	Define $X_i = (1/n)( \est_i - \itei )$ to be the error of the individual estimator.
	For any integer $k$, we have that 
	\[
	\E{ \abs{ X_i }^k } \leq \paren[\Big]{ \frac{2}{n} }^k \gamma^k \E[\big]{ \abs{\ooi}^k } \enspace.
	\]
\end{lemma}
\begin{proof}
	Observe that 
	\begin{align*}
		\E{ \abs{ X_i }^k }
		&= \frac{1}{n^k}\E[\Big]{ \abs{ \est_i - \efvi{\ooi} }^k }
			&\text{(def of $X_i$)} 
		\\
		&= \frac{1}{n^k}\E[\Big]{ \abs{ \ooi \rri - \E{\ooi \rri} }^k }
			&\text{(def of $\est_i$ and positivity)} 
		\\
		&\leq \frac{2^{k-1}}{n^k} \E[\Big]{ \abs{ \ooi \rri }^k + \abs{ \E{\ooi \rri}^k}  }
			&\text{$(a+b)^k \leq 2^{k-1} (\abs{a}^k + \abs{b}^k)$}
		\\ 
		&\leq \frac{2^{k-1}}{n^k} \paren[\Big]{ \E{ \abs{\ooi \rri} }^k + \E{ \abs{\ooi \rri} ^k} }
			&\text{(Jensen's inequality)} 
		\\
		&=\paren[\Big]{ \frac{2}{n} }^k \E[\big]{ \abs{\ooi \rri}^k } 
		\\
		&\leq \paren[\Big]{ \frac{2}{n} }^k \gamma^k \E[\big]{ \abs{ \ooi }^k }
		\enspace.
		\qedhere
	\end{align*}
\end{proof}

We are now ready to prove the generic central limit theorem stated in Theorem~\ref{thm:clt-dep-graph}.

\begin{proof}[Proof of Theorem~\ref{thm:clt-dep-graph}]
	
	As above, define $X_i = (1/n)( \est_i - \itei )$ to be the error of the individual estimator.
	Observe that the normalized sum of $X_1 \dots X_n$ is precisely the normalized error of the Riesz estimator: if $\sigma = \Var{\sum_{i=1}^n X_i}$ then
	\[
	W \triangleq \sigma^{-1} \sum_{i=1}^n X_i 
	= \sigma^{-1} \frac{1}{n} \sum_{i=1}^n ( \est_i - \itei ) 
	= \frac{\est - \ate}{\sqrt{\Var{\est}}}
	\enspace.
	\]
	Now, applying the Proposition~\ref{prop:dep-graph-clt-general} together with the moment bounds of Lemma~\ref{lemma:individual-error-moments} yields that 
	\begin{align*}
		d_{W}(W,Z) 
		&\leq 
		\frac{\depmax^2}{\sigma^3} \sum_{i=1}^n \E{ \abs{X_i}^3 }
		+
		\sqrt{ \frac{26}{\pi}} \frac{\depmax^{3/2}}{\sigma^2} \sqrt{ \sum_{i=1}^n \E{ X_i^4 } }
		\\
		&\leq 2^3 \frac{\depmax^2 \gamma^3}{\sigma^3 n^2} \cdot \paren[\Big]{ \frac{1}{n} \sum_{i=1}^n \E{ \abs{\ooi}^3 } } 
		+
		\sqrt{\frac{2^4 \cdot 26}{\pi}} \frac{\depmax^{3/2}  \gamma^2}{\sigma^2 n^{3/2}} \cdot \sqrt{ \frac{1}{n} \sum_{i=1}^n \E{ \ooi^4 } }
		\\
		&\leq \bigO[\Big]{ \frac{\depmax^2 \gamma^3}{\sigma^3 n^{2}} } 
		+ \bigO[\Big]{ \frac{\depmax^{3/2}  \gamma^2}{\sigma^2 n^{1/2}} }
		\intertext{where the last inequality follows from Bounded 4th moment assumption. Next, using non-superefficiency, we obtain that}
		&\leq \bigO{ \frac{\depmax^2 \gamma^3}{n^{1/2}} } + \bigO{ \frac{\depmax^{3/2} \gamma^2}{n^{1/2}} }
		\\
		&= \littleO{1} \enspace,
	\end{align*}
	where the last inequality follows by the bounded dependence assumption, noting that the first term is dominating the second one.
\end{proof}

%% file: tex/supp-simulations.tex
\section{Simulation Results} 

\subsection{Continuous Network Experiments} \label{sec:supp-network-experiment}

In this section, we review additional details of the simulation results presented in Section~\mainref{sec:num-illustrations} of the main body.
We begin by describing how the network and the potential outcomes were generated for each choice of $(d, r)$.
Recall that $d-1$ is the number of neighbors each node has and $r$ is the order of interaction terms.
Next, we present additional QQ plots to assess the appropriateness of the normal approximation at the sample sizes we consider.

\newcommand{\makecontinuousplots}[2]{
\begin{figure}[h!]
	\centering
	\parbox[t]{0.30\textwidth}{%
		\centering
		\includegraphics[width=0.30\textwidth]{\simspath/continuous/sim-res-100-#1-#2.pdf}%
		\\
		(a) $n=100$
	}
	\quad
	\parbox[t]{0.30\textwidth}{%
		\centering
		\includegraphics[width=0.30\textwidth]{\simspath/continuous/sim-res-1000-#1-#2.pdf}%
		\\
		(b) $n=1,000$
	}
	\quad
	\parbox[t]{0.30\textwidth}{%
		\centering
		\includegraphics[width=0.30\textwidth]{\simspath/continuous/sim-res-10000-#1-#2.pdf}%
		\\
		(c) $n=10,000$
	}
	\caption{QQ Plot of Sampling Distribution Relative to Normal, when $d=#1$, $r=#2$} \label{fig:qq-#1-#2}
\end{figure}
}

\phantom{.}

\makecontinuousplots{3}{3}

\makecontinuousplots{4}{3}

\makecontinuousplots{4}{4}

In order to construct a simulated network $G$, we sample $X_i \in [0, 1]$ uniformly for each unit.
For all pairs $(i,j) \in [n]^2$, we calculate $\delta_{ij} = U_{ij} (X_i - X_j)^2$, where $U_{ij}$ is uniform on $[0, 1]$.
The neighbors of unit $i$ is then the $d - 1$ units $j \in [n] \setminus \braces{i}$ with smallest $\delta_{ij}$.
This process induces homophily, where units that are similar in terms of $X_i$ tend to be neighbors.

We generate potential outcome functions which feature heterogeneity.
We order the coefficients $\alpha_{i,1}, \alpha_{i,2}, \dots$ in lexographical order and set them as 
$\alpha_{i,s} = 1 + 0.5 \sin\paren{4 \pi s X_i / K} + V_{i,s}$,
where $V_{ik}$ is uniform on $[-0.1, 0.1]$, $X_i$ is the random variable used to construct the edges, and $K$ is the dimension of the model space.
The periodic part of the potential outcome coefficients, $0.5 \sin\paren{4 \pi k X_i / K}$, is such that it goes from $-0.5$ to $0.5$ for the different basis functions with a frequency decided by $X_i$.
This part is constant when $X_i = 0$, and it completes two periods over the basis functions when $X_i = 1$.
This means that units with similar values of $X_i$ will have similar potential outcome functions.
The independent component $V_{ik}$ ensures that no units have identical potential outcome functions.
All random variables used to construct the graph and potential outcome functions, such as $X_i$ and $V_{ik}$, are drawn once and kept fixed between Monte Carlo rounds in the simulation.

The QQ plots of the estimator are given below in Figures~\ref{fig:qq-3-3}, \ref{fig:qq-4-3}, and \ref{fig:qq-4-4}.
In each of these figures, we see that the normal approximation becomes more appropriate as $n$ increases.
We can also see that the normal approximation does not hold as well for $n=100$ when $d=4$, $r=4$.
This reflects the fact that the model space is largest in this setting, so the Riesz representor is therefore more complex.

\subsection{Spatial Point Process Experiments} \label{sec:supp-spatial-experiments}

Our second simulation implements a point process experiment with the model space described in Section~\ref{sec:worked-spatial-point-process}.
We take the spatial region to be the unit square: $\mathcal{X} = [0,1]^2$.
We use a Poisson point process where treatment points are drawn uniformly on $\mathcal{X}$ and the number of points $m$ is fixed.
The $n$ outcome units are arranged on an equally spaced grid covering $\mathcal{X}$.
As above, we run three sample sizes $n$, being set to $10^2$, $10^3$ and $10^4$.
We set the bandwidth to $\sigma = 1 / \sqrt{n}$ and the number of treatment points to $m = n$, ensuring that observed outcomes between neighboring units are strongly correlated also for large $n$.

The coefficients of the potential outcome functions are set as
$$
\alpha_{i}^{(0)} = a_i,
\qquad
\alpha_{i}^{(1)} = 1 - b_i,
\qquadand
\alpha_{i}^{(2)} = \sin(4 a_i \pi) \cos (4 b_i \pi),
$$
where $a_i = \norm{u_i} / \sqrt{2}$ is the normalized distance from the unit's location to the origin and $b_i = \sqrt{2} \norm{u_i - (0.5, 0.5)}$ is the normalized distance to the center of the square.
We use these coefficients because they introduce sufficient heterogeneity between the units to make the estimation problem challenging.
The casual effect of interest is the coefficient $\alpha_{i}^{(2)}$ associated with the second interaction function.

We use numerical integration methods to construct the Riesz point and variance estimators.
There are no closed form expression for the moments used in the construction of the Riesz representors, so numerical approaches are necessary, and they will introduce slight approximation errors.
This setting is one in which the variance in principle can be estimated without bias, so a variance bound as in the previous subsection is not needed.
However, we modify the variance estimator slightly by only estimating covariance terms for pairs of units that are close to each other: $\norm{u_i - u_j} \leq 2 \sigma \sqrt{\log(1 / \sigma)}$.
Units that are far from each other in $\mathcal{X}$ will have negligible covariance, and including these terms in the estimator will introduce imprecision in the variance estimator and greatly increase the computational time.
While these excluded covariance terms are all close to zero, they are also all negative, so omitting them will introduce a slight positive bias of the variance estimator.
The choice of cutoff for estimating the covariance terms in this setting can therefore be seen as a trade-off between bias on the one hand and variance and computational time on the other hand.
We can make the bias arbitrary small by including more covariance terms.

{
\begin{table}[ht]
	\centering
	\caption{Simulation Results: Point Process Experiments}\label{tab:ppe-res}
	\begin{tabular}{rrrrrrrr}
		\toprule
		$n$ & MSE & Bias & $\Varsym$ & $\evb$ & CI $\Varsym$ & CI $\evb$ & Width \\ \midrule
		\input{\simspath/ppe/ppe-table-100000-samples.tex}\\ \bottomrule
	\end{tabular}
\end{table}
}

Table~\ref{tab:ppe-res} presents the results from the simulation study based on $100,000$ Monte Carlo rounds at each sample size.
The table follows a similar structure as in the previous subsection.
The first column gives the sample size.
The column labeled ``MSE'' gives the mean squared error normalized by the second moment of the outcomes, and the ``Bias'' and ``$\Varsym$'' columns give the squared bias and variance normalized by the MSE.
We find that the effect estimator is unbiased and that the mean square error decreases roughly proportionally with the sample size, which is expected because the number of neighbors that are strongly correlated with any unit remains constant as $n$ grows in these simulations.

Properties of the variance estimator and confidence intervals are presented in the remaining columns.
Column ``$\evb$'' gives the expectation of the variance estimator normalized by the variance: $\E{\evb} / \Var{\est}$.
The slight bias introduced by the covariance term cutoff discussed above is shown here.
Columns ``CI $\Varsym$'' and ``CI $\evb$'' give the coverage of Wald-type intervals using the true and estimated variance, respectively.
The confidence intervals based on the true variance covers at the nominal rate, indicating that the normal approximation is appropriate also in this setting.
The confidence intervals based on the estimated variance severely undercovers when $n$ is small, and slightly overcovers when $n$ is large.
The overcoverage for large $n$ is explained by the slight positive bias of the variance estimator, as shown in column $\evb$.
The undercoverage for small $n$ is explained by variability of the variance estimator itself.
When $n$ is small, many pairs of units are strongly correlated in this setting, so the variance estimator will be imprecise, even if it is close to unbiased.
And the imprecision of the variance estimator affects the coverage rates.
Put differently, Wald-type intervals rely on the convergence of the variance estimator, and the variance estimator is not sufficiently stable in this setting when $n = 100$.
However, this is resolved for the larger sample size, as evident from the remaining simulation results.
Finally, the ``Width'' column gives the relative width of the Wald-type intervals, showing that the width decreases with the sample size as expected.

%% file: sims/ppe/ppe-table-100000-samples.tex
  100  &  0.043  &  0.000  &  1.000  &  1.068  &  0.951  &  0.740  &  1.000 \\ 
 1000  &  0.007  &  0.000  &  1.000  &  1.074  &  0.946  &  0.943  &  0.570 \\ 
10000  &  0.001  &  0.000  &  1.000  &  1.081  &  0.948  &  0.959  &  0.211